\documentclass[article,prc,epsfig,floatfix,nofootinbib,aps]{revtex4}
\usepackage{epsfig}
\usepackage{epsf}
\begin{document}

\title{Cold Nuclear Matter Effects on $J/\psi$ and $\Upsilon$ Production
at the LHC}

\author{R. Vogt}

\address{{
Lawrence Livermore National Laboratory, Livermore, CA 94551, USA \break
and \break
Physics Department, University of California at Davis, Davis, CA 95616, 
USA} \break}

\begin{abstract}
The charmonium yields are expected to be 
considerably suppressed if a deconfined medium is formed in high-energy 
heavy-ion collisions.  In addition, the bottomonium states, with the 
possible exception
of the $\Upsilon(1S)$ state, are also expected to be suppressed in heavy-ion
collisions.  However, in proton-nucleus collisions the
quarkonium production cross sections, even those of the $\Upsilon(1S)$, 
are also suppressed.  These ``cold nuclear
matter'' effects need to be accounted for before signals of the high
density QCD medium can be identified in the measurements made in
nucleus-nucleus collisions.
We identify two cold nuclear matter effects important for
midrapidity quarkonium production: ``nuclear absorption'', typically 
characterized as a final-state effect on the produced quarkonium state and
shadowing, the modification of the parton densities in nuclei relative to the
nucleon, an initial-state effect.  We characterize these effects and study
the energy, rapidity, and impact-parameter dependence of initial-state 
shadowing in this paper.
\end{abstract}

\maketitle

\section{Baseline Total Cross Sections}

To better understand quarkonium suppression, it is necessary to have a good
estimate of the expected yields.  However, there are still a number of 
unknowns about quarkonium production in the primary
nucleon-nucleon interactions.  In this section, we discuss models of 
quarkonium production and give predictions for the yields in a number of 
collision systems.

\begin{table}[htbp]
\begin{center}
\begin{tabular}{|ccc|ccc|ccc|c|} \hline
$A$ & $E_A$ (TeV) & $y_A$ & $\sqrt{s_{_{NN}}}$ (TeV) & $y_{\rm diff}^{pA}$ &
$\Delta y_{\rm cm}^{pA}$  & $\sqrt{s_{_{NN}}}$ (TeV) & 
$y_{\rm diff}^{{\rm d}A}$ & $\Delta y_{\rm cm}^{{\rm d}A}$ 
& $\sqrt{s_{_{NN}}}$ (TeV) \\ \hline
& & & \multicolumn{3}{|c|}{$p+A$} & \multicolumn{3}{c|}{d$+A$} & $A+A$ \\ \hline
O  & 3.5  & 8.92 & 9.9  & 0.690 & 0.345 & 7 & 0 & 0 & 7 \\
Ar & 3.15 & 8.81 & 9.39 & 0.798 & 0.399 & 6.64 & 0.052 & 0.026 & 6.3  \\
Kr & 3.07 & 8.79 & 9.27 & 0.824 & 0.412 & 6.48 & 0.077 & 0.038 & 6.14 \\
Sn & 2.92 & 8.74 & 9.0  & 0.874 & 0.437 & 6.41 & 0.087 & 0.043 & 5.84 \\
Pb & 2.75 & 8.67 & 8.8  & 0.934 & 0.467 & 6.22 & 0.119 & 0.059 & 5.5  \\ \hline
\end{tabular}
\end{center}
\caption[]{For each ion species at the LHC, we give
the maximum beam energy per nucleon
and the corresponding beam rapidity.  Using the maximum proton or deuteron
beam energy: $E_p = 7$ TeV and $y_p = 9.61$; $E_{\rm d} = 3.5$ TeV and 
$y_{\rm d} = 8.92$ respectively, we present the maximum center-of-mass
energy per nucleon; rapidity difference, $y_{\rm diff}^{iA} = y_i - y_A$ 
($i = p,$ d); and 
center-of-mass rapidity shift, $\Delta y_{\rm cm}^{iA} = y_{\rm diff}^{iA}/2$, 
for $p+A$, d$+A$ and $A+A$ collisions.  Note that there is no rapidity shift in
the symmetric $A+A$ case.}
\label{delytable}
\end{table}

Since the LHC can collide either symmetric ($A+A$) or asymmetric ($A+B$) 
systems, we present results for $p+p$, $p+A$, d$+A$ and $A+A$ collisions.  
We consider d$+A$ collisions since the d$+A$ center-of-mass energy is closer 
to the $A+A$ collision
energy than top energy $p+A$ collisions.  The maximum ion beam energy per 
nucleon is the
proton beam energy, $E_p = 7$ TeV, times the charge-to-mass ratio, $Z/A$, of
the ion beam.  Thus the maximum deuteron beam energy is half that of the proton
beam, $E_{\rm d} = 3.5$ TeV.  The ion beam energies are given on the left-hand 
side of Table~\ref{delytable} for five reference nuclei: oxygen, $^{16}_8$O;
argon, $^{40}_{18}$Ar; krypton, $^{84}_{36}$Kr; tin, $^{119}_{50}$Sn; and lead,
$^{208}_{82}$Pb.  Note that we use the average elemental $A$ since a sample may
contain an admixture of several isotopes of different $A$.

In addition to the $A+A$ center-of-mass energy, we also show the maximum 
$p+A$ and
d$+A$ per nucleon center-of-mass energies, $\sqrt{s_{_{NN}}} = \sqrt{4 E_{p, \,
{\rm d}} E_A}$.  Because $E_{p, \, {\rm d}}$ is typically greater than $E_A$,
the center-of-mass rapidity can shift away from $y=0$.  The total shift is
$y_{\rm diff}^{iA} = y_i - y_A$ ($i = p,$ d) 
while the center of mass shifts by half
this amount, $\Delta y_{\rm cm}^{iA} = y_{\rm diff}^{iA}/2$.  
Table~\ref{delytable} shows
the maximum nucleon-nucleon center-of-mass energy per nucleon, the rapidity 
difference between the two beams, $y_{\rm diff}^{iA}$, 
and the center-of-mass shifts
for $p+A$ and d$+A$ collisions.  (The $Z/A$ ratio is the same for d and O thus
$\Delta y^{\rm dO} = 0$.)  Only $\sqrt{s_{_{NN}}}$ is given
for symmetric $A+A$ collisions since there is no rapidity shift.

If there were no cold nuclear matter effects on the production cross sections
at a given energy, the per nucleon cross sections would all be equal.  However,
the nuclear parton distributions (nPDFs) are known to be modified with respect
to the free proton PDFs as a function of parton momentum fraction $x$.  At 
low $x$, $x < 0.05$ (shadowing region), and high $x$, $x > 0.2$ (EMC region), 
the nuclear structure function, $F_2^A(x)$, the weighted sum of the charged
parton distributions,
is suppressed relative to that of the deuteron, $F_2^{\rm d}(x)$, while, 
in the intermediate $x$ region, the ratio $2F_2^A/AF_2^{\rm d}$ is enhanced 
(antishadowing) in nuclear deep-inelastic scattering (nDIS).  We refer to the 
modification of the parton densities in the nucleus as a function of $A$, $x$
and $\mu^2$ in general as shadowing.  While a 
combination of nDIS and Drell-Yan data can separate the nuclear valence and
sea quark densities, there is no direct probe of the nuclear gluon density,
rather it is inferred from the $\mu^2$ scaling violation. 

\begin{figure}[htbp] 
\centering
\resizebox{0.48\textwidth}{!}{\rotatebox{0}{%
\includegraphics*{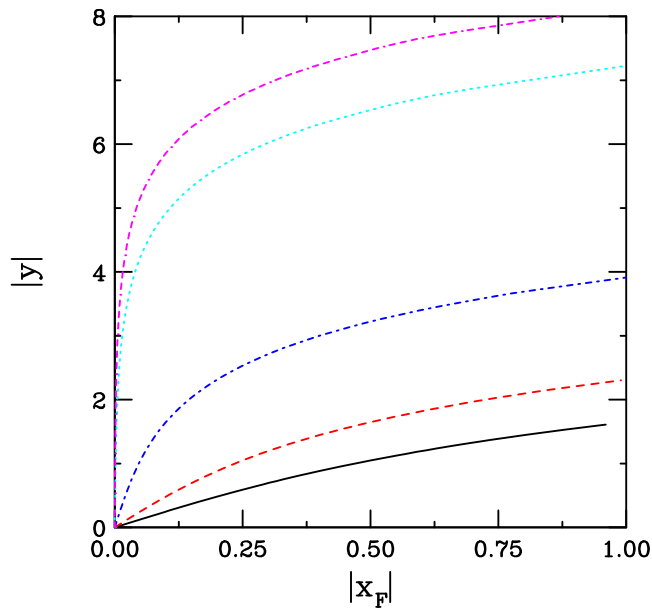}}}
\vglue-2mm
\caption[]{(Color online) For $M = 4$ and $\sqrt{s_{_{NN}}}
= 20$ (solid), 40 (dashed), 200 (dot-dashed), 5500 (dotted) and 14000 
(dot-dot-dot-dashed) GeV, 
we give the average value of the center-of-mass rapidity, $y$, 
in $p+p$ collisions as a 
function of $x_F$. Note the absolute values on $y$ and $x_F$: in the 
center-of-mass frame, $|x_F|<1$ and $|y| \leq y_{\rm max}$ so that the curves 
shown here are diagonally reflected around $x_F = y = 0$.}
\label{xfy}
\end{figure}

Gluon fusion dominates quarkonium production up to $x_F \sim 0.7$ already
at fixed-target energies \cite{rv866}, including over the entire accessible
rapidity range at the LHC, see Fig.~\ref{xfy}.  
Thus while the modification of the gluon distributions in nuclei is the most
important for quarkonium studies, it is unfortunately the most poorly known.
There are, however, a number of indirect constraints on the gluon density.
The scale evolution of $F_2^A$ and momentum conservation provide two important 
constraints.  Most of the low-$x$ nDIS data are at relatively low scales, below
the minimum scale of a number of PDF sets and therefore less useful for
studies of perturbative evolution.  RHIC data on hadron production are an
exception since intermediate $p_T$ hadron production occurs at relatively
low $x$ and at perturbative scales.  
At relatively high $x$, the shape of the PHENIX midrapidity $\pi^0$ data
\cite{PHENIXpi0} helps pin down the nuclear gluon density in the EMC region.

Quarkonium production occurs at sufficiently large scales to provide further 
constraints on the nuclear gluon PDFs.  There are some drawbacks however:
the quarkonium production mechanism is not fully understood, even in $p+p$
collisions, and the energy dependence of nuclear absorption is not
well known.  In the remainder of this section, we discuss the quarkonium
yields in various collision systems; the implementation of modified PDFs for
the nuclear parton densities; and quarkonium absorption by nucleons.

Early studies of high energy quarkonium production, particularly at high $p_T$,
were performed in the context of the
color singlet model (CSM) which calculates direct production of
a quarkonium state with definite total spin, parity and charge conjugation.
The CSM predicted that the $\chi_{c1}$ state, produced directly from $gg$ 
fusion, would have a much larger cross section than direct color-singlet
$J/\psi$ production which requires a 3-gluon vertex \cite{Baier}.  Instead, 
measurements of direct $J/\psi$ and
$\chi_c$ production showed that the $J/\psi$ cross section was, in fact,
larger than the $\chi_c$ cross section
\cite{Tevatron}.  However, the CSM can describe charmonium
production in cleaner environments such as photoproduction \cite{Klasen}
and more recent formulations of the CSM \cite{artoisenet}
can reproduce the magnitude of the $J/\psi$ data at RHIC \cite{Brodsky:2009cf}.
The different kinematics of the modified CSM probes larger values of $x$
and thus reduces the shadowing effect \cite{lansberg2}.

Nonrelativistic QCD (NRQCD) is an effective field theory
in which short-distance partonic interactions produce $Q \overline Q$ 
pairs in color singlet or color octet states which then evolve into a quarkonium
state, as characterized by nonperturbative matrix 
elements \cite{bodwin}.  The first term in the NRQCD
expansion is equivalent to the CSM.  The octet contributions 
are sufficient to explain the $J/\psi$ yield at the Tevatron.  
However, the NRQCD approach has so far failed to 
describe quarkonium polarization \cite{polar}.

Perhaps the simplest approach to quarkonium production is the color 
evaporation model (CEM) which treats
heavy flavor and quarkonium production on an equal footing.  The quarkonium 
production cross section is some fraction, $F_C$, of 
all $Q \overline Q$ pairs below the $H \overline H$ threshold where $H$ is
the lowest mass heavy-flavor hadron.  Thus the CEM cross section is
simply the $Q \overline Q$ production cross section with a cut on the pair mass
but without any contraints on the 
color or spin of the final state. The color of the
octet $Q \overline Q$ state is
`evaporated' through an unspecified process which does not change the momentum.
The additional energy needed to produce
heavy-flavored hadrons when the partonic center of mass energy, 
$\sqrt{\hat s}$, is less than $2m_H$, the $H \overline H$
threshold energy, is nonperturbatively obtained from the
color field in the interaction region.
Thus the quarkonium yield may be only a small fraction of the total $Q\overline 
Q$ cross section below $2m_H$.
At leading order, the production cross section of quarkonium state $C$ in
an $A+B$ collision is
\begin{eqnarray}
\frac{d\sigma_C^{\rm CEM}(s_{_{NN}})}{d^2r d^2b} & = & F_C \sum_{i,j} 
\int_{4m_Q^2}^{4m_H^2} d\hat{s}
\int dx_1 \, dx_2~ \int dz'\, dz \nonumber \\
&  & \mbox{} \times f_i^A(x_1,\mu^2,\vec{r},z)~
f_j^B(x_2,\mu^2,\vec{b} - \vec{r}, z')~ 
\hat\sigma_{ij}(\hat{s})~\delta(\hat{s} - x_1x_2s_{_{NN}})\, 
\, , \label{sigtil}
\end{eqnarray} 
where $A$ and $B$ can be any hadron or nucleus,
$ij = q \overline q$ or $gg$ and $\hat\sigma_{ij}(\hat s)$ is the
$ij\rightarrow Q\overline Q$ subprocess cross section.  
If one or both of the collision partners, $A$ and $B$, is a proton, 
then the transverse, $\vec{r}$, and longitudinal, $z$, spatial parameters 
may be replaced by delta functions, $\int d^2r dz \delta(\vec r) \delta (z)$, 
and the parton densities are simply $f_i^A(x_1,\mu^2,\vec{r},z) \equiv 
f_i^A(x_1,\mu^2)$.
Our calculations use the NLO $Q \overline Q$ code of Mangano {\it et al.} 
\cite{MNRcode} with
the $2m_H$ mass cut in Eq.~(\ref{sigtil}) and use the same 
parameters as in Refs.~\cite{hvqyr,rhicii} with the MRST parton densities
\cite{mrst}, optimized to obtain agreement with the $Q \overline Q$ cross
section, as described in Ref.~\cite{Gavai:1994in}.  
The factor $F_C$ can also be fit with other parton densities such as CTEQ6M
\cite{cteq}.  When the same mass and scale parameters are used, the energy
dependence of the cross section is very similar, see {\it e.g.}
Ref.~\cite{rhicii}.

All these formulations: CSM; NRQCD; and CEM assume the validity of collinear
factorization which relies on the separation of the initial and final states.
Collinear factorization was proven to be effective at all orders for 
the Drell-Yan process some time ago \cite{CSSdy}.  A subsequent paper by
Collins, Soper and Sterman showed that the factorization process was correct
for heavy flavor production up to corrections of order $(1/M)$
\cite{CSS}.  Thus while
higher-order corrections to the charm cross section are large, collinear
factorization is generally assumed to hold and, indeed, the scale dependence
of the approximate NNLO-NNLL charm cross section is seen to stabilize and
the next-order corrections are not as large \cite{nickrvcharm}.  Higher-twist
effects that might signify factorization breaking, such as intrinsic charm,
are generally most important at forward rapidities in the light-cone 
formulation \cite{intc,Pumplin}.  Factorization has not been strictly
proven for quarkonium where the final quarkonium state
may be connected to the initial state by soft gluons.  
In the CSM, the color singlet matrix
element is derived from quarkonium decays where the initial state plays no role.
Factorization is most difficult to prove for NRQCD.  It depends on the 
universality of the nonperturbative matrix elements.  However, recent works have
shown that a redefinition of these matrix elements allows factorization to be
restored \cite{qiu1,qiu2}.
The CEM is closest
in spirit to the calculation of the open heavy flavor cross section so that
collinear factorization should work equally well in the two approaches.
Since our calculations are in the CEM, we use collinear factorization to
calculate quarkonium production at the LHC. 

To go beyond $p+p$ collisions, 
the proton parton densities must be replaced by those of
the nucleus.  Then the collision geometry and the spatial dependence of the
shadowing parameterization also need to be considered.
We assume that if $A$ is a nucleus, the nuclear parton densities, 
$f_i^A(x_1,\mu^2,\vec r,z)$, factorize into
the nucleon density in the nucleus, $\rho_A(\vec{r},z)$, independent of the 
kinematics; the nucleon parton density, $f_i^p(x_1,\mu^2)$, independent of $A$;
and a shadowing ratio, $S^i_{{\rm P},{\rm S}}(A,x_1,\mu^2,\vec{r},z)$ that
parameterizes the modifications of the nucleon parton densities in the nucleus.
The first subscript, P, refers to the choice of shadowing parameterization,
while the second,  S, refers to the spatial dependence.
Thus,
\begin{eqnarray}
f_i^A(x_1,\mu^2,\vec{r},z) & = & \rho_A(s) 
S^i_{{\rm P},{\rm S}}(A,x_1,\mu^2,\vec{r},z) f_i^p(x_1,\mu^2) \, \, ,
\label{fanuc} \\ 
f_j^B(x_2,\mu^2,\vec{b} - \vec{r},z') & = & \rho_B(s') 
S^j_{{\rm P},{\rm S}}(B,x_2,\mu^2,\vec{b} - \vec{r},z') f_j^p(x_2,\mu^2) 
\, \, ,  \label{fbnuc} 
\end{eqnarray}
where $s = \sqrt{r^2 + z^2}$ and $s' = \sqrt{|\vec{b} - \vec{r}|^2 + z'^2}$.  

The nucleon densities of the heavy
nucleus are assumed to be Woods-Saxon distributions 
\cite{Vvv} and are normalized 
so that $\int d^2r dz \rho_A(s) = A$.  With no nuclear modifications, 
$S^i_{{\rm P},{\rm S}}(A,x,Q^2,\vec{r},z)\equiv 1$ and integration of the
nuclear parton densities over
the spatial variables gives
\begin{eqnarray}
\int d^2b \, d^2s \, dz \, dz' \,
f_i^A(x_1,\mu^2,\vec{r},z) f_j^B(x_2,\mu^2,\vec{b} - 
\vec{r},z') = AB  f_i^p(x_1,\mu^2) f_j^p(x_2,\mu^2) \, \, .
\end{eqnarray}  
The impact-parameter averaged
shadowing parameterization measured in nDIS is recovered by integrating 
$S_{{\rm P},{\rm S}}$ over the volume, weighted by the nuclear density,
\begin{eqnarray}
\frac{1}{A} \int d^2r dz \rho_A(s)
S^i_{{\rm P},{\rm S}}(A,x,\mu^2,\vec{r},z) = S^i_{\rm P}(A,x,\mu^2) \, \, . 
\label{snorm}
\end{eqnarray}
We discuss more details of the spatial dependence of $S_{{\rm P},{\rm S}}$ in 
Section~2.3.  Most available shadowing parameterizations, 
including the ones used here, 
ignore the small effects in deuterium.  However, we take the proton and neutron
numbers of both nuclei into account. The impact-parameter integrated
up and down quark distributions, 
needed for the $q \overline q$ 
contribution to quarkonium production, are calculated as 
\begin{eqnarray} f_q^A(x,\mu^2) =
(Z_A S_{{\rm P}\, p}^q(A,x,\mu^2) f_q^p(x,\mu^2) + N_A 
S_{{\rm P}\, n}^q(A,x,\mu^2) f_q^n(x,\mu^2)) 
\end{eqnarray}
for $q = u$ and $d$, assuming that, as for the proton and neutron parton 
densities, $S_{{\rm P}\, n}^u = S_{{\rm P}\, p}^d$ and 
$S_{{\rm P}\, n}^d = S_{{\rm P}\, p}^u$ and similarly for 
the antiquarks.  

To obtain the rapidity distribution from the total cross section, 
an additional delta function, $\delta (y - 
0.5\ln(x_1/x_2))$, is included in Eq.~(\ref{sigtil}).  At leading order, the
parton momentum fractions $x_1$ and $x_2$ are simply $x_{1,2} = (\sqrt{\hat{s}/
s_{_{_{NN}}}}) \exp(\pm y)$.  In this notation then, in the forward rapidity
region of a $p+A$ collision,
$x_1$, the proton momentum fraction, is larger and $x_2$, the parton momentum
fraction in the nucleus, is smaller than the midrapidity value, 
$x = \sqrt{\hat{s}/s_{_{_{NN}}}}$.

Some of the uncertainties in the production model may be overcome by studying
ratios, {\it e.g.} $(p+A)/(p+p)$, at the same center-of-mass
energy since the dominance of $gg$
processes means that the $(p+A)/(p+p)$ ratio is, to a good approximation, the
ratio of the gluon distribution in the nucleus relative to the gluon 
distribution in the proton.  We have chosen to use the CEM because it allows
predictions of the total cross section and the $p_T$-integrated rapidity
distributions where nuclear effects are more prominent.  Measuring the $J/\psi$
and $\Upsilon$ ratios simultaneously also provides a means of determining the 
scale evolution of the nuclear gluon distribution at relatively large,
perturbative scales if shadowing is the only cold nuclear matter effect in
$p+A$ and d$+A$ collisions.

At fixed-target energies, the $x_F$ dependence clearly shows that 
shadowing is not the only contribution to the $J/\psi$ nuclear 
dependence as a function of $x_F$ \cite{e866,HERA-Bnew}.  
Indeed, the characteristic decrease 
of $\alpha(x_F)$ for $x_F \geq 0.25$ cannot be explained
by shadowing alone \cite{rv866}.  In fact, the data so far suggest
approximate scaling with $x_F$, not the target momentum fraction $x_2$ 
\cite{Mike}, 
indicating the possible importance of higher-twist effects \cite{HVS}.
The preliminary PHENIX data show an increasing suppression at forward
rapidity \cite{Tony}, similar to that seen in fixed-target experiments at
large $x_F$.

Effects we have not considered here which may result in $x_F$ rather than
$x_2$ scaling and affect the high $x_F$ region are energy loss in cold matter
and intrinsic charm, both discussed extensively in Ref.~\cite{rv866}.
We do not consider these effects here because, at heavy-ion colliders, the 
relationship between $x_F$, rapidity, and $\sqrt{s_{_{NN}}}$ suggests that this
interesting $x_F$ region is pushed to far forward rapidities.  The onset of
initial-state energy loss should, in fact, appear at higher $x_F$ at larger
$\sqrt{s_{_{NN}}}$ if it depends on the momentum fraction $x_1$.
Figure~\ref{xfy} shows the relationship between $x_F$ and $y$ in the 
center-of-mass frame for $M = 4$ GeV and $20 \leq \sqrt{s_{_{NN}}} \leq
14000$ GeV.
Since $x_F = (2m_T /\sqrt{s_{_{NN}}})\sinh y$, 
the large center-of-mass energies at the LHC guarantees
that the forward $x_F$ region will not be accessible in the central rapidity
region of the LHC.  Instead, the $x_F$ distribution becomes narrowly peaked
with increasing energy while the rapidity distribution becomes broad and flat.
At $y=5$, the largest $x_F$ accessible (at the lowest
$\sqrt{s_{_{NN}}}$) is 0.081 for the $J/\psi$ and 0.25 for the $\Upsilon$.
The large $x_F$ region is therefore not probed by quarkonium production in 
$|y| \leq 5$.
Thus shadowing and absorption are likely the most
important cold nuclear matter effects at the LHC.

To implement nuclear absorption on quarkonium
production in
$p+A$ and d$+A$ collisions, the production cross section 
is weighted by the
survival probability, $S^{\rm abs}_C$, so that 
\begin{eqnarray} 
S^{\rm abs}_C(\vec b - \vec s,z^\prime) = \exp \left\{
-\int_{z^\prime}^{\infty} dz^{\prime \prime} 
\rho_A (\vec b - \vec s,z^{\prime \prime}) 
\sigma_{\rm abs}^C(z^{\prime \prime} - z^\prime)\right\} \, \, 
\label{nsurv} 
\end{eqnarray}
where $z^\prime$ is the longitudinal production point, 
as in Eq.~(\ref{fbnuc}), and 
$z^{\prime \prime}$ is the point at which the state is absorbed. 
The nucleon absorption cross section, $\sigma_{\rm abs}^C$, typically 
depends on the spatial location at which the
state is produced and how far it travels through the medium.

We could also consider absorption by comover interactions but this cross 
section is typically smaller than the nuclear absorption cross section.
In addition, in $A+A$ collisions, the higher temperatures and larger particle
densities would rule out hadronic comovers in the early stages.  Therefore,
we do not consider hadronic comovers as a cold matter effect in this paper.

If absorption alone is active, {\it i.e.}\ $S^i_{{\rm P}, \, 
{\rm S}}(A,x,\mu^2,\vec r,z) \equiv 1$, then an effective minimum bias 
$A$ dependence is obtained after integrating
Eqs.~(\ref{sigtil}) and (\ref{nsurv}) over the spatial coordinates.
If $S^{\rm abs}_C = 1$ also, $\sigma_{pA} \approx A \sigma_{pp}$ 
without any cold
nuclear matter effects.  (Note that for $gg$-dominated processes, such as
quarkonium production,
the relationship would be exact.  When $q \overline q$ or 
$q \overline q'$ interactions 
dominate, as in gauge boson production, the different 
relative proton and neutron numbers
make the above relationship approximate.)  If $S^i_{{\rm P}, \, 
{\rm S}}(A,x,\mu^2,\vec r,z) \equiv 1$ and $S^{\rm abs}_C \neq 1$, 
$\sigma_{pA} = A^\alpha \sigma_{pp}$ where the
exponent $\alpha$ can be related to the absorption cross section, as studied 
in detail for $J/\psi$ and $\psi'$ production by NA50 \cite{NA50psipsip}.
For a constant $\sigma_{\rm abs}^C$ with a sharp surface spherical nucleus of
density $\rho_A = \rho_0 \theta(R_A - b)$, it can be shown that
\begin{eqnarray}
\alpha = 1 - \frac{9\sigma_{\rm abs}^C}{16 \pi r_0^2} \label{alfdef}
\end{eqnarray} 
where $r_0 = 1.2$ fm \cite{RVphysrep}.  The relationship between $\alpha$
and $\sigma_{\rm abs}^C$ is less straightforward in more realistic geometries.

The NA50 \cite{NA50psipsip} and E866 \cite{e866} experiments measured 
a non-negligible difference in the effective $J/\psi$ and
$\psi'$ absorption cross sections at $\sqrt{s_{_{NN}}} = 23-29$ GeV and
$\sqrt{s_{_{NN}}} = 38.8$ GeV respectively.  In addition, the difference
between $\sigma_{\rm abs}^{J/\psi}$ and $\sigma_{\rm abs}^{\psi'}$ seems to 
decrease with $\sqrt{s_{_{NN}}}$.  The NA50 collaboration
measured $\Delta \sigma = \sigma_{\rm abs}^{\psi'} - 
\sigma_{\rm abs}^{J/\psi} = 4.2 \pm 1$ mb at 400 GeV and $2.8 \pm 0.5$ mb at
450 GeV \cite{NA50psipsip}.  At $x_F \sim 0$, the E866 results imply
$\Delta \alpha = \alpha_{J/\psi} - \alpha_{\psi'} < 0.2$ or, using
Eq.~(\ref{alfdef}), $\Delta \sigma < 1.6$ mb \cite{e866}.
This suggests that absorption is a 
final-state effect since an initial-state effect such as shadowing would not 
discriminate between the asymptotic $J/\psi$ and $\psi'$ final states.
Comparing the effective absorption cross sections determined at central 
rapidities from the CERN SPS to RHIC, absorption seems to decrease with 
energy \cite{LVW}.

Fewer $\Upsilon$ $p+A$ data are available.  The E772 experiment \cite{E772ups}
measured the $A$ dependence of the three $S$ states and found a reduced $A$
dependence relative to $J/\psi$ absorption.  The $A$ dependence of the three
$S$ states was indistinguishable within the uncertainties.  No $\Upsilon$ $A$
dependence was presented by the E866 collaboration.  The STAR d+Au/$p+p$ ratio
from RHIC suggests that, within large uncertainties, the $\Upsilon$ $A$ 
dependence is linear \cite{Haidong} and production is not significantly 
suppressed.
Thus absorption seems to be weaker overall for $\Upsilon$ production but there
is not clear indication so far of how much weaker it is or whether it has the
same energy dependence as the $J/\psi$.

If conventional shadowing parameterizations, such as the ones used in
this paper, are included, the effective $J/\psi$ absorption cross section 
may seem to decrease with energy due to the
increased effect of shadowing at low $x$.  
A decrease in absorption concurrent with
increased shadowing as $\sqrt{s_{_{NN}}}$ increases seems to 
approximately hold, even without shadowing, at fixed-target 
energies \cite{LVW}.  Such a decrease is consistent with the $J/\psi$
traversing the nucleus as a color singlet.  If the nuclear crossing time
is shorter than the $J/\psi$ formation time, the effective absorption
decreases with $\sqrt{s_{_{NN}}}$ as an ever smaller state passes through the
target.

If the effective absorption cross section indeed decreases with energy, 
then absorption should be a relatively
small contribution to the total $A$ dependence at the LHC.  This prediction is
easy to check: if absorption is negligible, the $J/\psi$ and 
$\psi'$ $(p+A)/(p+p)$ 
ratios should depend only on shadowing and should thus be equivalent.  The yield
is then related to the ratio of the nuclear to proton gluon densities
since $gg$ fusion dominates quarkonium production at these energies.
In this work, we have assumed that absorption is negligible so that the 
$(p+A)/(p+p)$, (d$+A)/(p+p)$ and $(A+A)/(p+p)$ $J/\psi$ and $\Upsilon$ ratios 
presented here are the same for all charmonium
and bottomonium states respectively.

If both the $p+A$ and $p+p$ data are taken at the same $\sqrt{s_{_{NN}}}$, the
same $x$ values of the gluon densities will be probed in the nucleus and
in the proton.  Such same energy comparison runs would be an excellent probe
of the nuclear gluon distributions because 
\begin{eqnarray} \frac{\sigma_{pA} (\sqrt{s_{_{NN}}})}{\sigma_{pp}
(\sqrt{s_{_{NN}}})} \propto \frac{1}{A}
\frac{f_g^A(x,\mu^2)}{f_g^p(x,\mu^2)} \, \, .  
\end{eqnarray}
However, if the $p+A$ and $p+p$ data are recorded
at different energies (and $x$ values), the extraction of the nuclear gluon
density is less straightforward since  
\begin{eqnarray}
\frac{\sigma_{pA}(\sqrt{s_{_{NN}}})}{\sigma_{pp}(\sqrt{s})} \propto \frac{1}{A}
\frac{f_g^A(x',\mu^2)}{f_g^p(x,\mu^2)} \, \, .
\end{eqnarray}

In both cases, the $p_T$-integrated ratios provide an additional uncertainty
because the scale evolution of the gluon density is not well known but is
expected to be strong 
\cite{Eskola:1998iy,Eskola:1998df,deFlorian:xxx,Hirai:xxx,Eskola:2008,Eskola:2009}.  
However, the quarkonium
$p_T$ distribution is steeply falling for $p_T \geq m$ so that the 
$p_T$-integrated ratios ratios are a good representation of $\mu^2 = \langle 
m_T \rangle^2$.  

The scale evolution of the gluon densities can
be probed in part by relative studies of low $p_T$ or $p_T$-integrated
$J/\psi$ ($m_\psi = 3.097$ GeV) and
$\Upsilon (1S)$ ($m_{\Upsilon(1S)} = 9.46$ GeV) production.  To more precisely
obtain the scale evolution of shadowing, it would be preferable to bin the
$J/\psi$ and $\Upsilon(1S)$ $(p+A)/(p+p)$ ratios in $p_T$.   One must be careful
in the interpretation of such ratios, particularly at $p_T < m$, since, at
fixed-target energies, the $p_T$-dependent $(p+A)/(p+p)$ ratios show that the
$J/\psi$ and $\Upsilon$ $p_T$ distributions are broader in $p+A$
than in $p+p$ interactions \cite{NA3,E772}.  This broadening has been
attributed to intrinsic parton $p_T$ kicks accrued by the interacting parton
as it traverses the nucleus before interacting \cite{GG,Pirner}.  The 
magnitude of the average $p_T$ kick increases with $A$ so that the 
$p_T$-dependent $(p+A)/(p+p)$ ratio is less than unity at low $p_T$ and increases 
above one with increasing $p_T$.  This effect is important at low center-of-mass
energies where the average $p_T$ of the produced quarkonium state is not large.
By LHC energies, while the $p_T$ kick may be rather small relative to
$\langle p_T^2 \rangle$, it may still
affect the analysis of shadowing effects in $p_T$-binned ratios but not
in $p_T$-integrated ratios.  We focus on the $p_T$-integrated results here and
will present $p_T$-dependent calculations elsewhere.

\begin{table}[htbp]
\begin{center}
\begin{tabular}{|cc||cccc||cc|} \hline
 & & \multicolumn{4}{c||}{$\sigma^{\rm dir}$/nucleon pair ($\mu$b)} &
\multicolumn{2}{c|}{$B \sigma^{\rm inc} AB$ ($\mu$b)} \\ \hline
System & $\sqrt{s_{_{NN}}}$ (TeV) & $J/\psi$ & $\chi_{c1}$ & 
$\chi_{c2}$ & $\psi'$ & $J/\psi$ & $\psi'$
\\ \hline
$p+p$   & 14   & 32.9 & 31.8 & 52.5 & 7.43 & 3.15    & 0.055 \\ \hline
$p+p$   & 10   & 26.8 & 26.0 & 43.3 & 6.06 & 2.57    & 0.044 \\ \hline
$p+p$   & 9.9  & 26.6 & 25.8 & 42.6 & 6.02 & 2.55    & 0.044 \\
$p+$O   & 9.9  & 23.8 & 23.0 & 38.0 & 5.37 & 36.5    & 0.632 \\ \hline
$p+p$   & 9.39 & 25.8 & 25.0 & 41.3 & 5.83 & 2.48    & 0.043 \\
$p+$Ar  & 9.39 & 22.0 & 21.2 & 35.1 & 4.96 & 84.1    & 1.46  \\ \hline
$p+p$   & 9.27 & 25.6 & 24.8 & 40.9 & 5.79 & 2.46    & 0.043 \\
$p+$Kr  & 9.27 & 20.9 & 20.2 & 33.4 & 4.73 & 168.4   & 2.92  \\ \hline
$p+p$   & 9    & 25.2 & 24.4 & 40.2 & 5.69 & 2.41    & 0.042 \\
$p+$Sn  & 9    & 20.2 & 19.6 & 32.3 & 4.56 & 230.4   & 3.99  \\ \hline
$p+p$   & 8.8  & 25.0 & 24.2 & 39.9 & 5.65 & 2.40    & 0.042 \\
$p+$Pb  & 8.8  & 19.5 & 18.9 & 31.1 & 4.40 & 388.8   & 6.75  \\ \hline
$p+p$   & 7    & 21.8 & 21.1 & 34.9 & 4.93 & 2.09    & 0.036 \\
$p+$O   & 7    & 19.5 & 19.0 & 31.3 & 4.42 & 30.0    & 0.520 \\
d+O     & 7    & 19.5 & 19.0 & 31.3 & 4.42 & 60.0    & 1.04  \\
O+O    & 7    & 17.6 & 17.0 & 28.1 & 3.98 & 432.4   & 7.51  \\ \hline
$p+p$   & 6.64 & 21.2 & 20.5 & 33.8 & 4.78 & 2.02    & 0.035 \\ 
d+Ar    & 6.64 & 18.1 & 17.5 & 28.9 & 4.09 & 138.5   & 2.39  \\ \hline
$p+p$   & 6.48 & 20.9 & 20.2 & 33.3 & 4.71 & 2.00    & 0.034 \\
d+Kr    & 6.48 & 17.2 & 16.6 & 28.0 & 3.95 & 281.3   & 4.85  \\ \hline
$p+p$   & 6.41 & 20.7 & 20.1 & 33.1 & 4.68 & 1.98    & 0.034 \\
d+Sn    & 6.41 & 16.8 & 16.2 & 26.8 & 3.78 & 378.3   & 6.52  \\ \hline
$p+p$   & 6.3  & 20.5 & 19.9 & 32.8 & 4.63 & 1.97    & 0.034 \\
$p+$Ar  & 6.3  & 17.6 & 17.0 & 28.1 & 3.97 & 67.3    & 1.17  \\ 
Ar+Ar  & 6.3  & 15.0 & 14.5 & 23.9 & 3.38 & 2300    & 40.0  \\ \hline
$p+p$   & 6.22 & 20.4 & 19.7 & 32.5 & 4.60 & 1.95    & 0.34  \\ 
d+Pb    & 6.22 & 16.0 & 15.5 & 25.6 & 3.62 & 637.3   & 10.98 \\ \hline
$p+p$   & 6.14 & 20.2 & 19.6 & 32.3 & 4.56 & 1.94    & 0.034 \\
$p+$Kr  & 6.14 & 16.6 & 16.1 & 26.6 & 3.76 & 134.0   & 2.32  \\
Kr+Kr  & 6.14 & 13.7 & 13.2 & 21.8 & 3.08 & 9245    & 160.6 \\ \hline
$p+p$   & 5.84 & 19.6 & 19.0 & 31.3 & 4.42 & 1.88    & 0.033 \\
$p+$Sn  & 5.84 & 15.9 & 15.4 & 25.4 & 3.59 & 181.3   & 3.14  \\
Sn+Sn  & 5.84 & 12.8 & 12.4 & 20.4 & 2.89 & 17391   & 302.0 \\ \hline
$p+p$   & 5.5  & 18.9 & 18.3 & 30.2 & 4.26 & 1.81    & 0.032 \\
$p+$Pb  & 5.5  & 14.9 & 14.4 & 23.8 & 3.37 & 297.6   & 5.16  \\
Pb+Pb  & 5.5  & 11.7 & 11.3 & 18.7 & 2.64 & 48500   & 842   \\ \hline
\end{tabular}
\end{center}
\caption[]{The direct cross section per nucleon pair (central columns) and 
the dilepton yield per nucleon multiplied by $AB$.  The results are given for 
the MRST PDFs \cite{mrst} with $m_c = 1.2$ GeV, $\mu_F = \mu_R = 2m_T$.}
\label{psisigs}
\end{table}

As an example of the possible cross sections for quarkonium production at the
LHC, we present the total cross sections in $p+p$, $p+A$, d$+A$ and $A+A$ collisions
at the relevant energies.  To illustrate the effects of shadowing on the total
cross section, calculated to next-to-leading order in the CEM 
\cite{Gavai:1994in}, we use the EKS98 parameterization 
\cite{Eskola:1998iy,Eskola:1998df}.  For each possible maximum $p+A$, 
d$+A$ and $A+A$
center-of-mass energy, we also give the $p+p$ cross section at that same energy.
In addition, for the $A+A$ center-of-mass energies, we also give the $p+p$ and 
$p+A$ cross sections at that energy.  The results are given in 
Tables~\ref{psisigs} and \ref{upssigs}.  The central columns are the 
direct cross sections per nucleon pair for all charmonium and bottomonium 
states.  The effects are largest
for charmonium (lower $x$ and $\mu^2$ than the $\Upsilon$ states) and for 
the heaviest nuclei (lowest
energies -- highest $x$ -- but largest $A$).  
On the right-hand side of the tables, the inclusive (direct plus feed down)
cross sections are multiplied by the dilepton decay branching ratios.  
They are also multiplied by $AB$ to obtain the minimum bias total
cross sections.  

The approximate $A$ dependence of the total cross section relative to the
$p+p$ cross section at the same center-of-mass energy, assuming no other cold
matter effects, can be obtained from the $(AB)^\alpha$ parameterization so that,
per nucleon,
\begin{eqnarray}
\alpha(p+A/p+p) & \sim & 1 + \frac{\ln[f_g^A(x_2',\mu^2)/f_g^p(x_2,\mu^2)]}{\ln A}
\\
\alpha(A+B/p+p) & \sim & 1 + \frac{\ln[f_g^A(x_1',\mu^2)f_g^B(x_2',\mu^2)/
(f_g^p(x_1,\mu^2)f_g^p(x_2,\mu^2))]}{\ln (AB)}
\end{eqnarray}
where $x_2' = x_2$ and $x_1' = x_1$ if the center-of-mass energies are the same
for the two systems.
For $J/\psi$ production in $p$Pb and Pb+Pb collisions relative to $p+p$
collisions at 5.5 TeV, $\alpha \sim  0.76$ and 0.52 respectively.  In the
case of $\Upsilon$ production, we have $\alpha\sim 0.88$ and 0.76
respectively.

\begin{table}[htbp]
\begin{center}
\begin{tabular}{|cc||ccccc||ccc|} \hline
 & & \multicolumn{5}{c||}{$\sigma^{\rm dir}$/nucleon pair ($\mu$b)} &
\multicolumn{3}{c|}{$B \sigma^{\rm inc} AB$ ($\mu$b)} \\ \hline
System & $\sqrt{s_{_{NN}}}$ (TeV) & $\Upsilon$ & $\Upsilon'$ & 
$\Upsilon''$ & $\chi_b(1P)$ & $\chi_b(2P)$ & 
$\Upsilon$ & $\Upsilon'$ & $\Upsilon''$ 
\\ \hline
$p+p$   & 14   & 0.43 & 0.27 & 0.16  & 0.89 & 0.69 & 0.020 & 0.0074 & 0.0036 \\
\hline
$p+p$   & 10   & 0.33 & 0.21 & 0.12  & 0.70 & 0.54 & 0.016 & 0.0059 & 0.0028 
\\ \hline
$p+p$   & 9.9  & 0.32 & 0.20 & 0.12  & 0.66 & 0.51 & 0.015 & 0.0055 & 0.0026 \\
$p+$O   & 9.9  & 0.30 & 0.19 & 0.11  & 0.62 & 0.48 & 0.23  & 0.082  & 0.040  \\
\hline
$p+p$   & 9.39 & 0.30 & 0.19 & 0.12  & 0.63 & 0.49 & 0.014 & 0.0052 & 0.0025 \\
$p+$Ar  & 9.39 & 0.28 & 0.17 & 0.11  & 0.57 & 0.44 & 0.53  & 0.19   & 0.092  \\
\hline
$p+p$   & 9.27 & 0.30 & 0.19 & 0.11  & 0.62 & 0.48 & 0.014 & 0.0052 & 0.0025 \\
$p+$Kr  & 9.27 & 0.27 & 0.17 & 0.10  & 0.55 & 0.43 & 1.06  & 0.39   & 0.19   \\
\hline
$p+p$   & 9    & 0.29 & 0.18 & 0.11  & 0.61 & 0.47 & 0.014 & 0.0050 & 0.0024 \\
$p+$Sn  & 9    & 0.26 & 0.16 & 0.099 & 0.53 & 0.42 & 1.46  & 0.53   & 0.26   \\
\hline
$p+p$   & 8.8  & 0.29 & 0.18 & 0.11  & 0.60 & 0.47 & 0.014 & 0.0059 & 0.0024 \\
$p+$Pb  & 8.8  & 0.25 & 0.16 & 0.097 & 0.52 & 0.41 & 2.51  & 0.96   & 0.45   \\
\hline
$p+p$   & 7    & 0.23 & 0.15 & 0.090 & 0.48 & 0.38 & 0.011 & 0.0043 & 0.0019 \\
$p+$O   & 7    & 0.22 & 0.14 & 0.085 & 0.46 & 0.36 & 0.17  & 0.061  & 0.029  \\
d+O     & 7    & 0.22 & 0.14 & 0.085 & 0.46 & 0.36 & 0.34  & 0.12   & 0.058  \\
O+O    & 7    & 0.21 & 0.13 & 0.081 & 0.44 & 0.34 & 2.57  & 0.97   & 0.46   \\
\hline
$p+p$   & 6.64 & 0.22 & 0.14 & 0.085 & 0.46 & 0.36 & 0.011 & 0.0038 & 0.0019 \\
d+Ar    & 6.64 & 0.20 & 0.13 & 0.079 & 0.42 & 0.33 & 0.78  & 0.27   & 0.13   \\
\hline
$p+p$   & 6.48 & 0.22 & 0.14 & 0.083 & 0.45 & 0.35 & 0.010 & 0.0037 & 0.0018 \\
d+Kr    & 6.48 & 0.20 & 0.12 & 0.076 & 0.41 & 0.32 & 1.57  & 0.56   & 0.28   \\
\hline
$p+p$   & 6.41 & 0.21 & 0.14 & 0.082 & 0.44 & 0.35 & 0.010 & 0.0036 & 0.0018 \\
d+Sn    & 6.41 & 0.19 & 0.12 & 0.074 & 0.40 & 0.34 & 2.34  & 0.77   & 0.41   \\
\hline
$p+p$   & 6.3  & 0.21 & 0.14 & 0.082 & 0.44 & 0.34 & 0.010 & 0.0038 & 0.0018 \\
$p+$Ar  & 6.3  & 0.20 & 0.12 & 0.075 & 0.41 & 0.32 & 0.37  & 0.13   & 0.065  \\
Ar+Ar  & 6.3  & 0.18 & 0.12 & 0.070 & 0.38 & 0.29 & 13.8  & 5.29   & 2.43   \\
\hline
$p+p$   & 6.22 & 0.21 & 0.13 & 0.080 & 0.43 & 0.34 & 0.010 & 0.0035 & 0.0017 \\
d+Pb    & 6.22 & 0.18 & 0.12 & 0.071 & 0.38 & 0.30 & 3.68  & 1.31   & 0.65   \\
\hline
$p+p$   & 6.14 & 0.21 & 0.13 & 0.080 & 0.43 & 0.33 & 0.0099& 0.0038 & 0.0017 \\
$p+$Kr  & 6.14 & 0.19 & 0.12 & 0.072 & 0.39 & 0.30 & 0.75  & 0.27   & 0.13   \\
Kr+Kr  & 6.14 & 0.17 & 0.11 & 0.066 & 0.35 & 0.28 & 57.4  & 21.8   & 10.1   \\
\hline
$p+p$   & 5.84 & 0.20 & 0.12 & 0.076 & 0.41 & 0.32 & 0.0094& 0.0035 & 0.0017 \\
$p+$Sn  & 5.84 & 0.18 & 0.11 & 0.068 & 0.37 & 0.29 & 1.01  & 0.36   & 0.18   \\
Sn+Sn  & 5.84 & 0.16 & 0.10 & 0.062 & 0.33 & 0.26 & 108.1 & 41.3   & 19.0   \\
\hline
$p+p$   & 5.5  & 0.19 & 0.12 & 0.070 & 0.39 & 0.30 & 0.0090& 0.0029 & 0.0016 \\
$p+$Pb  & 5.5  & 0.17 & 0.11 & 0.064 & 0.34 & 0.27 & 1.65  & 0.60   & 0.29   \\
Pb+Pb  & 5.5  & 0.15 & 0.094& 0.057 & 0.31 & 0.24 & 304   & 116.1  & 53.5   \\
\hline
\end{tabular}
\end{center}
\caption[]{The direct cross section per nucleon pair (central columns) and 
the dilepton yield per nucleon multiplied by $AB$.  The results are given for 
the MRST PDFs \cite{mrst} with $m_b = 4.75$ GeV, $\mu_F = \mu_R = m_T$.}
\label{upssigs}
\end{table}

\begin{figure}[htbp] 
\centering
\resizebox{0.95\textwidth}{!}{\rotatebox{0}{%
\includegraphics*{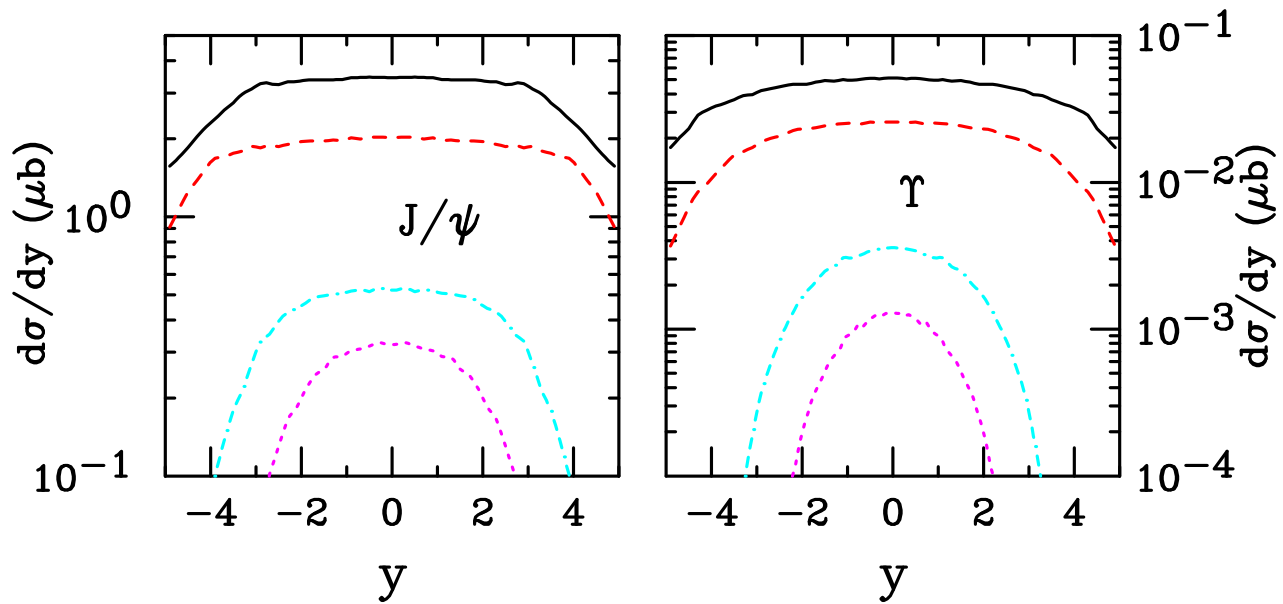}}}
\vglue-2mm
\caption[]{(Color online)
The $J/\psi$ (left-hand side) and $\Upsilon$ (right-hand side)
rapidity distributions at $\sqrt{s_{_{NN}}} = 200$ (dotted), 500 (dot-dashed),
5500 (dashed) and 14000 (solid) GeV calculated as in 
Refs.~\protect\cite{hvqyr,rhicii}.  The kinks in the $J/\psi$ distributions
at LHC energies are the point where $x<10^{-5}$.
Since the $\Upsilon$ factorization scale is larger, the $\Upsilon$ rapidity 
distributions are smoother.  Note the different scales on the $y$-axes.}
\label{pprapdists}
\end{figure}

The $p+p$ rapidity distributions for $J/\psi$ and $\Upsilon$ production 
at $\sqrt{s_{_{NN}}} = 5.5$ and 14 TeV are compared to RHIC distributions
at $\sqrt{s_{_{NN}}} = 200$ and 500 GeV in Fig.~\ref{pprapdists}.  
The LHC distributions are relatively constant over a range of 5 or
more units of rapidity, demonstrating that the cross sections are high enough 
to obtain good statistics for quarkonium states, even for forward production
and detection, provided that the decay leptons are of sufficiently high $p_T$
to reach the detectors\footnote{This will be more difficult for CMS
and ATLAS than for ALICE since the minimum muon $p_T$ for detection in the
large $p+p$ experiments is 3.5 GeV/$c$.}.

\section{Cold Nuclear Matter Effects}

In this section, we describe the cold nuclear matter effect of initial-state
shadowing on $J/\psi$ and
$\Upsilon$ production at the LHC.  We first discuss the shadowing 
parameterizations used in our calculations.  We then show the effect of 
shadowing on the rapidity distributions in the $p+A$, d$+A$ and $A+A$
collisions available at the LHC.  Finally, we discuss the collision centrality
dependence on a simple model of inhomogeneous shadowing where the effect 
depends on the path length of the parton through the nucleus.

\subsection{Shadowing parameterizations}

We use several parameterizations of the nuclear modifications in the parton
densities to probe the possible range of
gluon shadowing effects: EKS98 \cite{Eskola:1998iy,Eskola:1998df}, nDSg
\cite{deFlorian:xxx}, HKN \cite{Hirai:xxx}, EPS08 \cite{Eskola:2008}
and EPS09 \cite{Eskola:2009}.  
All sets involve fits to data, typically nDIS data
with additional constraints from other observables such as Drell-Yan dimuon
production.  Since these provide no
direct constraint on the nuclear gluon density, it is obtained through
fits to the $\mu^2$ dependence of the nuclear structure function, $F_2^A$,
and momentum conservation.  The useful perturbative $\mu^2$ range of the
nDIS data is rather limited since these data are only available at fixed-target
energies.  Thus the reach in momentum fraction, $x$, is also limited and
there is little available data for $x < 10^{-2}$ at perturbative values of
$\mu^2$.  This situation is likely
not to improve until an $eA$ collider is constructed \cite{thomasqm08}.

The EKS98 parameterization, by Eskola and 
collaborators, available for
$A > 2$, is a leading order fit using the GRV LO \cite{Gluck:1991ng} 
proton parton densities as a baseline \cite{Eskola:1998iy,Eskola:1998df}.  
The kinematic range is $2.25 \leq
\mu_{\rm EKS98}^2 \leq 10^4$~GeV$^2$ and $10^{-6} \leq x < 1$.  
deFlorian and Sassot
produced the nDS and nDSg parameterizations \cite{deFlorian:xxx} at both
leading and next-to-leading order for $4<A<208$.  
The weak gluon shadowing of the nDS
parameterization appears to be ruled out by the rapidity dependence of $J/\psi$
production at RHIC \cite{RHICdanew}.  The stronger gluon shadowing of nDSg
is used here.  Calculations with the nDS parameterization predict negligible
shadowing effects.  The kinematic reach in $x$ is the same as EKS98 while the
$\mu^2$ range is larger, $1 < \mu_{\rm nDSg}^2 <10^6$~GeV$^2$.  
Hirai and collaborators
produced the leading order HKN parameterization by fitting 
parton densities for protons,
deuterons and 16 heavier nuclei, typically those most commonly used in nDIS
experiments.  If a particular value of $A$ needed for our calculations is
not included, a set with a similar value of $A$ is substituted.  The HKN
parameterization goes lower in $x$ than the other parameterizations, $10^{-9}
< x < 1$, and higher in scale, $1<\mu_{\rm HKN}^2 <10^8$~GeV$^2$.  
The EPS08
parameterization, a fit by Eskola and collaborators that includes the
BRAHMS d+Au data on forward rapidity hadron production at RHIC \cite{BRAHMS},
is designed to maximize the possible gluon shadowing\footnote{It has been
suggested that the BRAHMS data should not be used to calculate gluon shadowing
effects and that the strong shadowing of EPS08 violates a unitarity bound at the
minimum scale \cite{Boristhedork}.}.  
The EPS08 $x$ range is the same as EKS98, 
$10^{-6} \leq x < 1$, while the $\mu^2$ range was extended, $1.96 \leq
\mu_{\rm EPS08}^2 \leq 10^6$ GeV$^2$.  
Very recently, the EPS09 \cite{Eskola:2009} parameterization, which excludes
the BRAHMS data from the fits, was introduced.  The EPS09 parameterization 
includes uncertainties on the global analyses, both at LO and NLO, by varying
one of the 15 fit parameters within its extremes while holding the 
others fixed.  The upper and lower bounds on EPS09 shadowing are obtained by
adding the resulting uncertainties in quadrature \cite{Eskola:2009}.  The EPS09
central LO results are in quite good agreement with the older
EKS98 parameterization while the maximum possible gluon shadowing effect
resulting from their uncertainty analysis is similar to the EPS08 gluon ratio.
The minimal amount
of gluon shadowing is nearly negligible, similar to nDS \cite{deFlorian:xxx}
and even leaves room for some antishadowing in light ions.
We present the central EPS09 ratio as well as the ratios corresponding to the
maximum and minimum range of the shadowing effect, obtained by adding the
relative differences in quadrature, as prescribed in Ref.~\cite{Eskola:2009}.
For computational convenience, we use the LO version of the nPDF 
parameterizations since the NLO CEM calculations give similar shadowing
results \cite{SQM04}.  This is to be expected since, even though the LO and
NLO values of the cross section and the shadowing parameterization are 
different, when convoluted, they give the same ratios by design, see {\it e.g.}
Ref.~\cite{deFlorian:xxx}.

While the $x$ values probed at midrapidity are
$\approx 10^{-4}$ for the $J/\psi$ and $\approx 10^{-3}$ for the $\Upsilon$, 
well within
the $x$ range of the parameterizations, this is not necessarily the case
away from midrapidity.  At the largest values of $\sqrt{s_{_{NN}}}$, 
$x$ values lower than the minimum valid $x$ of the parameterization
may be reached within the rapidity range of the
LHC detectors.  In these cases, the shadowing parameterizations are 
unconstrained by data.  However, when $x < 10^{-6}$ 
the EKS98, nDSg, EPS08 and EPS09 parameterizations return the value of the
shadowing ratio at $x = 10^{-6}$.  The minimum
$x$ value, $10^{-9}$, for the HKN parameterization is small enough that this
minimum is not reached, even for the highest energies.  

\begin{figure}[htbp]
\begin{center}
\includegraphics[width=0.5\textwidth]{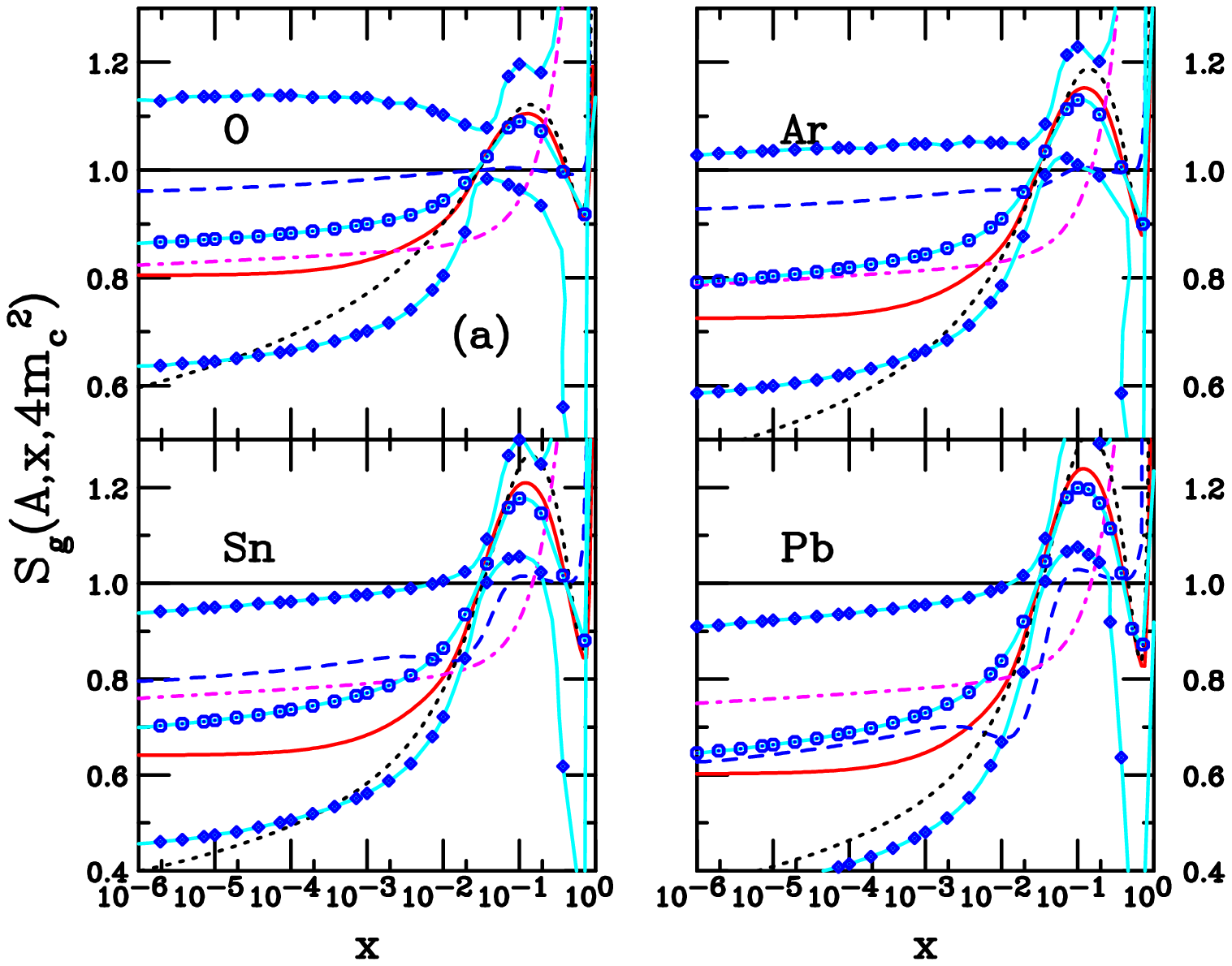}
\\
\includegraphics[width=0.5\textwidth]{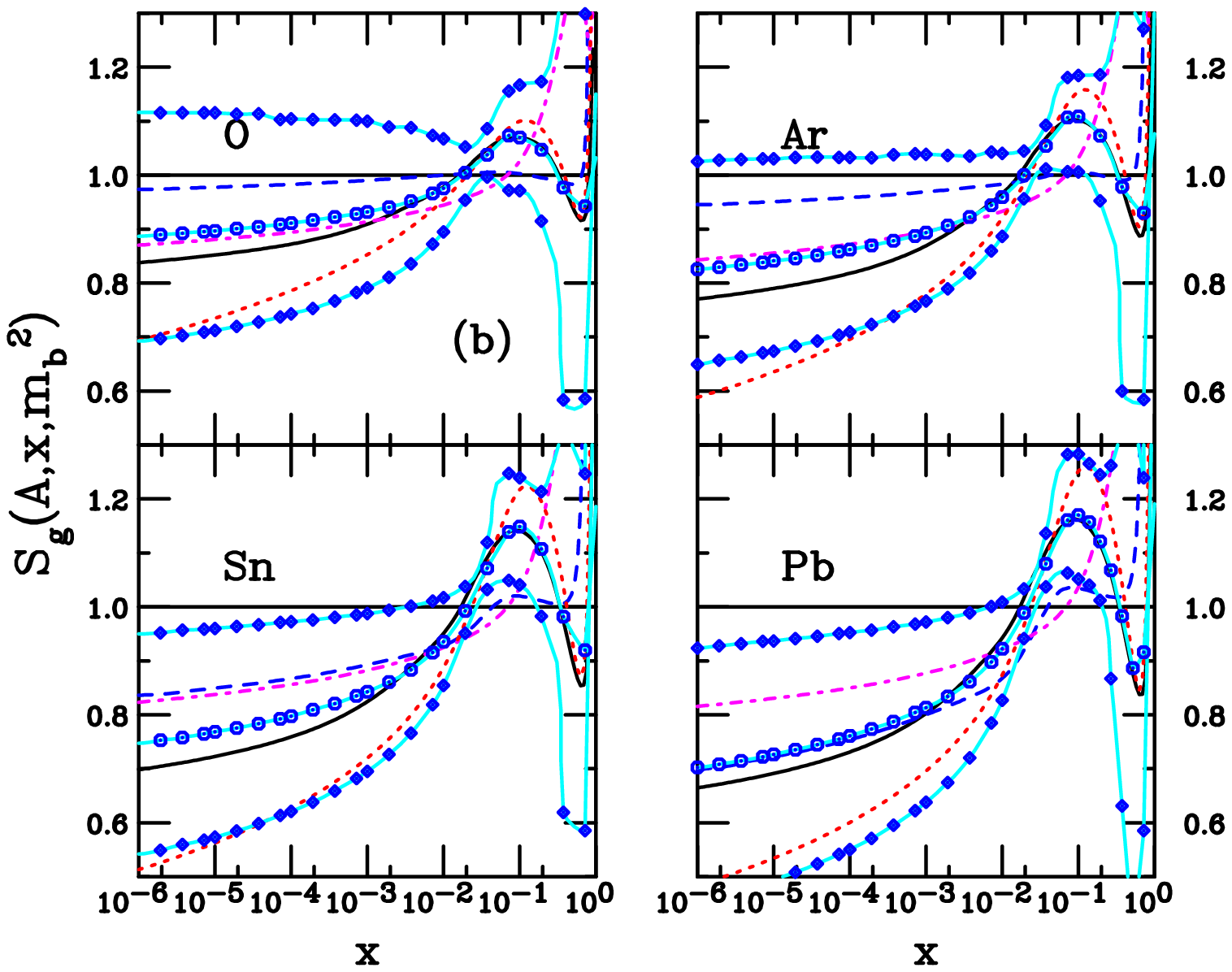}
\end{center}
\caption[]{(Color online)
The LO shadowing parameterizations for $J/\psi$ (a) and $\Upsilon$
(b) scales for O (upper left), Ar (upper right), Sn (lower left) and Pb
(lower right) nuclei.  The parameterizations are EKS98 (solid), nDSg (dashed),
HKN (dot-dashed), EPS08 (dotted) and EPS09 (solid lines with symbols).  
Note that the lower limit on the
$y$-axis is changed for Sn and Pb on the left-hand side.
}
\label{nglue}
\end{figure}

The ratios of the nuclear gluon densities relative to the gluon density
in the proton are shown in
Fig.~\ref{nglue} for four different ion species available at the LHC: 
$A = {\rm O}$, Ar, Sn and Pb.  The calculations for $A = {\rm Kr}$, 
an alternative intermediate mass 
ion species, are not shown.  Results for scales appropriate for
$J/\psi$, Fig.~\ref{nglue}(a), and $\Upsilon$, Fig.~\ref{nglue}(b), production
illustrate the scale dependence of the 
parameterizations.
The scales correspond to those used in the calculations of the cross sections
in Tables~\ref{psisigs} and \ref{upssigs} with $\mu = 2m_c$ for charm
and $m_b$ for bottom
respectively.  If a lower scale, $\mu = m_c$, is used for
charm, the shadowing effect is stronger since $\mu^2$ 
is then closer to the minimum
scale of the parameterization.  Note that in all cases
shadowing increases with decreasing $x$ and increasing $A$ while decreasing
with scale, $\mu$, as seen by comparing Fig.~\ref{nglue}(a) and (b).  
For example, the EKS98, nDSg and HKN ratios appear to be 
approximately independent of $x$ for $x < 10^{-3}$ at the $J/\psi$ scale
but not at the $\Upsilon$ scale.

The EKS98, EPS08 and EPS09 parameterizations (solid and dotted curves and 
solid curves with symbols respectively)
exhibit large antishadowing, $S^g > 1$, in
the region $0.02<x<0.2-0.3$, becoming more pronounced for larger $A$.
The nDSg parameterization (dashed curves) show very weak antishadowing around
$x \sim 0.1$.  At $x < 10^{-2}$, the nDSg ratios are weakest for 
$A= {\rm O}$ and Ar, similar to HKN for $A= {\rm Sn}$ 
and compatible with EKS98 for $x < 10^{-3}$.  The 
HKN parameterization (dot-dashed curves), on the
other hand, is similar to EKS98 for $A = {\rm O}$ but has a weak $A$ dependence
so that HKN shadowing is the weakest at low $x$ and large $A$.  
The EPS08 parameterization is
similar to EKS98 for $x > 0.01$ but exhibits stronger antishadowing at 
large $A$.  It also has the strongest shadowing at low $x$ since the low-$p_T$
forward-rapidity BRAHMS data was included in the fit.  The scale
dependence of nDSg and HKN appears to be weaker than EKS98.
The EPS09 band is obtained by calculating the deviations from the central value
for the 15 parameter variations on either side of the central set and adding
them in quadrature.  The range of the LO EPS09 uncertainty band encompasses
all other shadowing ratios, similar to EPS08 for the maximum effect and even
leading to antishadowing for lighter ions.  
(The central ratio is shown with circular symbols
on the solid curve while the bounds include diamond symbols.) 
For smaller nuclei, the upper edge of the
EPS09 uncertainty (minimal shadowing effect) gives a bound above unity for
$S^g$.

All the parameterizations increase at large $x$ with $S^g >1$ for $x>0.1$ (HKN
and nDSg) and $x>0.7$ (EKS98 and EPS08).
The rise in the HKN parameterization is steepest and occurs at the lowest $x$,
beginning at the $x$ value of the antishadowing peak in the 
EKS98 and EPS08 ratios.
This high $x$ region will not be explored by the LHC detectors since it
is only reached at rapidities outside their acceptance.  

Finally, we note that since our $p+A$ calculations
assume the ion beam travels in the negative $z$ direction, 
low $x$ corresponds to large 
forward rapidity while high $x$ corresponds to large backward rapidity.

\begin{figure}[htbp]
\begin{center}
\includegraphics[width=0.5\textwidth]{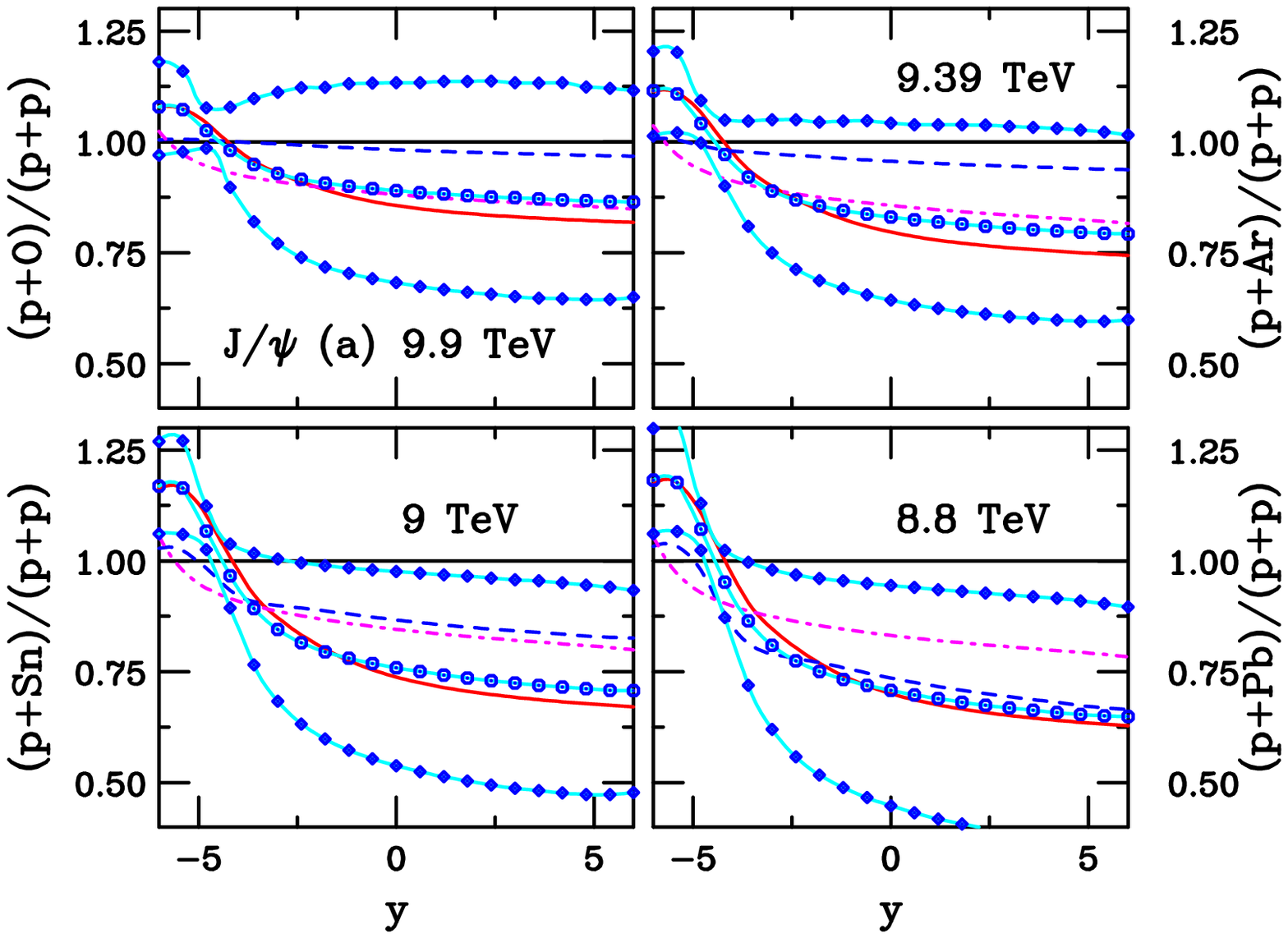} \\
\includegraphics[width=0.5\textwidth]{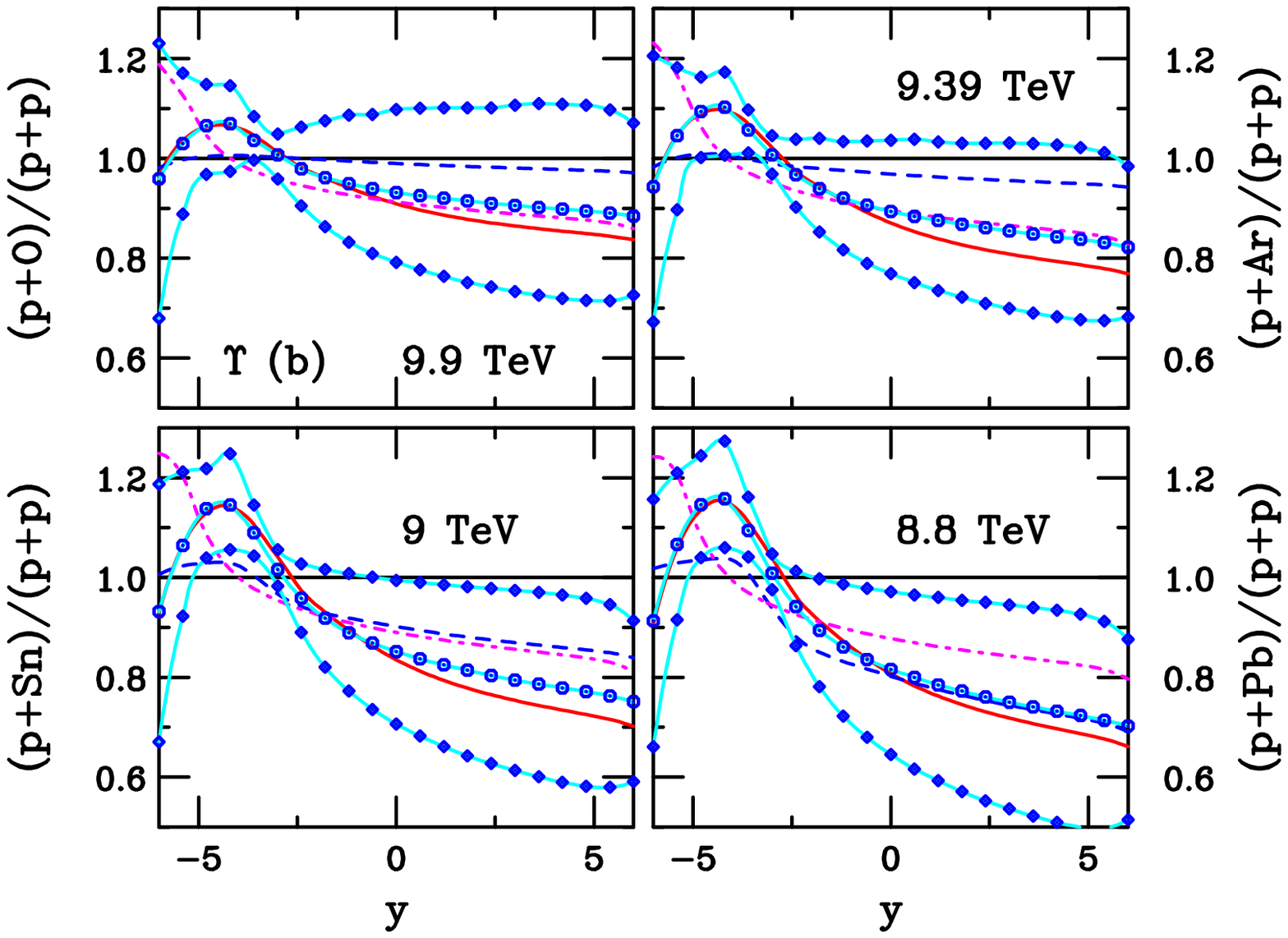}
\end{center}
\caption[]{(Color online)
The $(p+A)/(p+p)$ ratios with both $p+A$ and $p+p$ collisions at the $p+A$
energy in the equal-speed frame.  
No rapidity shift has been taken into account.  The effect of 
shadowing on $J/\psi$ (a, upper 4 panels) and $\Upsilon$ (b, lower 4 panels) 
production is shown.  Each panel displays the production ratios
for $p+$O at $\sqrt{s_{_{NN}}} = 9.9$ TeV (upper left), $p+$Ar at 
$\sqrt{s_{_{NN}}} = 9.39$ TeV (upper right), $p+$Sn at $\sqrt{s_{_{NN}}} = 9$ 
TeV  (lower left) and $p+$Pb at $\sqrt{s_{_{NN}}} = 8.8$ TeV (lower right).  
The calculations are with CTEQ6 
and employ the EKS98 (solid),
nDSg (dashed), HKN (dot-dashed) and EPS09 (solid curves with
smbols) shadowing parameterizations.
}
\label{fig8hi}
\end{figure}

\subsection{Rapidity dependence}
\label{rapdepsec}

We now show predictions of the $J/\psi$ and $\Upsilon$ production ratios as
a function of rapidity for cold nuclear matter, CNM, effects
at the LHC.  If
$h+A$ data  (where $h = p$ or d) can be taken at the same energy 
as the $p+p$ and/or $A+A$ data, as
at RHIC, it is easier to make comparisons.  However, the setup of the LHC makes
this ideal situation more difficult.  At the nominal injection energy, the 
proton beam has an energy of 7 TeV while the nuclear beam energy per nucleon
is lower by the nuclear charge-to-mass ratio, $Z/A$.  To make a $p+p$ 
comparison, if we are not to rely on calculations extrapolated to lower 
energy, the $p+p$ collisions have to be run at the $p+A$ or $A+A$ per nucleon 
energies.  For the proton
and ion beam energies to be the same, the proton beam must then circulate at
lower than optimal energy, decreasing the luminosity.  
Since sustained low energy $p+p$ runs are unlikely in early LHC running, 
especially for
sufficiently accurate quarkonium data as a function of rapidity, it may be
necessary to rely on higher energy $p+p$ reference data\footnote{Although the
startup LHC $p+p$ runs are at lower energies, it is not clear how much 
quarkonium data will be extracted during these runs.  Thus we show our 
results relative to the maximum $p+p$ energy of 14 TeV.}.
However, there is a catch.
In $p+A$ collisions where a 7 TeV proton beam collides with a $7(Z/A)$ TeV
per nucleon ion beam, the so-called equal-speed or equal-rigidity
frame, the center-of-mass 
rapidity is not fixed at $y=0$ but displaced by $\Delta y_{\rm cm}^{iA}$.
In $p$Pb collisions, the shift can be nearly 0.5 units, an important difference,
see Table~\ref{delytable} for the magnitude of the possible shifts.
To minimize the rapidity shift and to bring the $hA$ comparison energy
closer to that of the $A+A$ energy, d$+A$ collisions may be desirable since
$E_{\rm d} = 3.5$ TeV per
nucleon relative to $E_{\rm Pb} = 2.75$ TeV, see Table~\ref{delytable}.
Since d$+A$ collisions require a second ion source, this may not be realized
in the short term.

\begin{figure}[htbp]
\begin{center}
\includegraphics[width=0.5\textwidth]{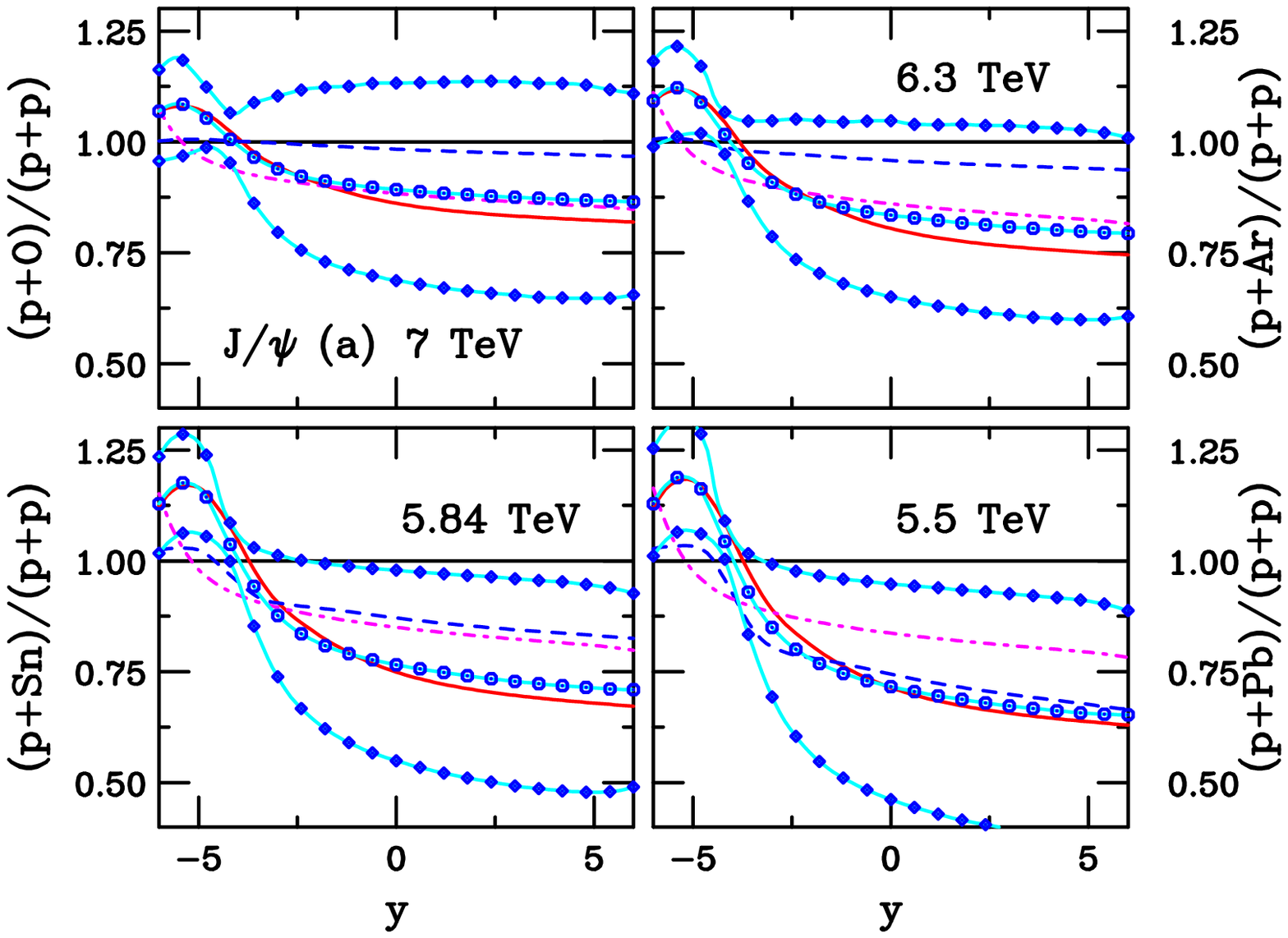} \\
\includegraphics[width=0.5\textwidth]{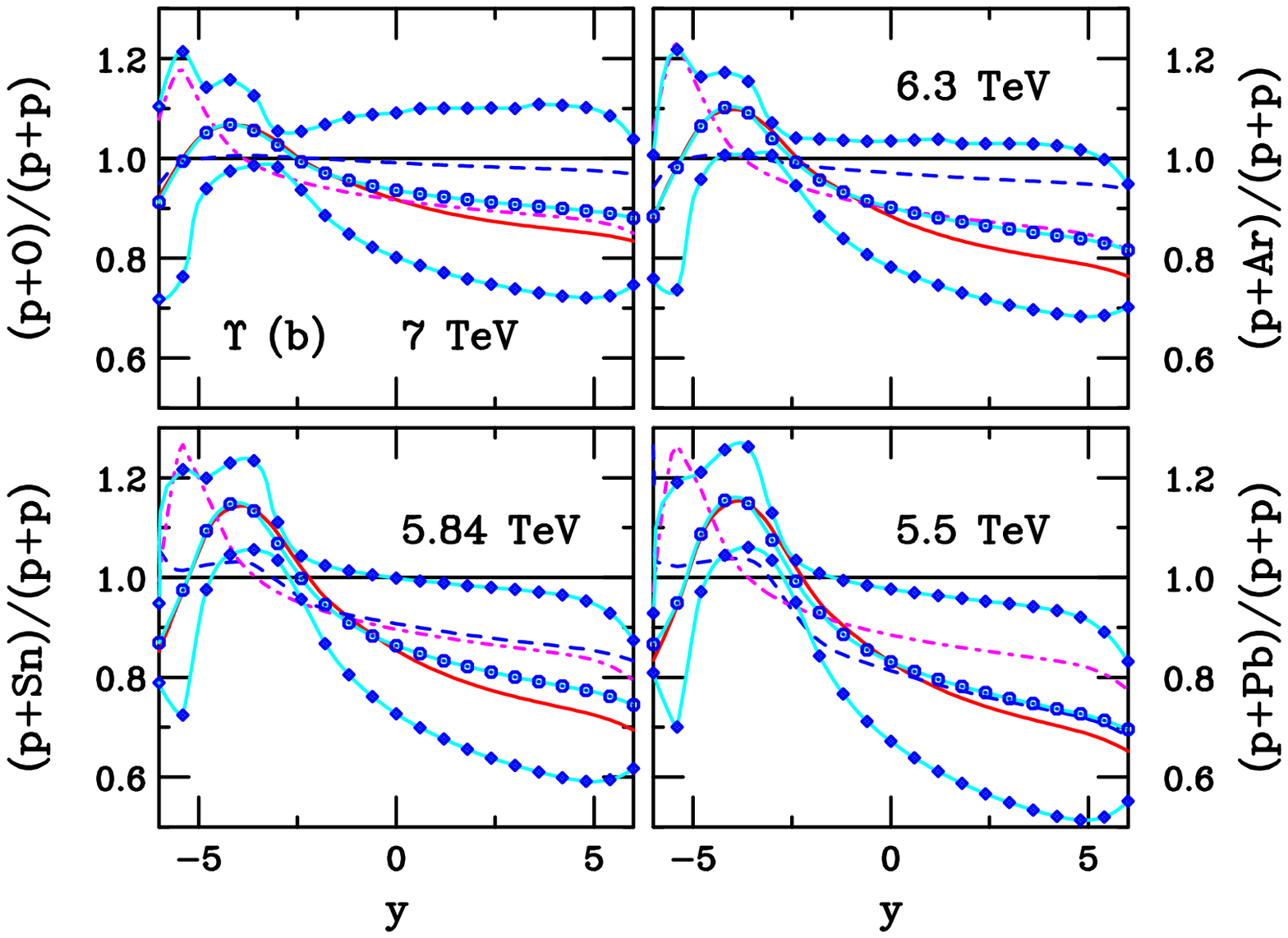}
\end{center}
\caption[]{(Color online)
The $(p+A)/(p+p)$ ratios with both $p+A$ and $p+p$ collisions at the $A+A$
center-of-mass energy.  The effect of 
shadowing on $J/\psi$ (a, upper 4 panels) and $\Upsilon$ (b, lower 4 panels) 
production is shown.  Each set of panels displays the production ratios
for $p+$O at $\sqrt{s_{_{NN}}} = 7$ TeV (upper left), $p+$Ar at 
$\sqrt{s_{_{NN}}} = 6.3$ TeV (upper right), $p+$Sn at $\sqrt{s_{_{NN}}} = 6.14$ 
TeV (lower left) and $p+$Pb at 
$\sqrt{s_{_{NN}}} = 5.5$ TeV (lower right).  The calculations are with 
CTEQ6 and employ the EKS98 (solid),
nDSg (dashed), HKN (dot-dashed) and EPS09 (solid curves with
symbols) shadowing parameterizations.
}
\label{fig8lo}
\end{figure}

\begin{figure}[htbp]
\begin{center}
\includegraphics[width=0.5\textwidth]{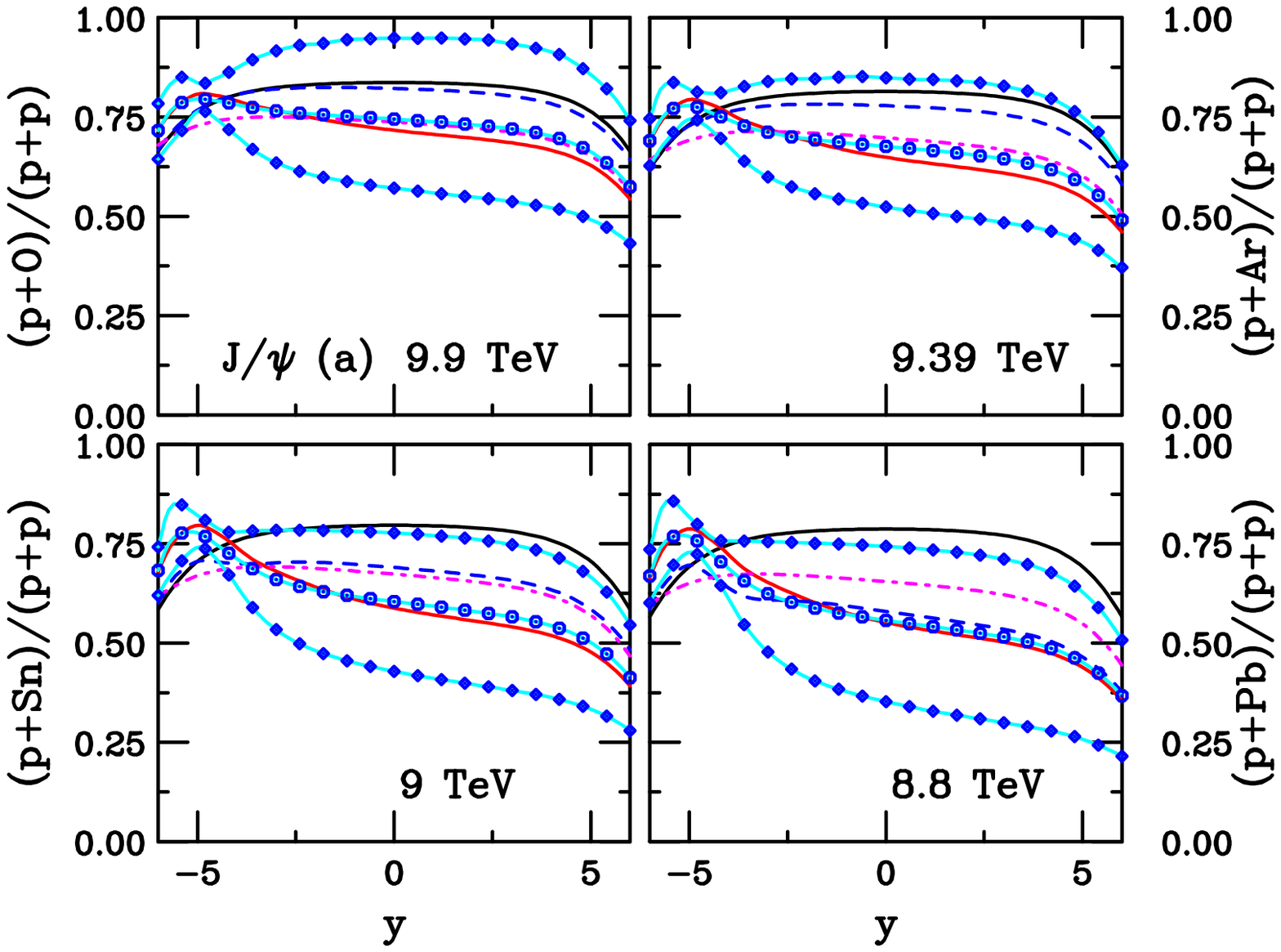} \\
\includegraphics[width=0.5\textwidth]{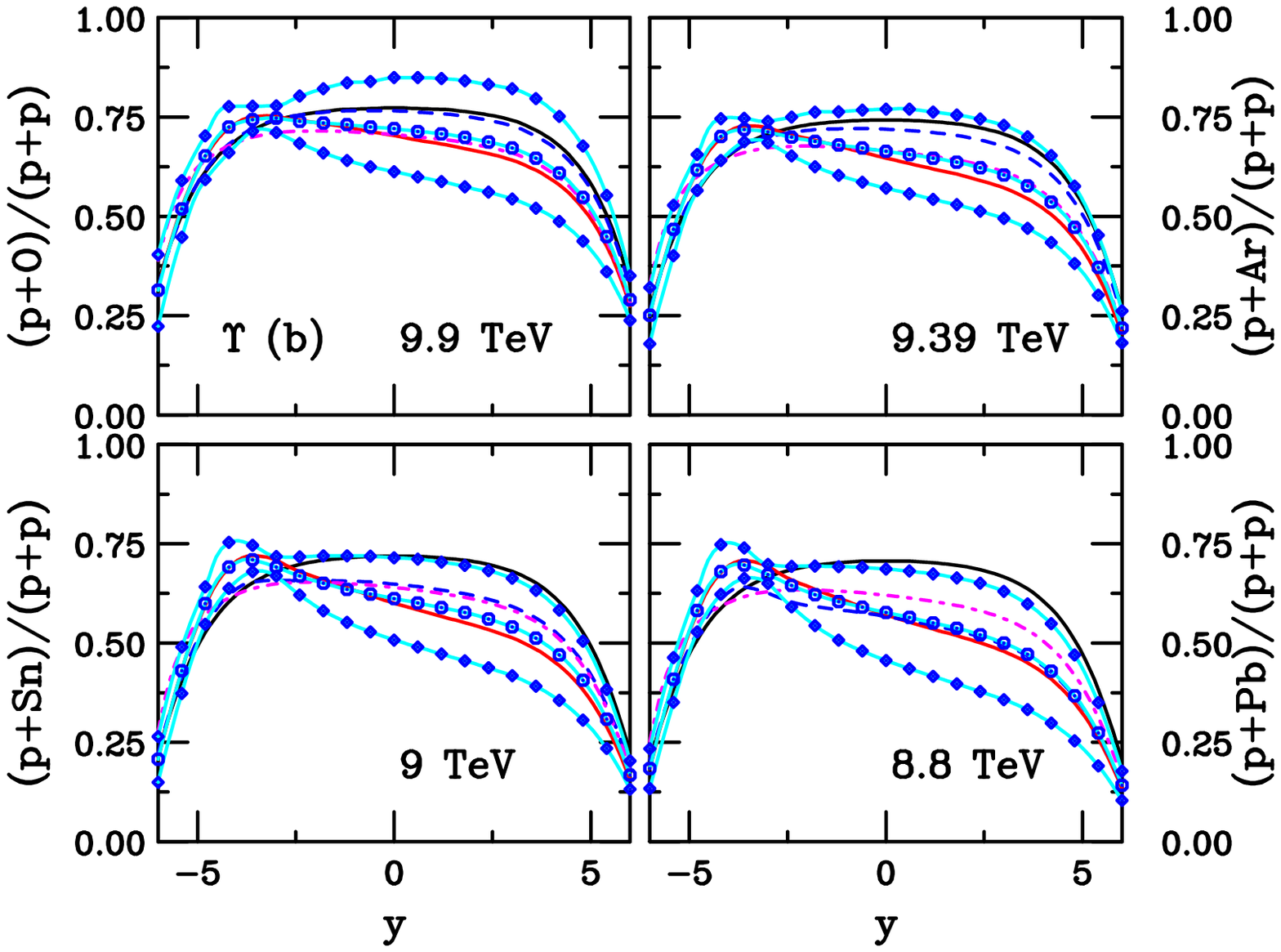}
\end{center}
\caption[]{The $(p+A)/(p+p)$ ratios with the $p+p$ rapidity distributions 
calculated 
at $\sqrt{s} = 14$ TeV.  While the $p+A$ distributions are calculated in the
equal-speed frame, no rapidity shift 
has been taken into account.  The effect of 
shadowing on $J/\psi$ (a, upper 4 panels) and $\Upsilon$ (b, lower 4 panels) 
production is shown.  Each set of panels displays the production ratios
for $p+$O at $\sqrt{s_{_{NN}}} = 9.9$ TeV (upper left), $p+$Ar at 
$\sqrt{s_{_{NN}}} = 9.39$ TeV (upper right), $p+$Sn at $\sqrt{s_{_{NN}}} = 9$ 
TeV (lower left) and $p+$Pb at 
$\sqrt{s_{_{NN}}} = 8.8$ TeV (lower right), all calculated in the equal-speed 
frame.  The calculations are with CTEQ6 and employ the EKS98 (solid), 
nDSg (dashed), HKN (dot-dashed) and EPS09 (solid curves with
symbols) shadowing parameterizations.
The solid curve symmetric around $y=0$ is the $(p+A)/(p+p)$ ratio 
without shadowing.
}
\label{fig9hi}
\end{figure}

\begin{figure}[htbp]
\begin{center}
\includegraphics[width=0.5\textwidth]{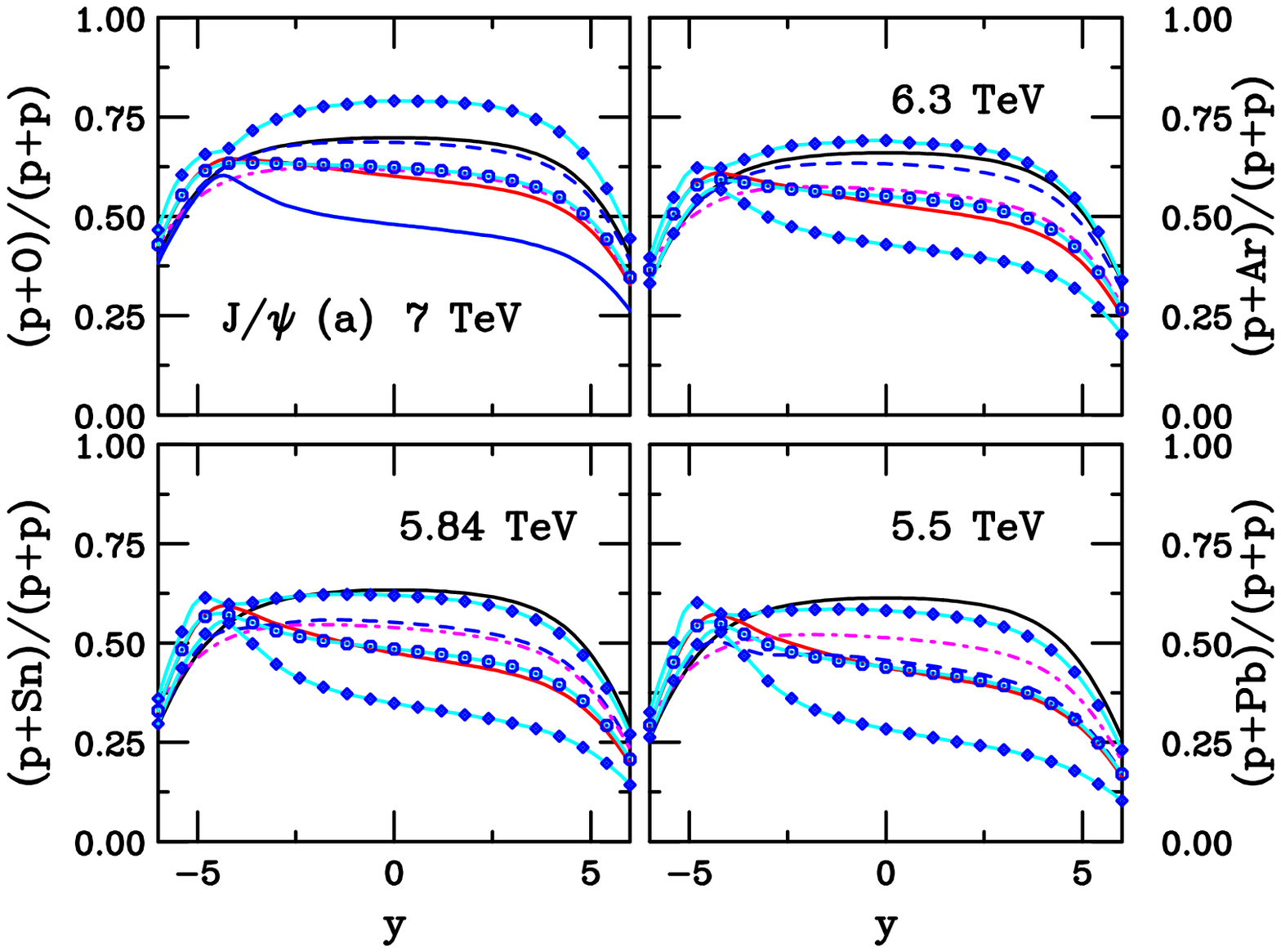}
\includegraphics[width=0.5\textwidth]{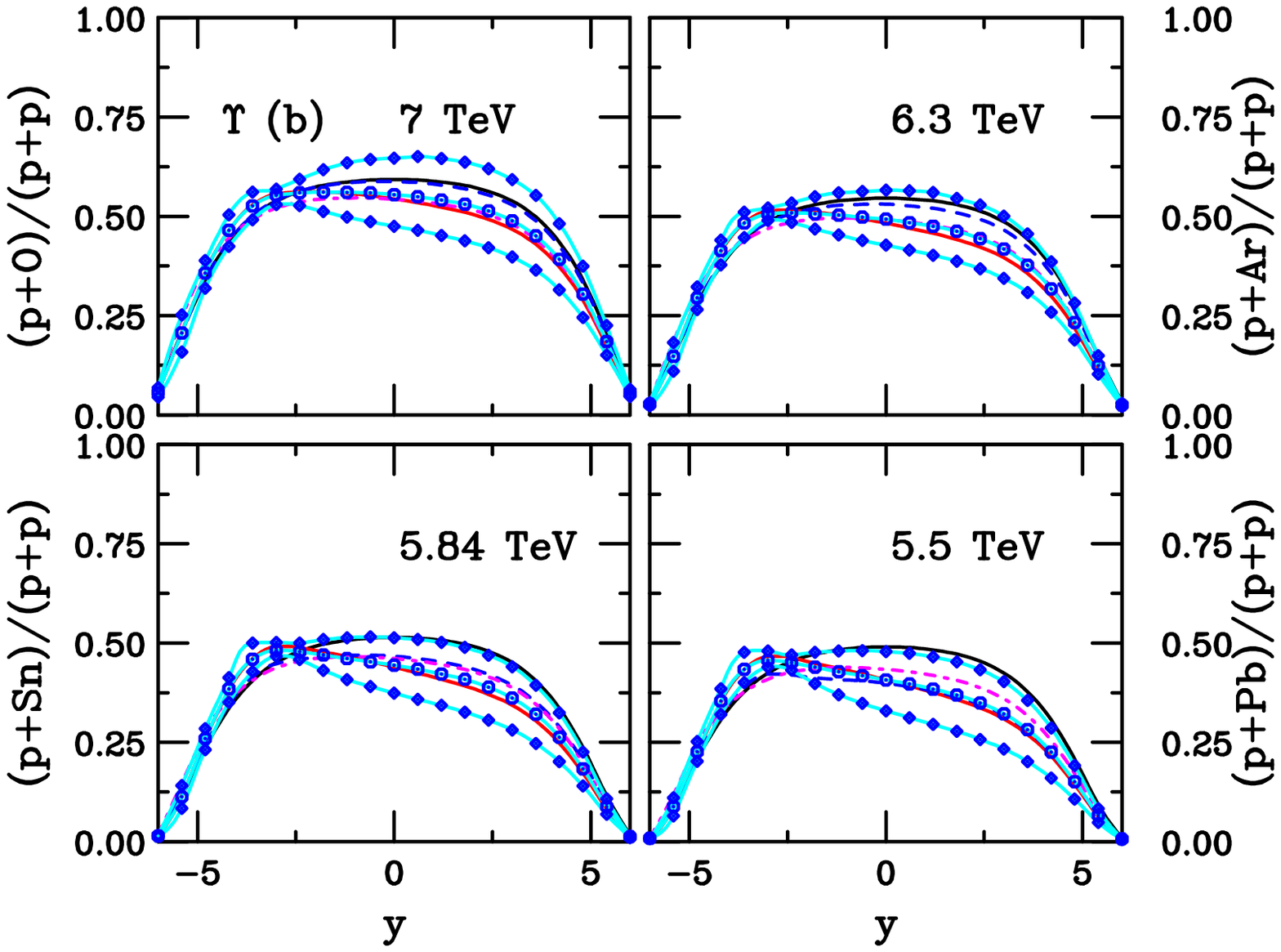}
\end{center}
\caption[]{The $(p+A)/(p+p)$ ratios with the $p+p$ rapidity distributions 
calculated 
at $\sqrt{s} = 14$ TeV.  The $p+p$ distributions are calculated at the $A+A$
center-of-mass energy.  The effect of 
shadowing on $J/\psi$ (a, upper 4 panels) and $\Upsilon$ (b, lower 4 panels) 
production is shown.  Each set of panels displays the production ratios
for $p+$O at $\sqrt{s_{_{NN}}} = 7$ TeV (upper left), $p+$Ar at 
$\sqrt{s_{_{NN}}} = 6.3$ TeV (upper right), $p+$Sn at $\sqrt{s_{_{NN}}} = 6.14$ 
TeV (lower left) and $p+$Pb at 
$\sqrt{s_{_{NN}}} = 5.5$ TeV (lower right).  The ratios are calculated at
the $A+A$ energy.  The calculations are with CTEQ6 and employ the EKS98 (solid), 
nDSg (dashed), HKN (dot-dashed) and EPS09 (solid curves with
symbols) shadowing parameterizations.
The solid curve symmetric around $y=0$ is the $(p+A)/(p+p)$ ratio without shadowing.
}
\label{fig9lo}
\end{figure}

\begin{figure}[htbp]
\begin{center}
\includegraphics[width=0.475\textwidth]{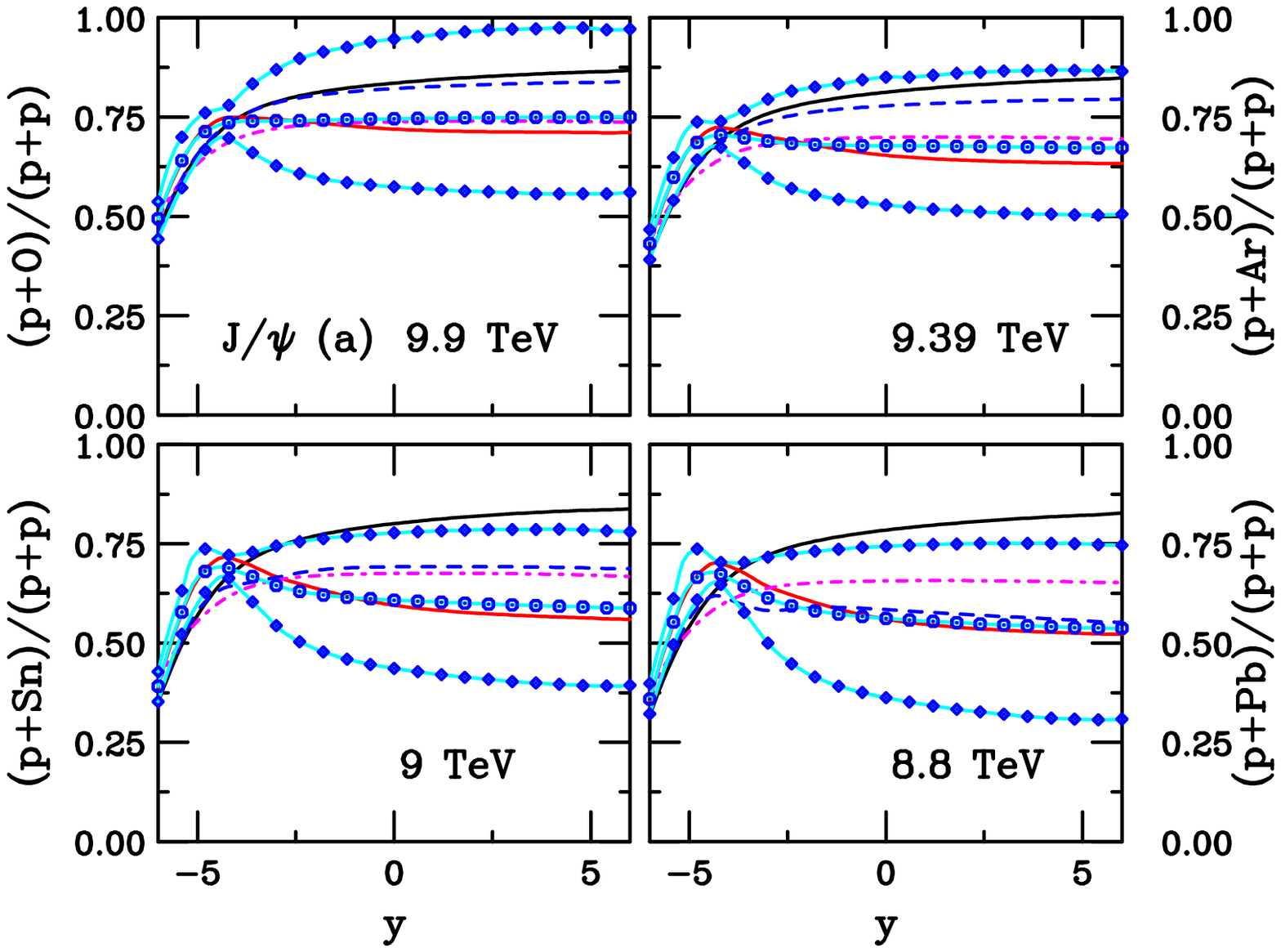} \\
\includegraphics[width=0.475\textwidth]{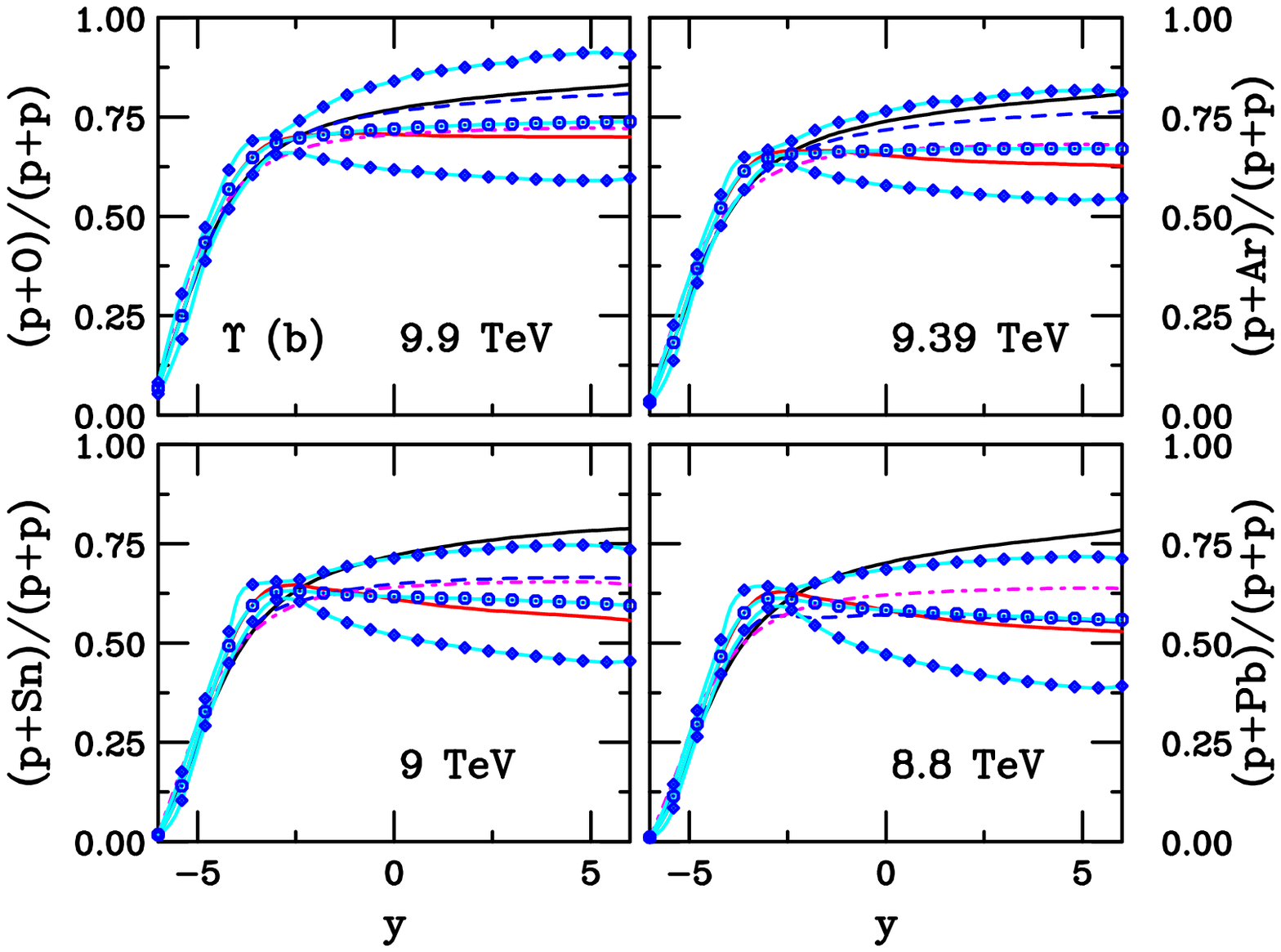}
\end{center}
\caption[]{(Color online)
The $(p+A)/(p+p)$ ratios with the $p+p$ rapidity distributions calculated 
at $\sqrt{s} = 14$ TeV.  The $p+A$ rapidity distributions are calculated in
the equal-speed frame with the rapidity shift taken into account.  
The effect of 
shadowing on $J/\psi$ (a, upper 4 panels) and $\Upsilon$ (b, lower 4 panels) 
production is shown.  Each set of panels displays the production ratios
for $p+$O at $\sqrt{s_{_{NN}}} = 9.9$ TeV (upper left), $p+$Ar at 
$\sqrt{s_{_{NN}}} = 9.39$ TeV (upper right), $p+$Sn at $\sqrt{s_{_{NN}}} = 9$ 
TeV (lower left) and $p+$Pb at $\sqrt{s_{_{NN}}} = 8.8$ TeV  (lower right).  
The calculations are with CTEQ6 and employ the EKS98 (solid), 
nDSg (dashed), HKN (dot-dashed) and EPS09 (solid curves with
symbols) shadowing parameterizations.
The upper solid curve at $y>0$ is the shifted $(p+A)/(p+p)$ ratio without shadowing.
}
\label{fig10}
\end{figure}

We thus study several different possibilities for 
determining cold nuclear matter
effects on nucleus-nucleus collisions at the LHC.  We go from ideal to more
realistic scenarios.  We first show the $(p+A)/(p+p)$ per nucleon
ratios at the same per nucleon
center-of-mass energy for both systems, assuming the appropriate $p+p$ energies
are available, Figs.~\ref{fig8hi} and \ref{fig8lo}.  
In the case where $p+A$ and $p+p$ interactions
are compared at the $A+A$
energy, we assume zero rapidity gap, $\Delta y_{\rm cm}^{pA} =0$, 
between the colliding 
beams since the proton beam energy is reduced to match that 
of the nucleus.  In the more likely scenario, Figs.~\ref{fig9hi} and 
\ref{fig9lo}, the $p+A$
data will be taken in the equal-speed frame at a higher energy than the
$A+A$ collisions.  Therefore, we next show the $(p+A)/(p+p)$ per nucleon ratios 
with respect to $p+p$
collisions at $\sqrt{s} = 14$ TeV with $\Delta y_{\rm cm}^{pA} = 0$ for $p+A$
collisions both in the equal-speed frame and at the
$A+A$ center-of-mass energy.  The final $p+A$ calculations shown are the most
realistic: the $p+A$ cross section in the equal-speed frame with finite $\Delta
y_{\rm cm}^{pA}$ is shown relative
to the $p+p$ cross section at 14 TeV.  In this case, Fig.~\ref{fig10},
the numerator and 
denominator are calculated with different energies and different center-of-mass
rapidities.  Next, the (d$+A)/(p+p)$ per nucleon
ratios are presented for two cases: with the
d$+A$ and $p+p$ collisions at the d$+A$ center-of-mass energy
and with d$+A$ collisions in the equal-speed
frame  with $\Delta y_{\rm cm}^{{\rm d}A} \neq 0$ and 
$p+p$ collisions at 14 TeV.  Finally, we present the baseline $(A+A)/(p+p)$ 
per nucleon ratios with the $p+p$ center-of-mass energy 
tuned to the $A+A$ energy and at the nominal 14 TeV $p+p$ energy.  In the 
case of symmetric $p+p$ and $A+A$ collisions, there is no rapidity gap.

We present the rapidity dependence of $p+A$, d$+A$, and
$A+A$ collisions for $A = {\rm O}$, Ar, Sn and Pb
relative to $p+p$ collisions, both at the same energy as the
nuclear system and at 14 TeV.  The $A+B/p+p$ ratios are shown for
the EKS98, nDSg and HKN parameterizations.  The EPS09 central
ratios and the associated uncertainty bands are also shown.  Since the
EPS08 ratios are similar to the lower limit (strongest shadowing) of the EPS09 
uncertainty band at small $x$, we do not show any further calculations with
EPS08.  We use the CTEQ6 parton densities to calculate the ratios shown in
the next two sections.  We have checked that the ratios with the MRST 
densities are essentially
identical since the same mass and scale parameters are used in the two
calculations\footnote{The ratios would only differ if these parameters were 
changed.  However, we leave them fixed since they were optimized to other 
data for a given set of parton densities.}.
Both $J/\psi$ and $\Upsilon$ results are
shown in each figure.  To guide the reader and
clarify the discussion, we first present
a table of the figures with the center-of-mass energy of the $A+B$
and $p+p$ collisions and the rapidity shift.
\begin{table}[htbp]
\begin{center}
\begin{tabular}{|c|c|c|cccc|c|} \hline
Figure & Cross Section Ratio & 
$\sigma_{pp}$ & \multicolumn{4}{|c|}{$\sigma_{BA}$} & $\Delta y_{\rm cm}^{BA}$ 
\\ 
number & $(B + A)/(p+p)$ & 
$\sqrt{s_{pp}}$ & $\sqrt{s_{B{\rm O}}}$ (TeV) & 
$\sqrt{s_{B{\rm Ar}}}$ (TeV) & $\sqrt{s_{B{\rm Sn}}}$ (TeV) &
$\sqrt{s_{B{\rm Pb}}}$ (TeV) & \\ \hline
\multicolumn{8}{|c|}{$B = p$} \\ \hline
\protect\ref{fig8hi} 
& $\sigma_{pA}(\sqrt{s_{pA}},y)/[A \sigma_{pp}(\sqrt{s_{pA}},y)]$ &
$\sqrt{s_{pA}}$ & 9.9 & 9.39 & 9 & 8.8 & 0 \\
\protect\ref{fig8lo} 
& $\sigma_{pA}(\sqrt{s_{AA}},y)/[A \sigma_{pp}(\sqrt{s_{AA}},y)]$ &
$\sqrt{s_{AA}}$ & 7 & 6.3 & 5.84 & 5.5 & 0 \\
\protect\ref{fig9hi} 
& $\sigma_{pA}(\sqrt{s_{pA}},y)/[A \sigma_{pp}(\sqrt{s_{pp}},y)]$ &
14 TeV & 9.9 & 9.39 & 9 & 8.8 & 0 \\
\protect\ref{fig9lo} 
& $\sigma_{pA}(\sqrt{s_{pA}},y)/[A \sigma_{pp}(\sqrt{s_{pp}},y)]$ &
14 TeV & 7 & 6.3 & 5.84 & 5.5 & 0 \\
\protect\ref{fig10} 
& $\sigma_{pA}(\sqrt{s_{pA}},(y-\Delta y_{\rm cm}^{pA})))/[A \sigma_{pp}
(\sqrt{s_{pp}},y)]$ &
14 TeV & 9.9 & 9.39 & 9 & 8.8 & $\Delta y_{\rm cm}^{pA}$ \\ \hline
\multicolumn{8}{|c|}{$B = {\rm d}$} \\ \hline
\protect\ref{fig11} 
& $\sigma_{{\rm d}A}(\sqrt{s_{{\rm d}A}},y)/[2A \sigma_{pp}(\sqrt{s_{{\rm 
d}A}},y)]$ &
$\sqrt{s_{{\rm d}A}}$ & 7 & 6.64 & 6.41 & 6.62 & 0 \\
\protect\ref{fig12} 
& $\sigma_{{\rm d}A}(\sqrt{s_{{\rm d}A}},(y-\Delta y_{\rm cm}^{{\rm d}A}))/
[2A \sigma_{pp}(\sqrt{s_{pp}},y)]$ &
14 TeV & 7 & 6.64 & 6.41 & 6.62 & $\Delta y_{\rm cm}^{{\rm d}A}$ \\ \hline
\multicolumn{8}{|c|}{$B = A$} \\ \hline
\protect\ref{fig13} 
& $\sigma_{AA}(\sqrt{s_{pA}},y)/[A^2 \sigma_{pp}(\sqrt{s_{pA}},y)]$ &
$\sqrt{s_{AA}}$ & 7 & 6.3 & 5.84 & 5.5 & 0 \\
\protect\ref{fig14} 
& $\sigma_{AA}(\sqrt{s_{AA}},y)/[A^2 \sigma_{pp}(\sqrt{s_{pp}},y)]$ &
14 TeV & 7 & 6.3 & 5.84 & 5.5 & 0 \\
\hline
\end{tabular}
\end{center}
\caption[]{Summary of the contents of Figs.~\protect\ref{fig8hi} - 
\protect\ref{fig14} in Section~\protect\ref{rapdepsec}.  
Here $B$ is the identity of the collision partner,
$B = p$ for $p+A$, d for d$+A$ and $A$ for $A+A$ collisions.  The value of
the center-of-mass energy for $p+p$ collisions used in the calculation of
the baseline $p+p$ cross section is given
in the third column: $\sqrt{s_{pp}} = \sqrt{s_{pA}}$ for $p+A$;
$\sqrt{s_{{\rm d}A}}$ for d$+A$; $\sqrt{s_{AA}}$ for $A+A$; and 14 TeV
for maximum energy $p+p$ collisions.  The center-of-mass energy for the $B+A$
cross sections are given in the next four columns.  Finally, whether or not
the rapidity shift is included is indicated in the last column.  The value of
$\Delta y_{\rm cm}^{BA}$ is given in Table~\protect\ref{delytable}.  Note that
all ratios are given for the per nucleon $B+A$ cross section.
}
\end{table}

The $(p+A)/(p+p)$ ratios with equal $p+A$ and $p+p$ center-of-mass energies,
shown in Figs.~\ref{fig8hi} and \ref{fig8lo},
illustrate the direct shadowing effect.  The
ratios are given both at the energy in the equal-speed frame, the likely
$\sqrt{s_{_{NN}}}$ for $p+A$ collisions (Fig.~\ref{fig8hi}), and at the same 
$\sqrt{s_{_{NN}}}$ as the corresponding $A+A$ collisions (Fig.~\ref{fig8lo}).  
The results are shown for all shadowing parameterizations.
The nuclear beam is assumed to be moving from positive to negative rapidity
so that the smallest values of $x$ probed in the nucleus are at large,
positive $y$.  

The LHC could be run as either a $p+A$ or an $A+p$ collider.
Since the ATLAS and CMS detectors are symmetric around $y=0$ with central
muon detectors in the range $|y| \leq 2.4$, ALICE is the only experiment
that could benefit from running in both modes because their dimuon spectrometer
covers $-4<y<-2.4$ in these coordinates \cite{upcyr}.  
However, since ALICE has muon coverage
in the largest $y$ region, running in both modes could be an advantage for
reconstructing the nuclear effects in quarkonium measurements, especially
since the $y$ distributions are rather flat over a broad rapidity range.
The large rapidity rates are thus non-negligible.

The $J/\psi$ ratios are shown in the upper half of the figures
while the $\Upsilon$ results
are on the lower half.  Since shadowing is an initial-state effect, the
same ratios would also be expected for the $\chi_c$ and $\psi'$ on the left and
the higher $\Upsilon$ states ($\Upsilon'$, $\Upsilon''$, $\chi_b(1P)$ and
$\chi_b(2P)$) on the right.  The ratios in Figs.~\ref{fig8hi}
and \ref{fig8lo} are 
stretched mirror images of the gluon shadowing ratios in Fig.~\ref{nglue}.  
The lowest $x$ values are probed by the lightest nuclei since the center of
mass energy is higher for nuclei with $Z/A \sim 0.5$ than heavier, neutron-rich
nuclei with lower $Z/A$.  
The differences in the shadowing ratios for a given parameterization
are greatest at large negative $y$ where $x$ is largest.  As $A$ increases
and $\sqrt{s_{_{NN}}}$ decreases, the antishadowing peak moves closer to
midrapidity (less negative $y$).  Increasing the scale from that appropriate
for the $J/\psi$ to that for the $\Upsilon$ also moves the antishadowing
peak closer to $y=0$.  For example, the EKS98 antishadowing peak is
fully visible for $\Upsilon$ production, ocurring at $y \sim -3.5$, while they
only appear at $y \leq -5$ for the $J/\psi$.  As $\mu^2$ increases, the
differences in the EPS09 sets becomes more pronounced at large $x$, leading
to the more irregular shapes of the upper and lower limits of the EPS09
uncertainty range at negative rapidity.  Note that the central ratio is smooth.
Thus, the results in 
Figs.~\ref{fig8hi} and \ref{fig8lo} suggest that by running the LHC in 
both $p+A$ and $A+p$ modes the modification of the nuclear gluon parton
density could be traced out over a wide $x$ range, taking 
advantage of the ALICE muon coverage.

Finally, we note that at $y=6$, corresponding to $x < 10^{-6}$, the EKS98 
and nDSg shadowing ratios are outside their range of validity.  This is also
near the region where DGLAP evolution of the parton densities is likely to break
down.  Nonlinear evolution of the proton
parton densities is expected at sufficiently small $x$.  The onset of these
nonlinearities is predicted to be at larger $x$ for nuclei.  However, it is
not obvious that nonlinear parton evolution automatically leads to a
{\em reduction} of the small $x$ gluon density even though the nonlinear term in
the gluon evolution has a negative sign \cite{MQ}.  The behavior of the low
$x$ gluon density cannot be determined without a complete re-evaluation of all 
the parton densities since the sea quark evolution is coupled to that of the
gluon and overall momentum conservation must be maintained along with the
integrity of the global fit.
See Ref.~\cite{EHKQS} for details of modified parton densities based on
nonlinear DGLAP evolution and
Refs.~\cite{EKV2,BDEKV,heralhcyr} for a discussion of the 
possible effect on charm production at the LHC.

Since it is more likely that the best $p+p$ reference data will be at
$\sqrt{s} = 14$ TeV or 10 TeV for the initial LHC run, Figs.~\ref{fig9hi}
and \ref{fig9lo}
show the $(p+A)/(p+p)$ ratios with the $p+p$ reference at 14 TeV.  The 
magnitude of the two ratios (for $p+A$ collisions in the equal-speed 
frame, Fig.~\ref{fig9hi}, and at the same energy as the corresponding 
$A+A$ collisions, Fig.~\ref{fig9lo})
is due to the difference in $\sqrt{s_{_{NN}}}$ relative to 14 TeV.  The
$p+A$ ratios with the $p+A$ center-of-mass energies equal to those of
$A+A$ collisions are lower.  The rapidity distributions narrow while their
magnitudes are reduced with decreasing $\sqrt{s_{_{NN}}}$.  Thus 
the $(p+A)/(p+p)$ ratios
without shadowing decrease steadily from 9.9 to 5.5 TeV while the narrowing
of the ratios becomes more pronounced.

The symmetric solid curves in Figs.~\ref{fig9hi} and
\ref{fig9lo} are the $(p+A)/(p+p)$ ratios without
shadowing.  Shadowing results in asymmetric ratios but since 
the $p+A$ phase space is narrower than that of 14 TeV $p+p$ collisions, 
the ratios in these figures turn over and drop to zero
at large $|y|$.  The narrower phase space has a bigger effect on the $\Upsilon$
production ratios since the full
$\Upsilon$ rapidity range is within $|y|<6$ while the $J/\psi$ $y$ distribution
is broader.
The antishadowing peak is lowered and broadened when dividing by the 14 TeV
$p+p$ rapidity distribution and is only really apparent for the EKS98 and
EPS09 parameterizations.  The maximum shadowing
allowed by EPS09 shows the most 
asymmetric curvature, especially for $J/\psi$.  The EPS09 ratios suggest that
the effect could either be large, as suggested by the EPS08 analysis, or 
small enough to be effectively indistinguishable from no shadowing.
It will thus be harder to 
differentiate between shadowing parameterizations when employing 
the higher energy $p+p$ reference.

As discussed previously, there is an additional complication due to the
rapidity shift of the $p+A$ center of rapidity in the equal-speed frame.
The shift increases with $A$ as $Z/A$ decreases, reducing the energy of the
ion beam relative to the proton beam.  This results in 
nearly half a unit rapidity shift
in $p$Pb collisions, as shown in the center part of Table~\ref{delytable},
labeled $p+A$.  The $(p+A)/(p+p)$ ratios including the rapidity shift 
and the maximum
energy $p+p$ reference are shown in Fig.~\ref{fig10}.  Note that only the
$p+A$ results in the equal-speed frame are shown.  Since the proton beam
momentum in $p+A$ collisions at the
$A+A$ center-of-mass energy must be the same as that
of the ion beam, $\Delta y_{\rm cm}^{pA} = 0$.  The $p+A$ rapidity distribution
is given a positive shift, to the right, since the proton beam, at higher $y$,
is assumed to come from the left and move to the right.  Thus, at large 
negative $y$, the ratios are lower than in Figs.~\ref{fig9hi} and \ref{fig9lo}
and are flatter as a function of rapidity.  While the nuclear effects
on the parton densities are most difficult to disentangle here, this scenario
is the most realistic.  (It may be possible to eliminate or reduce the effect
of the rapidity shift by employing different rapidity cuts to compare
distributions.)  If the LHC is run with the proton
and ion beam directions reversed, the antishadowing peak may be enhanced
and the large positive rapidity ratios decreased.  

\begin{figure}[htbp]
\begin{center}
\includegraphics[width=0.5\textwidth]{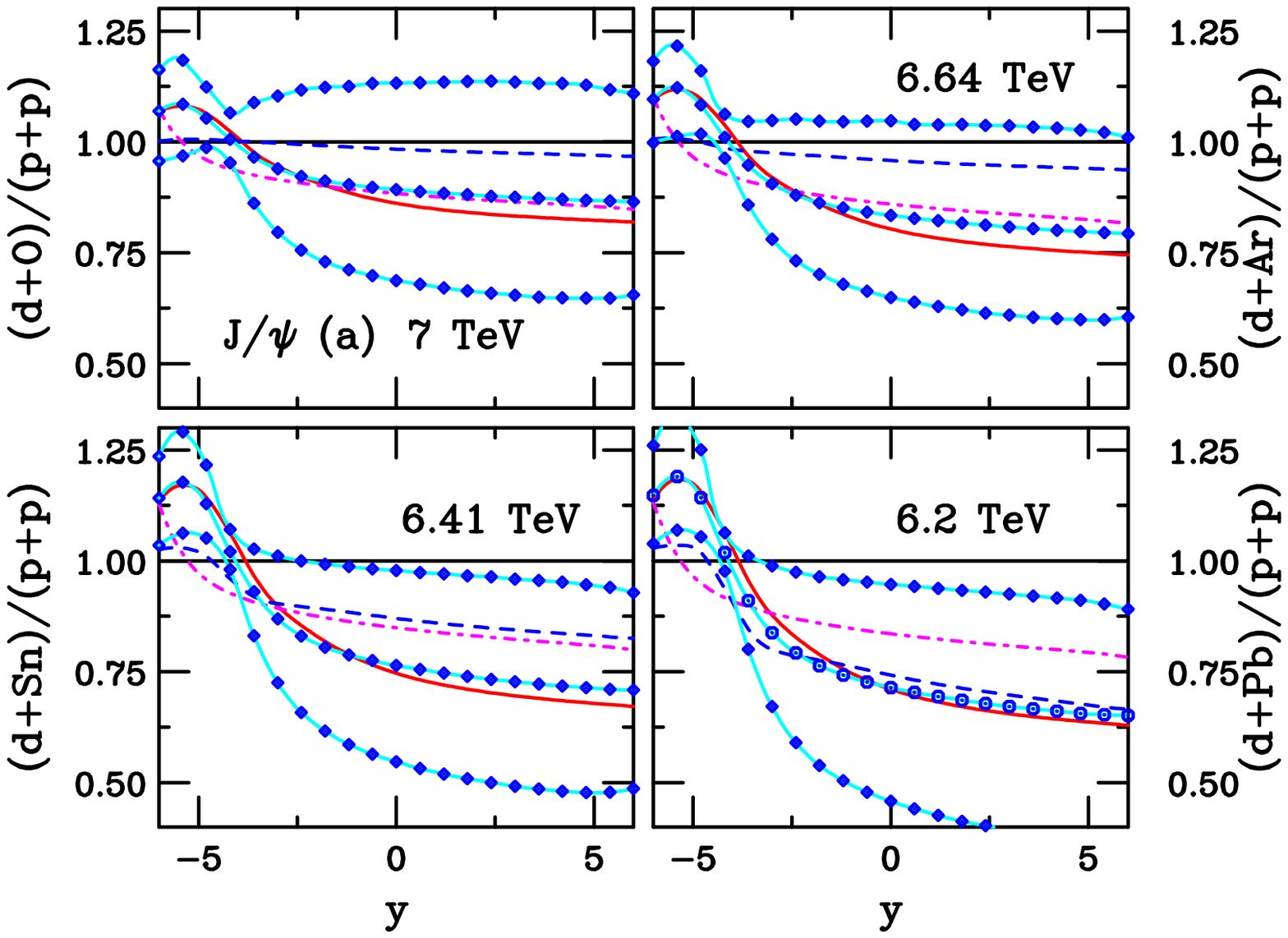} \\
\includegraphics[width=0.5\textwidth]{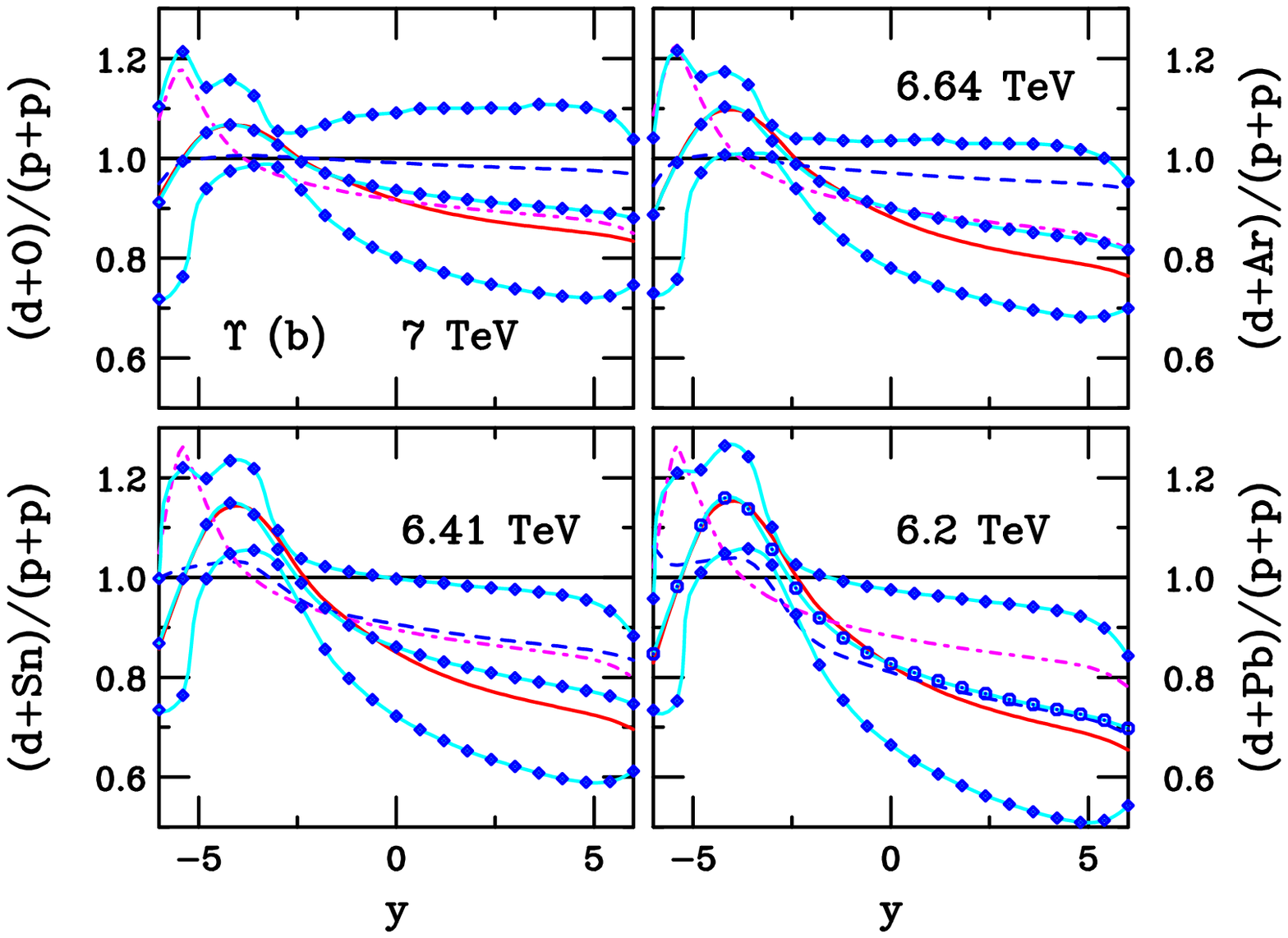}
\end{center}
\caption[]{(Color online)
The (d$+A)/(p+p)$ ratios with both d$+A$ and $p+p$ collisions at the d$+A$
energy in the equal-speed frame.  No rapidity shift has been taken into account.
The effect of 
shadowing on $J/\psi$ (a, upper 4 panels) and $\Upsilon$ (b, lower 4 panels) 
production is shown.  Each set of panels displays the production ratios
for d+O at $\sqrt{s_{_{NN}}} = 7$ TeV (upper left), d+Ar at $\sqrt{s_{_{NN}}} 
= 6.64$ TeV 
(upper right), d+Sn at $\sqrt{s_{_{NN}}} = 6.41$ TeV (lower left) and d+Pb at 
$\sqrt{s_{_{NN}}} = 6.2$ TeV (lower right).  The calculations are with 
CTEQ6 and employ the EKS98 (solid),
nDSg (dashed), HKN (dot-dashed) and EPS09 (solid curves with
symbols) shadowing parameterizations.
}
\label{fig11}
\end{figure}

The effect of the rapidity shift is reduced if d$+A$ collisions are run 
instead of $p+A$ collisions.  The d$+A$ center-of-mass energy is closer to that
of $A+A$ collisions since $Z/A < 1$ for the deuteron rather than equal to
1 as for protons.  The ratios with the d$+A$ and $p+p$ collisions at the same 
center-of-mass energy per nucleon are shown in 
Fig.~\ref{fig11} (similar to Fig.~\ref{fig8lo} for $p+A$).  
They are like those in Fig.~\ref{fig8lo}
with equal $p+A$ and $A+A$ center-of-mass energies
since $\sqrt{s_{_{NN}}}$ is similar for d$+A$ and $A+A$ collisions.
Shadowing effects on the deuteron are assumed to be negligible.

\begin{figure}[htbp]
\begin{center}
\includegraphics[width=0.5\textwidth]{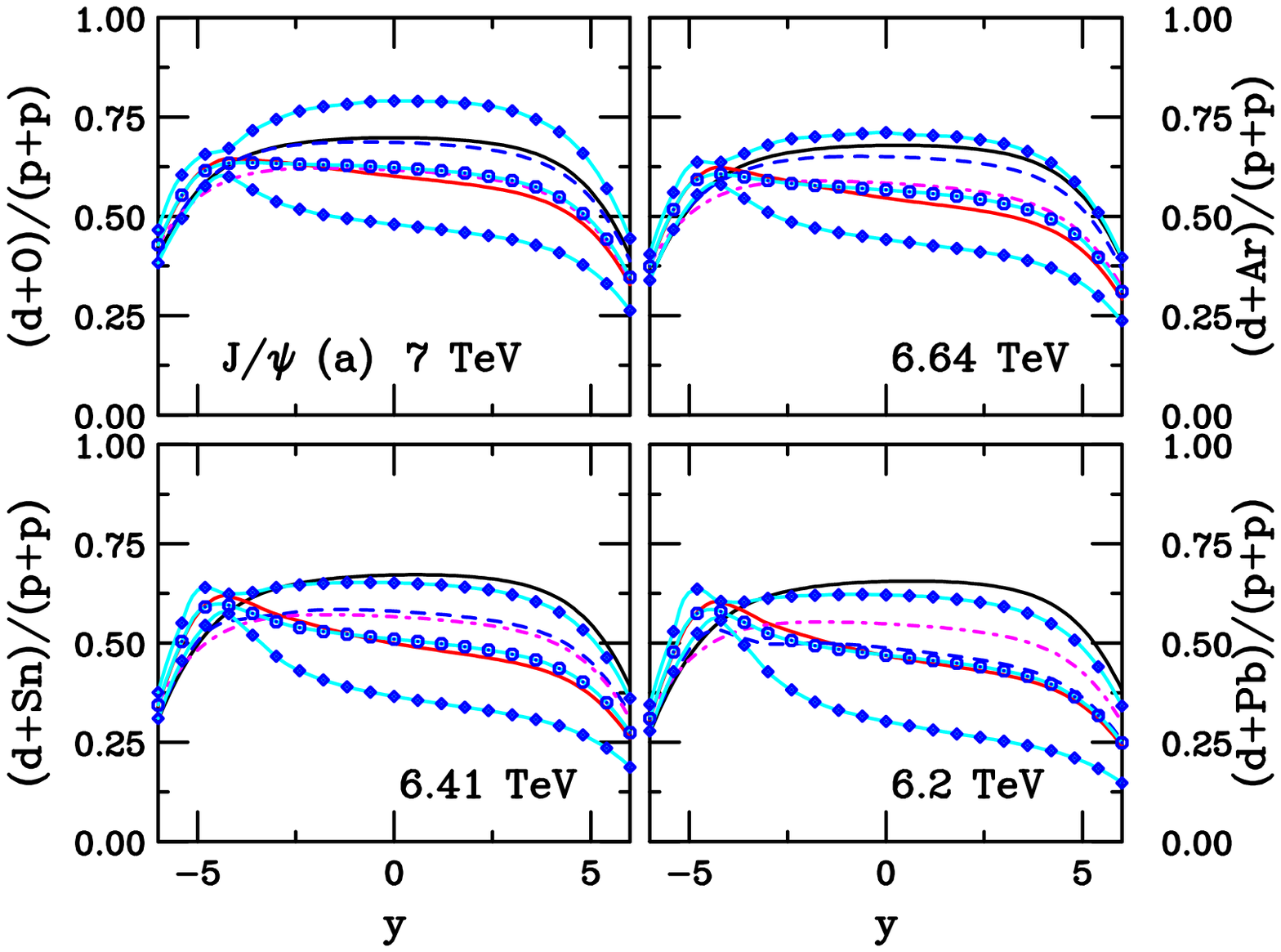} \\
\includegraphics[width=0.5\textwidth]{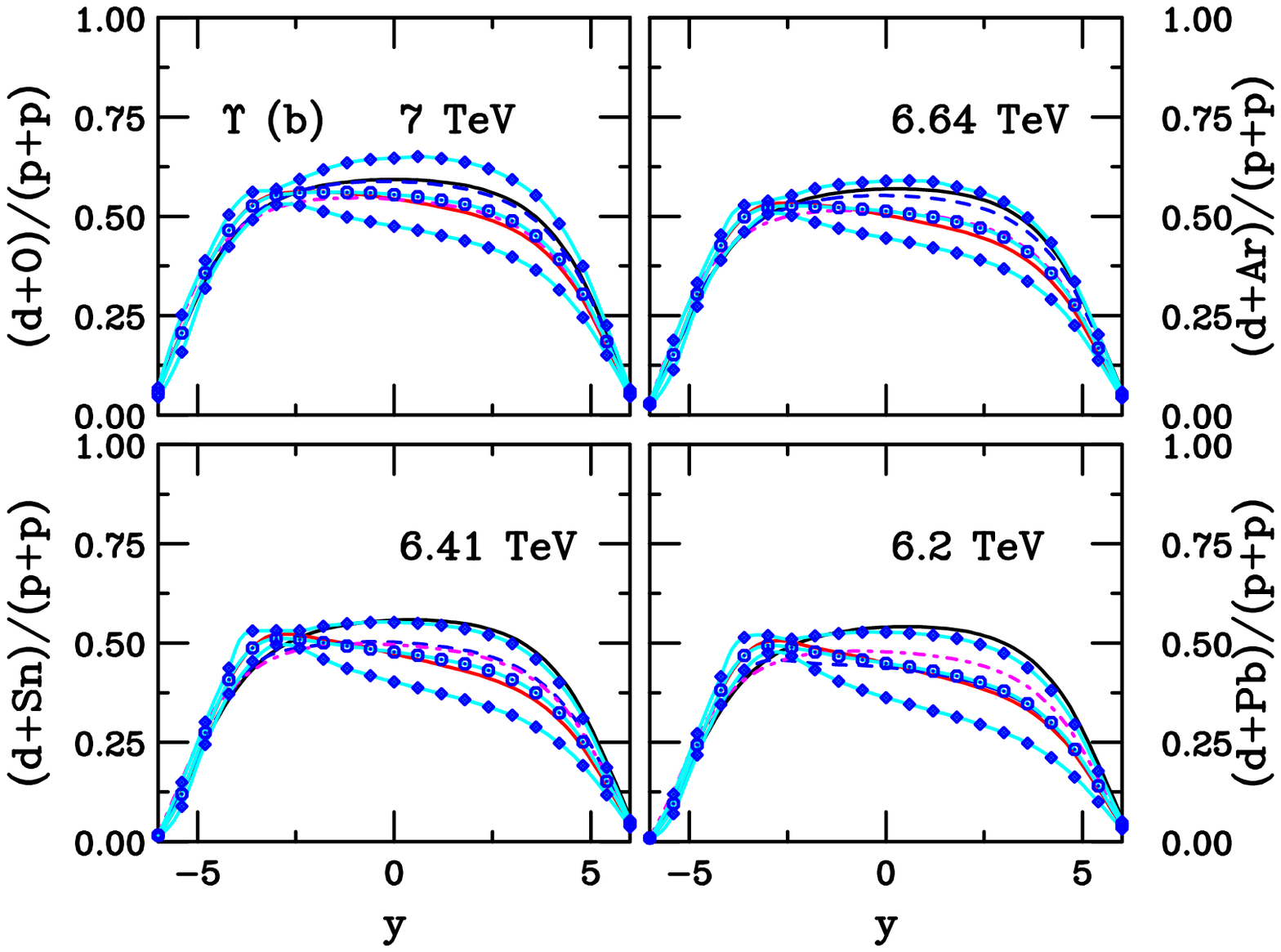}
\end{center}
\caption[]{(Color online) 
The (d$+A)/(p+p)$ ratios with the $p+p$ distributions calculated at 
$\sqrt{s} = 14$ TeV and the d$+A$ rapidity shift taken into account.  The effect
of shadowing on $J/\psi$ (a, upper 4 panels) and $\Upsilon$ (b, lower 4 panels) 
production is shown.  Each set of panels dispalys the production ratios
for d+O at $\sqrt{s_{_{NN}}} = 7$ TeV (upper left), d+Ar at $\sqrt{s_{_{NN}}} 
= 6.64$ TeV 
(upper right), d+Sn at $\sqrt{s_{_{NN}}} = 6.41$ TeV (lower left) and d+Pb at 
$\sqrt{s_{_{NN}}} = 6.2$ TeV (lower right).  The calculations are with 
CTEQ6 and employ the EKS98 (solid),
nDSg (dashed), HKN (dot-dashed) and EPS09 (solid curves
with symbols) shadowing parameterizations.
The symmetric solid curve is the result without shadowing.
}
\label{fig12}
\end{figure}

The results with a 14 TeV $p+p$ reference and the small rapidity shift
taken in account, see Table~\ref{delytable} for 
$\Delta y_{\rm cm}^{ {\rm d}A}$,
are shown in Fig.~\ref{fig12}.  Recall that there is no rapidity shift for 
d+O collisions since $Z_{\rm d}/A_{\rm d} = 
Z_{\rm O}/A_{\rm O} = 0.5$.  Thus the equal-speed
and center-of-rapidity frames coincide.  In d+Pb collisions, since
$\Delta y_{\rm cm}^{{\rm d}A} < 0.06$, the shift is negliglble.  Thus the 
d$+A$ rapidity distributions relative
to the 14 TeV $p+p$ reference {\em with} the rapidity shift, shown in 
Fig.~\ref{fig12}, are similar to those
in Fig.~\ref{fig9lo} with $\Delta y_{\rm cm}^{pA} = 0$ and 
the same $\sqrt{s_{_{NN}}}$ in
$p+A$ and $A+A$ collisions.  Note, however, that the
ratios in Fig.~\ref{fig12} are somewhat closer to unity
since the d$+A$ center-of-mass energy is larger.  Thus the more realistic d$+A$ 
scenario shown in Fig.~\ref{fig12} would be preferable for determining 
nuclear effects on the parton densities both because of the relatively
similar center-of-mass energies and the smaller rapidity shift.

\begin{figure}[htbp]
\begin{center}
\includegraphics[width=0.5\textwidth]{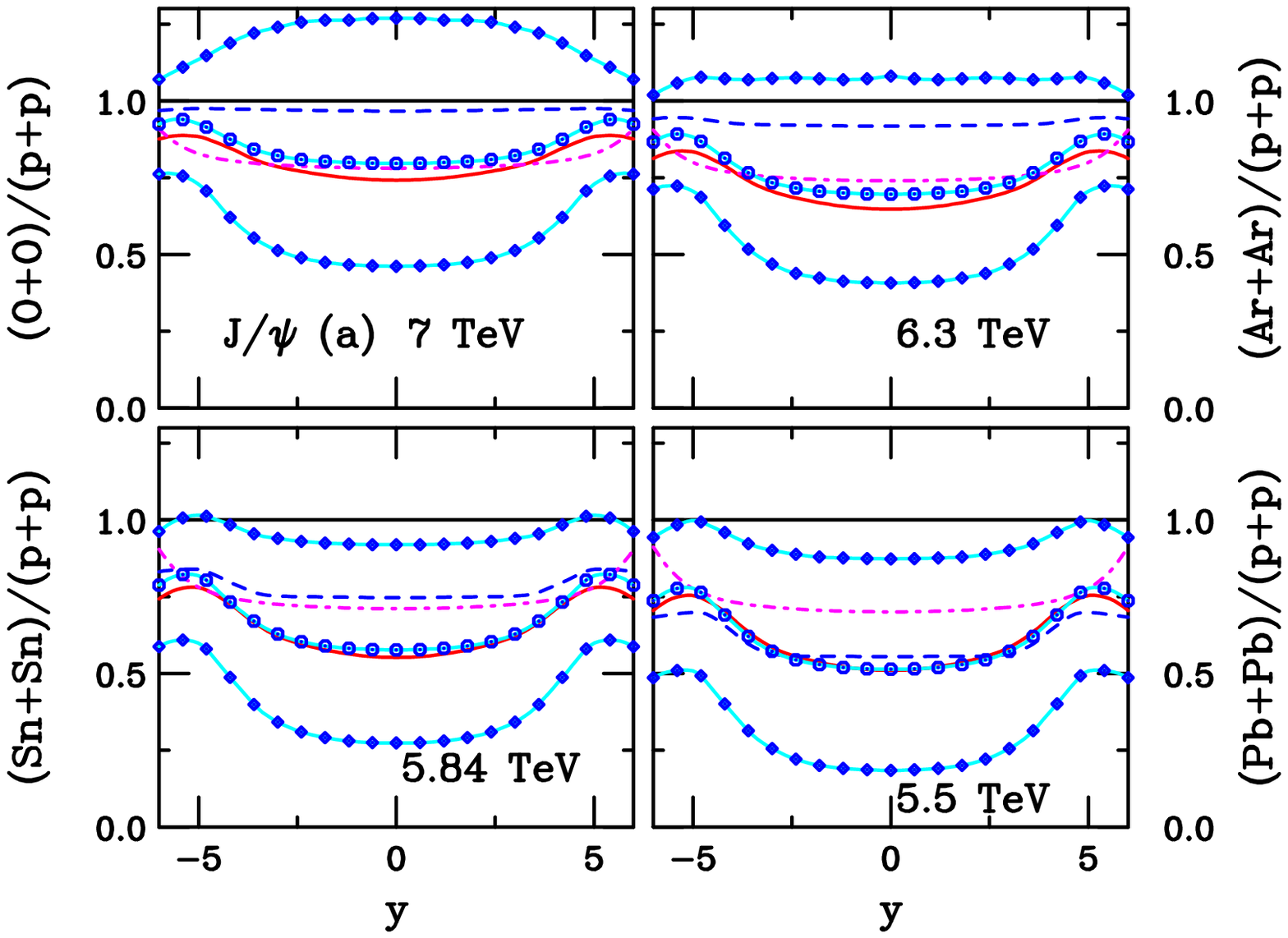} \\
\includegraphics[width=0.5\textwidth]{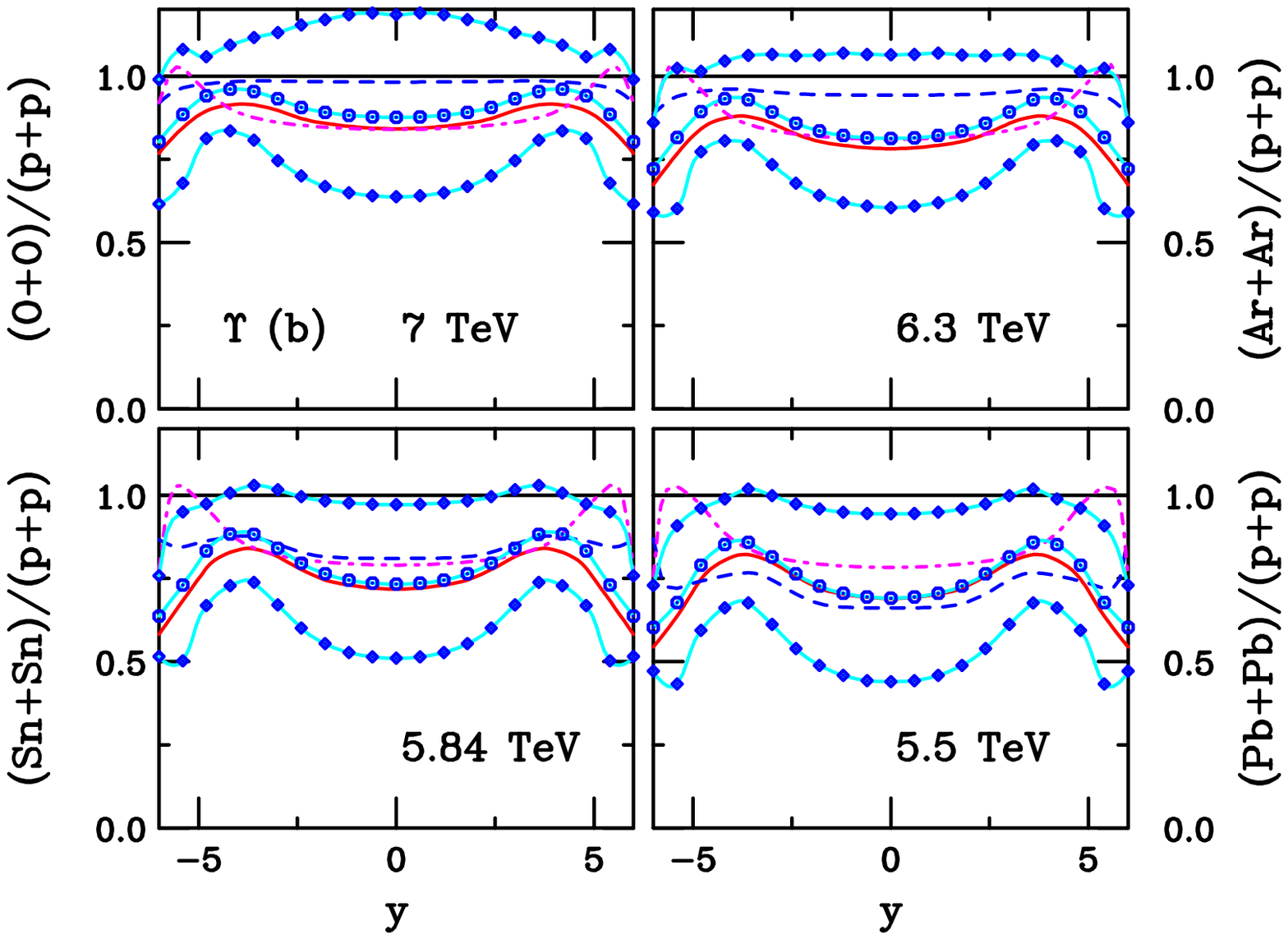}
\end{center}
\caption[]{(Color online) The $(A+A)/(p+p)$ ratios with both $A+A$ and 
$p+p$ collisions calculated at the $A+A$
center-of-mass energy. The effect of 
shadowing on $J/\psi$ (a, upper 4 panels) and $\Upsilon$ (b, lower 4 panels) 
production is shown.  Each set of panels displays the production ratios
for O+O at $\sqrt{s_{_{NN}}} = 7$ TeV (upper left), Ar+Ar at 
$\sqrt{s_{_{NN}}} = 6.3$ TeV (upper right), Sn+Sn at $\sqrt{s_{_{NN}}} = 
6.14$ TeV (lower left) and Pb+Pb at $\sqrt{s_{_{NN}}} = 5.5$ TeV (lower right).
The calculations are with CTEQ6 and employ the EKS98 (solid),
nDSg (dashed), HKN (dot-dashed) and EPS09 (solid curves with
symbols) shadowing parameterizations.
}
\label{fig13}
\end{figure}

\begin{figure}[htbp]
\begin{center}
\includegraphics[width=0.5\textwidth]{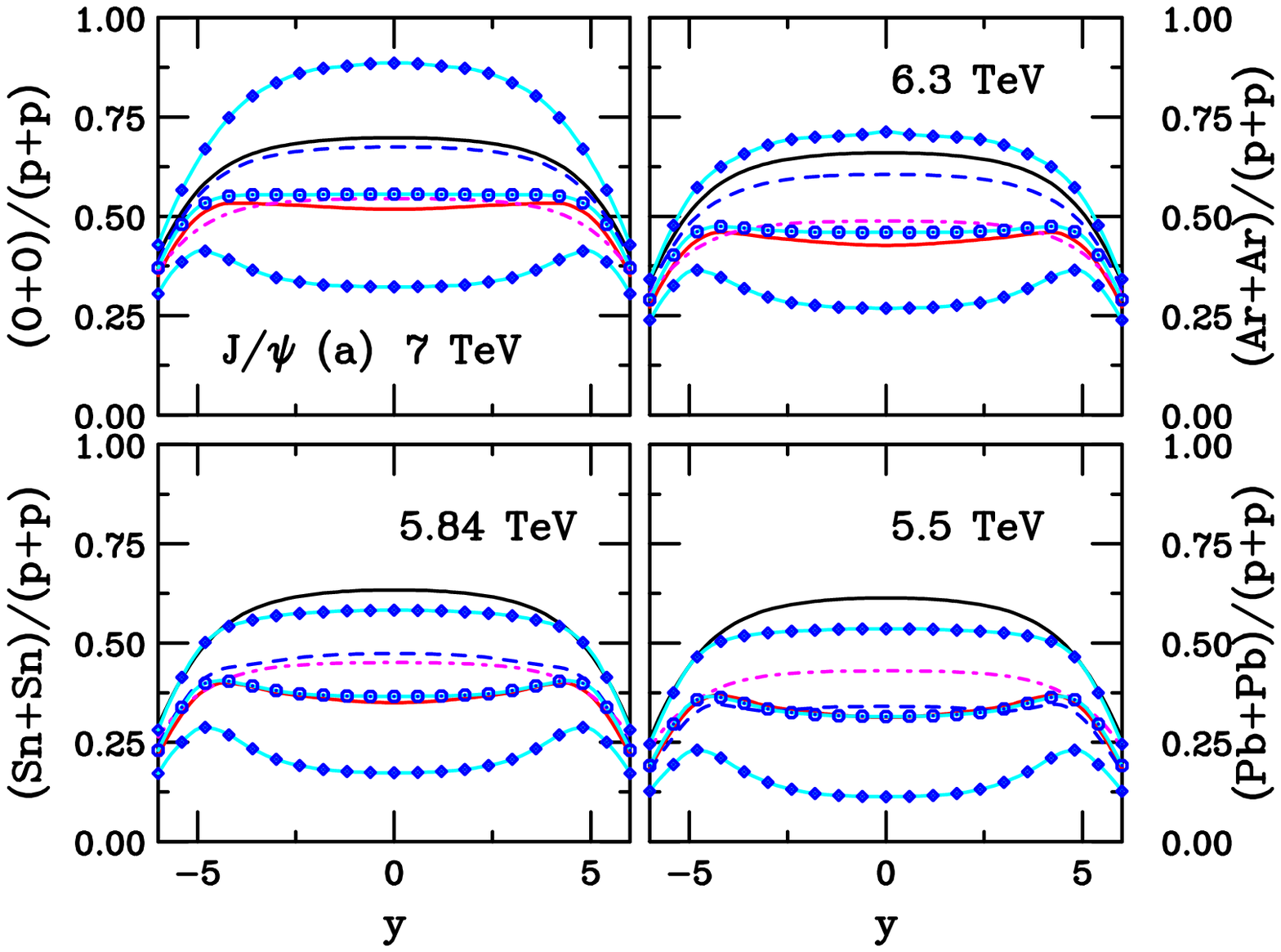} \\
\includegraphics[width=0.5\textwidth]{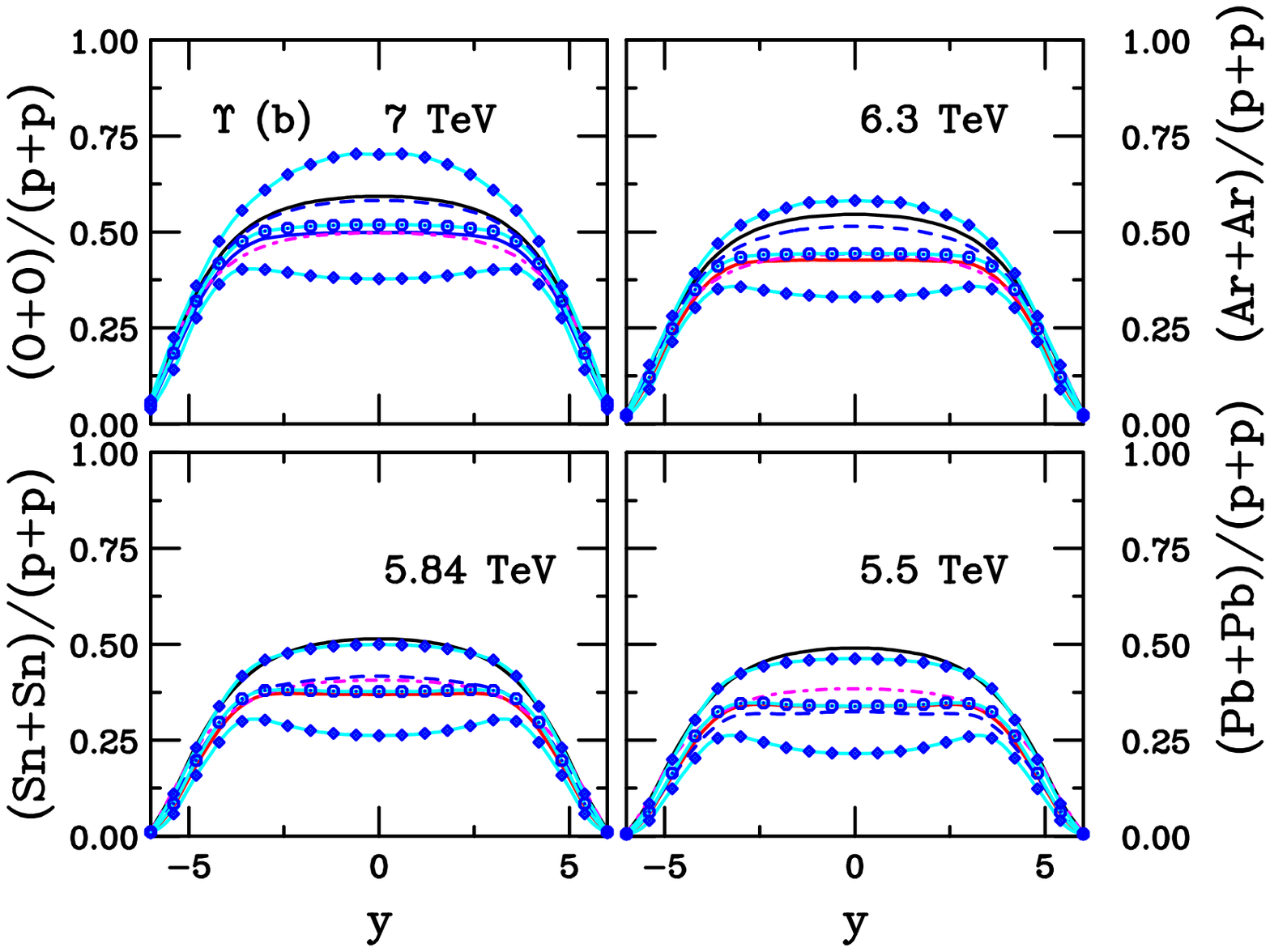}
\end{center}
\caption[]{(Color online)
The $(A+A)/(p+p)$ ratios with the $p+p$ rapidity distributions calculated 
at $\sqrt{s} = 14$ TeV.  The effect of 
shadowing on $J/\psi$ (a, upper 4 panels) and $\Upsilon$ (b, lower 4 panels) 
production is shown.  Each set of panels displays the production ratios
for O+O at $\sqrt{s_{_{NN}}} = 7$ TeV (upper left), Ar+Ar at $\sqrt{s_{_{NN}}} = 6.3$ TeV 
(upper right), Sn+Sn at $\sqrt{s_{_{NN}}} = 6.14$ TeV (lower left) and Pb+Pb at
$\sqrt{s_{_{NN}}} = 5.5$ TeV (lower right).  The calculations are with CTEQ6 
and employ the EKS98 (solid),
nDSg (dashed), HKN (dot-dashed) and EPS09 (solid curves
with symbols) shadowing parameterizations.
The upper solid curve is the $(A+A)/(p+p)$ ratio without shadowing.
}
\label{fig14}
\end{figure}

We now extrapolate to $A+A$ interactions to show the projected CNM effects from
shadowing alone.
The results for $A+A$ collisions are shown in Figs.~\ref{fig13} and 
\ref{fig14}.  The $(A+A)/(p+p)$ ratio with both systems calculated at the $A+A$
center-of-mass energy are shown in Fig.~\ref{fig13} while the 14 TeV $p+p$
reference is employed to obtain the ratios in Fig.~\ref{fig14}.  
The results in Fig.~\ref{fig13} are essentially the 
convolutions of the $(p+A)/(p+p)$ ratios (with the same $\sqrt{s_{_{NN}}}$ for
both systems and no rapidity shift) shown in
Fig.~\ref{fig8lo} with their mirror image $Ap/pp$ ratios.  While the $A+A$
ratios exhibit antishadowing peaks at $y \sim \pm (4-5) $, the 
$(A+A)/(p+p)$ ratios
are less than unity everywhere because the product of the $(p+A)/(p+p)$
ratios at postive $y$ and the $(A+p)/(p+p)$ 
ratios at negative $y$ is always smaller
than one, {\it e.g.} $[(p+A)/(p+p)]_{y \sim 5} \sim 0.6 -0.75$ while 
$[(A+p)/(p+p)]_{y \sim -5}
\sim 1.2$.  Thus, when all ratios are calculated at the $A+A$ center-of-mass
energy, assuming factorization of $A+A$ collisions into a convolution of $p+A$
and $A+p$ collisions,
\begin{eqnarray}
\frac{A+A}{p+p}|_{y \sim \pm 5} = 
\frac{p+A}{p+p}|_{y \sim 5} \times 
\frac{A+p}{p+p}|_{y \sim - 5} 
< 1 \, \, . 
\end{eqnarray} 

Calculations of color singlet $J/\psi$ interactions in matter using the 
dipole approximation of the $J/\psi$-hadron cross section suggest that
factorization is inapplicable due to the coherence of the interaction
\cite{tuchin}.  These gluon saturation models assume the dominance of
higher-twist effects enhanced by powers of $A^{1/3}$ in $p+A$ interactions.
If these models were valid, enhanced suppression of the $J/\psi$ should
set in at large rapidity.  This does indeed seem to be the case at RHIC
where $1.2< y < 2.2$ corresponds to $0.0045>x_2>0.0017$ \cite{rhicii}.  
However, the forward $x_F$ data at $\sqrt{s} = 38$ GeV [$0.2 < x_F 0.8$
and $0.027 > x_2 > 0.008$] and 17 GeV [$0.1<x_F<0.35$ and $0.13>x_2>0.07$], 
in particular, exhibit the same trends as at RHIC \cite{carlosect}.  These
fixed-target $x_2$ ranges lie in the transition region from antishadowing to
shadowing (38 GeV) and in the antishadowing region (17 GeV), see 
Fig.~\ref{nglue}, seemingly too
large to be in the saturation region, especially at $\sqrt{s} = 17$ GeV.

As is the case for the RHIC $A+A$ calculations at 
$\sqrt{s_{_{NN}}} =200$ GeV \cite{rvhip}, 
there is typically more suppression predicted at
$y=0$ than at more forward and backward rapidities for all the shadowing
parameterizations as well as for both $J/\psi$ and $\Upsilon$ production.
At RHIC, the $A+A$ data are more suppressed at forward rapidity than at
central rapidity, both in the minimum bias data as a function of rapidity
and as a function of collision centrality, as quantified by the number of
participant nucleons.  Standard models of shadowing alone or shadowing with
absorption by nucleons in cold nuclear matter or shadowing combined with
dissociation in a quark-gluon plasma leads to strong suppression at central
rapidities.  However, $J/\psi$ regeneration by coalescence of $c$ and 
$\overline c$ quarks in the medium \cite{hvqyr,bob}
is biased toward central rapidities and could
lead to more suppression at forward rapidity relative to central rapidity
since the rapidity distribution of $J/\psi$ production by coalescence is
expected to be narrower than the initial $J/\psi$ rapidity distribution 
\cite{bob}.  Thus, with coalescence, there should be more suppression at forward
$y$ than at midrapidity.  The same trend should hold at the LHC.  
Coalescence production of the $J/\psi$ should be even
more important than at RHIC since more $c \overline c$ pairs are created in a
central Pb+Pb collision at $\sqrt{s_{_{NN}}} = 5.5$ TeV.  We can also expect
that $\Upsilon$ production by coalescence may be similar to that expected
for the 
$J/\psi$ at RHIC since the $b \overline b $ production cross section at the LHC
will be similar to the $c \overline c$ production cross section at RHIC
\cite{rhicii}.

The $(A+A)/(p+p)$ ratios with the 14 TeV $p+p$ reference, shown in Fig.~\ref{fig14},
are relatively flat.  The dip around midrapidity has been washed out, except for
the $J/\psi$ ratios calculated with the EKS98 and EPS09 (central and
maximum shadowing) parameterizations
where some indication remains.
For comparison, the $(A+A)/(p+p)$ ratios without shadowing are shown in the upper 
solid curves.

\subsection{Impact parameter dependence}

We now discuss the impact parameter dependence of quarkonium production
at the LHC.
Unfortunately, there is little relevant data on the spatial dependence
of shadowing.  Fermilab experiment
E745 studied the spatial distribution of nuclear structure functions
with $\nu N$ interactions in emulsion.  The presence of one or more
dark tracks from slow protons is used to infer a more central
interaction \cite{E745}.  For events with no dark tracks, no shadowing
is observed while, for events with dark tracks, shadowing is enhanced
over spatially-independent measurements from other experiments.
Unfortunately, this data is too limited to be used in a fit of the
spatial dependence.

The minimum bias shadowing we have discussed up to now
is homogeneous, impact parameter-integrated shadowing.  The impact 
parameter-dependent results shown in 
this section portray inhomogeneous shadowing.
In central collisions, with small impact parameter, $b$, we can expect
inhomogeneous shadowing to be stronger than the 
homogeneous result.  
In peripheral (large impact parameter) collisions, inhomogeneous
effects are weaker than the homogeneous results but some
shadowing is still present due to the overlapping tails of the density
distributions.  The
stronger the homogeneous shadowing, the larger the difference between the
central and peripheral results.  

We assume that the shadowing is proportional to the parton path through
the nucleus \cite{KVpsi},
\begin{eqnarray} 
S^i_{{\rm P},\rho} (A,x,Q^2,\vec{r},z) = 1 + N_\rho (S^i_{\rm P}(A,x,Q^2) - 1) 
\frac{\int dz
\rho_A(\vec{r},z)}{\int dz \rho_A(0,z)} \label{rhoparam} \, \, ,
\end{eqnarray}
where $N_\rho$ is chosen to satisfy the normalization condition in
Eq.~(\ref{snorm}).
The integral over $z$ in Eq.~(\ref{rhoparam}) 
includes the material traversed by
the incident nucleon. At large distances, $s \gg R_A$, the nucleons
behave as free particles, while in the center of the nucleus, the
modifications are larger than the average value $S^i_{\rm P}$.

We calculate the nuclear suppression factor, $R_{AB}$, for $p+A$, d$+A$ and $A+A$
collisions.  The suppression factor is defined as the ratio
\cite{raadef}
\begin{equation}
R_{AB}(N_{\rm part};b)  = {d \sigma_{AB}/dy \over T_{AB}(b) d\sigma_{pp}/dy }
\label{rab}
\end{equation}
where $d\sigma_{AB}/dy$ and $d\sigma_{pp}/dy$ are the quarkonium rapidity
distributions in $A+B$ and $p+p$
collisions and $T_{AB}$ is the nuclear overlap function,
\begin{eqnarray}
T_{A B} (b) = \int d^2s dz dz' \rho_A(s,z) \rho_B(|\vec{b}-\vec{s}|,z') \, \, .
\label{tabdef}
\end{eqnarray}
In $p+A$ collisions, we assume that the proton has a negligible size,
$\rho_A(s,z) = \delta(s) \delta (z)$ so that $T_{AB}(b)$ 
collapses to the nuclear profile function $T_B(b) = \int dz' \rho_B(b,z')$.
The deuteron cannot be treated like a point particle since it is large
and diffuse.  We use the H\'{u}lthen
wave function \cite{hulthen} to calculate the deuteron density
distribution.  However, we do not include shadowing effects on the deuteron.

\begin{figure}[htbp]
\centering
\includegraphics[width=0.75\textwidth]{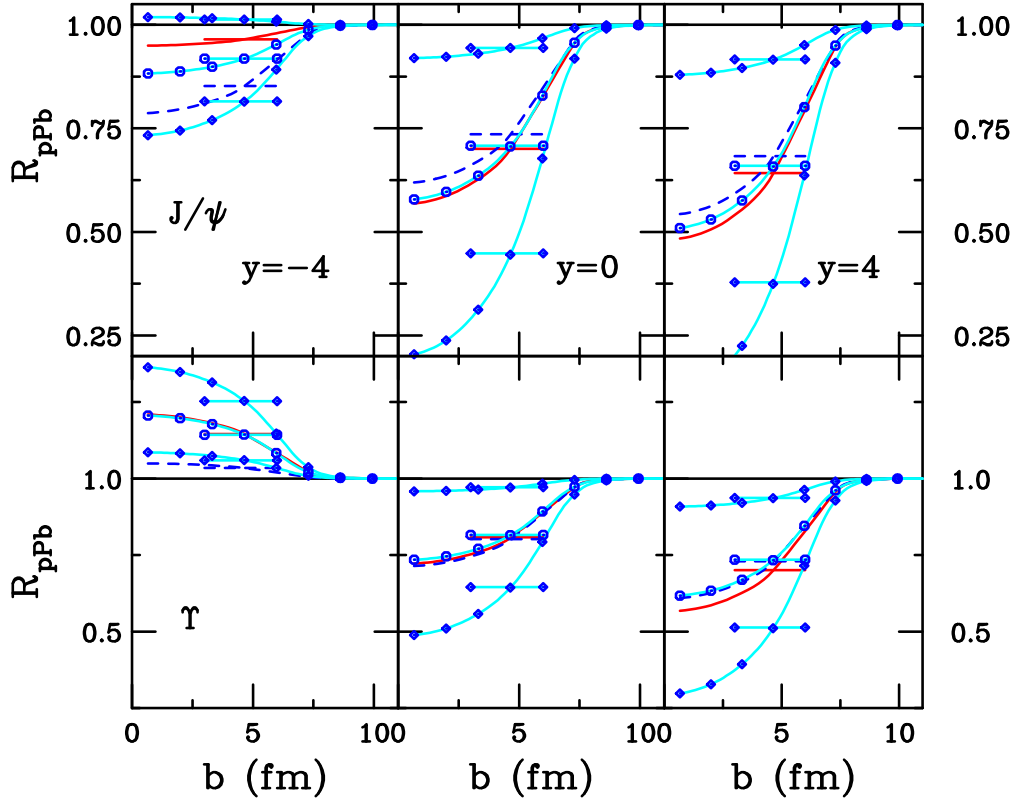}
\caption[]{(Color online)
The suppression factor $R_{p{\rm Pb}}$ at $y=-4$ (left), 0 (center)
and 4 (right) as a function of $b$.  The 
result is shown for $J/\psi$ (top) and $\Upsilon$ (bottom) in $p$+Pb 
relative to $p+p$ collisions at the same energy, $\sqrt{s_{_{NN}}} = 8.8$ TeV,
and employ the EKS98 (solid),
nDSg (dashed) and EPS09 (solid curves with
symbols) shadowing parameterizations.  The horizontal
lines show the impact-parameter integrated results.
}
\label{fig17}
\end{figure}

\begin{figure}[htbp]
\centering
\includegraphics[width=0.75\textwidth]{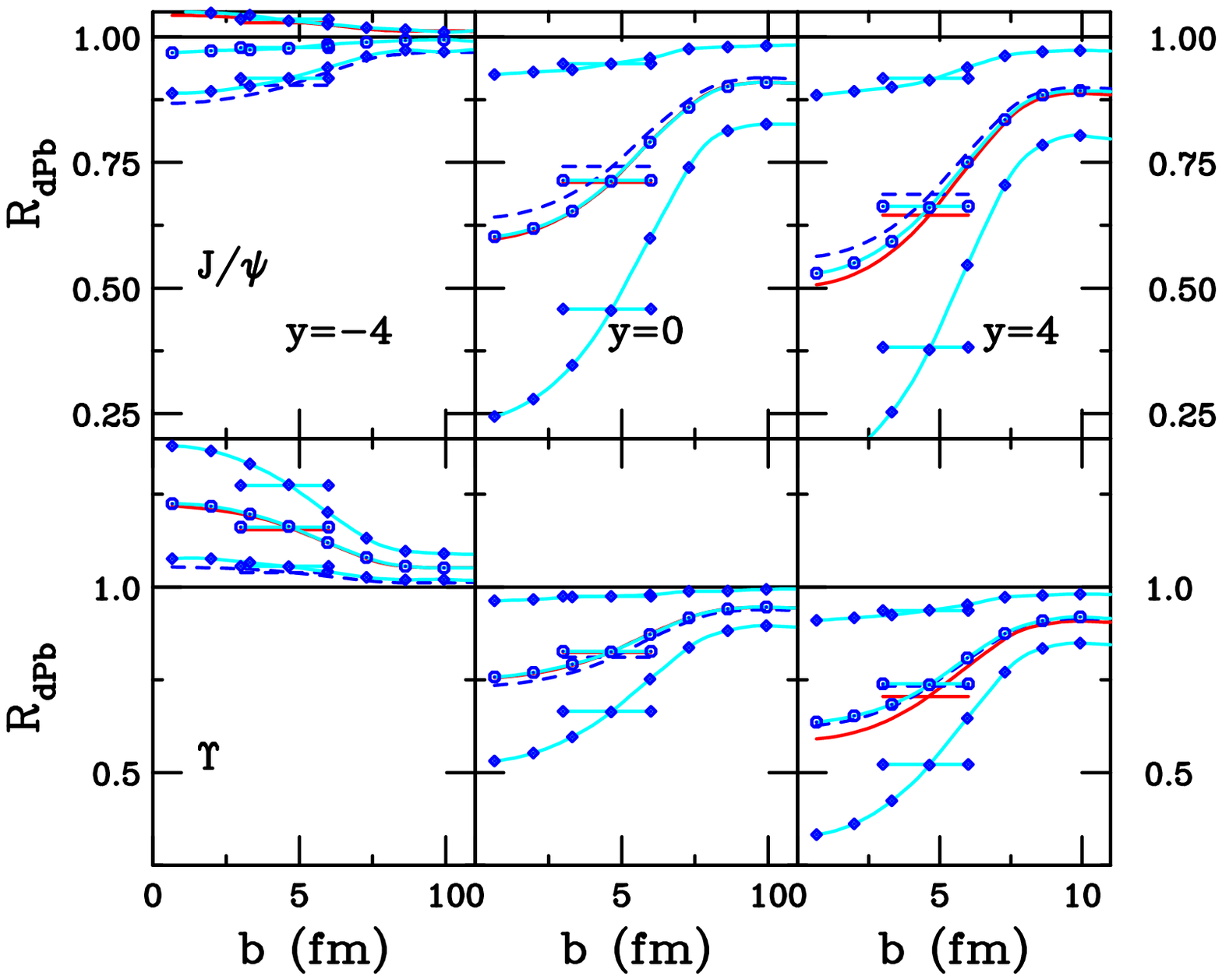}
\caption[]{(Color online)
The suppression factor $R_{\rm d Pb}$ at $y=-4$ (left), 0 (center)
and 4 (right) as a function of $b$.  The 
result is shown for $J/\psi$ (top) and $\Upsilon$ (bottom) in d+Pb 
relative to $p+p$ collisions at the same energy, $\sqrt{s_{_{NN}}} = 6.2$ TeV,
and employ the EKS98 (solid),
nDSg (dashed) and EPS09 (solid curves with
symbols) shadowing parameterizations.  The horizontal
lines show the impact-parameter integrated results.
}
\label{fig18}
\end{figure}

We show the $p+$Pb and d+Pb suppression factors as a function of impact
parameter with $\sqrt{s_{_{NN}}} = 8.8$ TeV and 6.2 TeV in the numerator and
denominator in Figs.~\ref{fig17} and \ref{fig18} respectively.  We concentrate
on the largest $A$ ion, Pb, to maximize the relevant impact parameter range.  
The results in Fig.~\ref{fig17}
are given for three values of rapidity: $y = -4$ (backward rapidity, in the
antishadowing region for $\Upsilon$), $y=0$ (midrapidity) and $y=4$ 
(forward rapidity, where fairly strong shadowing is expected).  
We present $J/\psi$ ratios on top and $\Upsilon$ ratios on the bottom.  For
comparison, the horizontal lines, centered around the
average path length through the lead nucleus,  $b \sim (3/4)R_{\rm Pb}$,
show the impact parameter-integrated ratios
in Fig.~\ref{fig8hi}.  The $b$ dependence is strong, resulting
in $R_{p{\rm Pb}} \sim 1$ for $b > R_{\rm Pb}$.  Shadowing
is stronger in central colisions than the average integrated value,
as expected.  Because the average decreases at forward rapidities while the
spatial dependence is relatively unchanged, the strongest $b$ dependence is
seen for the most forward
rapidity value, $y=4$.   The nDSg, EKS98, and EPS09 shadowing ratios in 
Fig.~\ref{nglue} are very similar for lead nuclei.  
Thus their suppression ratios
are also similar.  Since only one nuclear density profile is involved
in the calculation of $R_{pA}$, the impact parameter dependence reflects that 
of Eq.~\ref{rhoparam} rather directly.  

A weaker impact parameter dependence is seen for d+Pb collisions in 
Fig.~\ref{fig18}.  The overall shadowing effect is reduced since the
energy is lower, $\sqrt{s_{_{NN}}} = 6.2$ TeV relative to 8.8 TeV for $p$Pb
collisions.  In addition, shadowing persists to large values of impact 
parameter.  In a heavy nucleus, the density is large and approximately
constant except close to the surface, as expressed by the Woods-Saxon
density distributions \cite{Vvv}.  However, the diffuse wavefunction of 
the deuteron has a finite amplitude at surprisingly large distances.
These long tails produce some remnant shadowing effect even at very large
$b$, as seen in Fig.~\ref{fig18}.  See also the discussion in 
Ref.~\cite{KVpsi}.  

The d+Au results at RHIC have been presented as a function of the number of
binary nucleon-nucleon collisions, $N_{\rm coll}(s_{_{NN}};b) 
= \sigma_{\rm inel}(s_{_{NN}})
T_{AB}(b)$, rather than impact parameter
itself, see Ref.~\cite{RHICdanew} for details. 
The number of collisions is greatest for the most central collisions,
$b \approx 0$, and decreases with increasing $b$.  Since the inelastic
nucleon-nucleon cross section, $\sigma_{\rm inel}(s_{_{NN}})$,
is energy dependent, the number of collisions increases with energy even 
though $T_{AB}(b)$ does not.  Thus $N_{\rm coll}(s_{_{NN}};b)$ is significantly 
larger at the LHC than at RHIC for the same $A+B$ system because of the 
considerable increase in $\sigma_{\rm inel}(s_{_{NN}})$ (from 42 mb at
$\sqrt{s_{_{NN}}} = 200$ GeV to 75 mb at $\sqrt{s_{_{NN}}}
= 5.5$ TeV.

\begin{figure}[htbp]
\begin{center}
\includegraphics[width=0.5\textwidth]{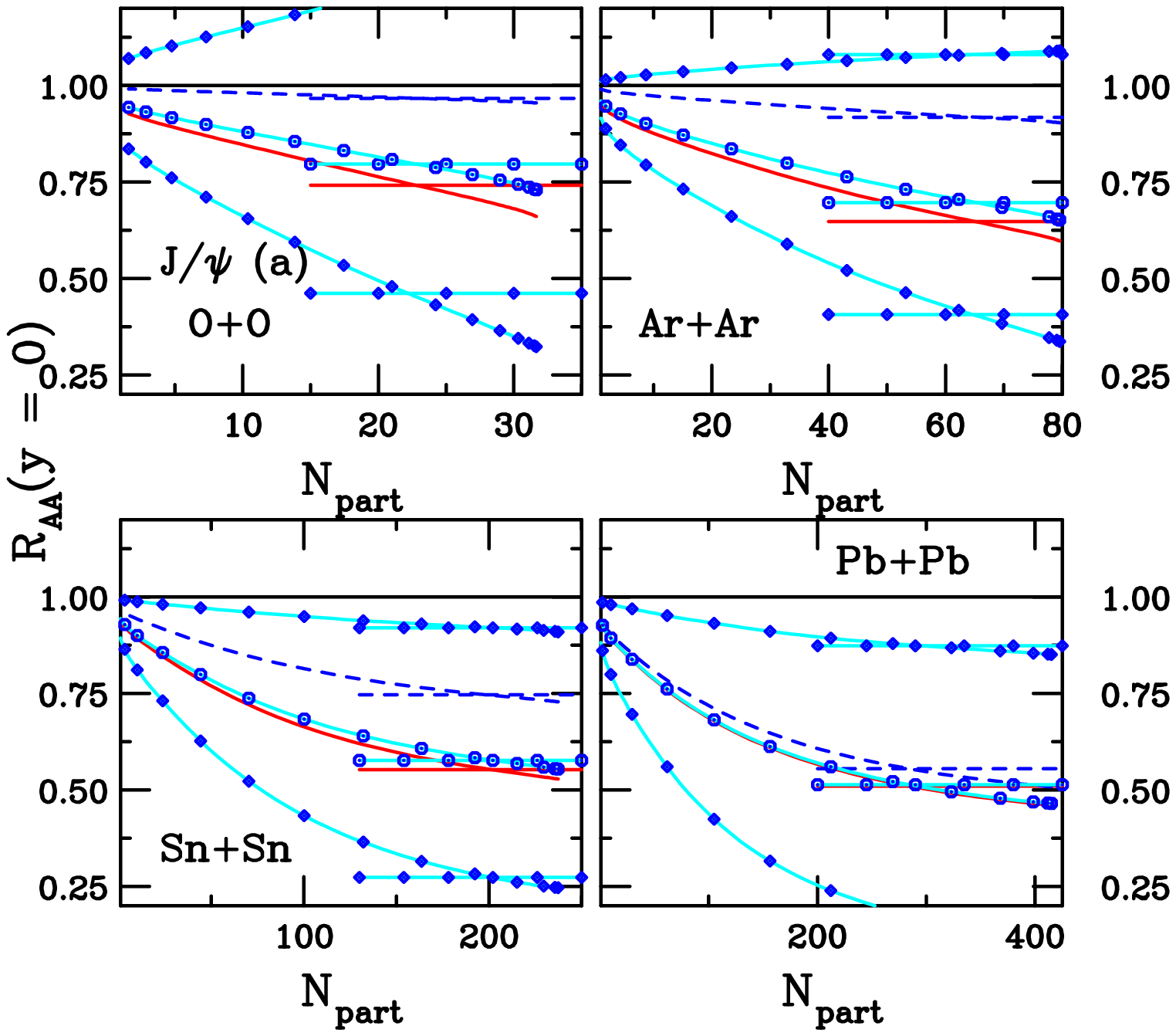} \\
\includegraphics[width=0.5\textwidth]{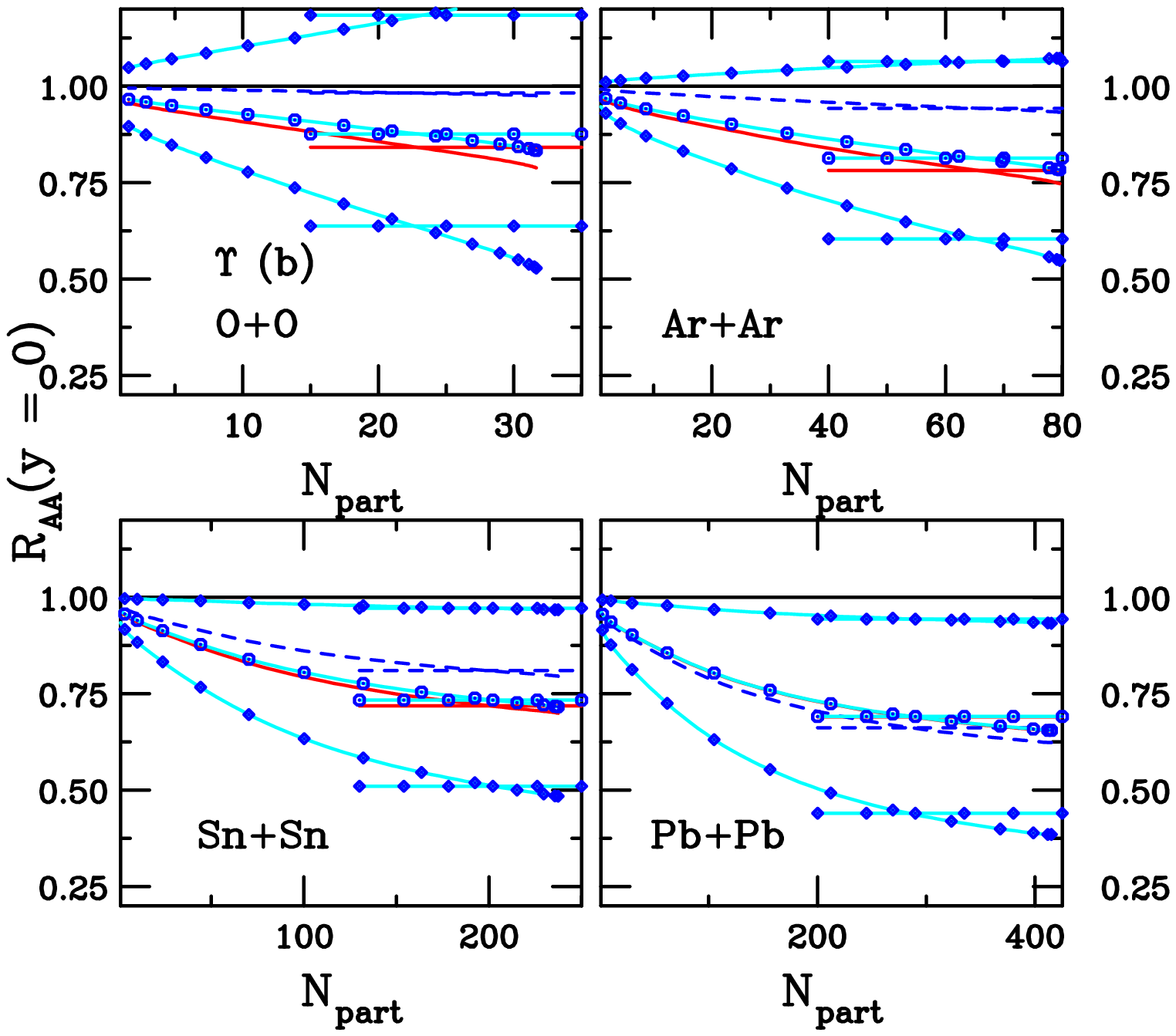}
\end{center}
\caption[]{(Color online)
The suppression factor $R_{AA}$ at $y=0$ as a function 
of $N_{\rm part}$.  The 
effect of shadowing on $J/\psi$ (a, upper 4 panels) and $\Upsilon$ (b, lower
4 panels) production is 
shown.  Each set of panels displays the suppression factor 
for O+O at $\sqrt{s_{_{NN}}} = 7$ TeV (upper left), Ar+Ar at 
$\sqrt{s_{_{NN}}} = 6.3$ TeV 
(upper right), Sn+Sn at $\sqrt{s_{_{NN}}} = 6.14$ TeV (lower left) and Pb+Pb at
$\sqrt{s_{_{NN}}} = 5.5$ TeV (lower right).  The calculations are with CTEQ6 
and employ the EKS98 (solid),
nDSg (dashed) and EPS09 (solid curves with
symbols) shadowing parameterizations.
}
\label{fig15}
\end{figure}

The results for nucleus-nucleus collisions are presented as a function of
the number of participant nucleons, $N_{\rm part}$, which depends on $b$ as
\begin{eqnarray}
N_{\rm part}(b) & = & \int \,d^2s \left[T_A(s)(1 - 
\exp[-\sigma_{\rm inel}(s_{_{NN}})
T_B(|\vec b - \vec s|)]) \right. \nonumber \\
&  & \mbox{} \left. + T_B(|\vec{b} - \vec{s}|)(1 - 
\exp[-\sigma_{\rm inel}(s_{_{NN}})
T_A(s)]) \right] \, \, . \label{npartdef}
\end{eqnarray}
Large values of $N_{\rm part}$ are obtained for small impact parameters with
$N_{\rm part}(b=0) = 2A$ for spherical nuclei.  Small values of $N_{\rm part}$
occur in very peripheral collisions.  
Figure~\ref{fig15} shows $R_{AA}(N_{\rm part})$, at $y=0$ for the
four $A+A$ systems where
the $p+p$ and $A+A$ rapidity distributions are calculated at the same 
center-of-mass energy.  A similar pattern is observed for other values of
$y$ since the $(A+A)/(p+p)$ ratios are approximately independent
of rapidity over a rather broad range.  The $(A+A)/(p+p)$ ratio at $y=0$ from
Fig.~\ref{fig13} is indicated by a horizontal line.  Note that $R_{AA}(N_{\rm
part})$ in Fig.~\ref{fig15} is equal to $(A+A)/(p+p)$ in Fig.~\ref{fig13} for
$N_{\rm part}(b \approx R_A)$.
In small systems, $R_{AA}(N_{\rm part})$
is almost linear with more curvature appearing for larger collision systems.

Since the $p+p$ reference is not likely to be immediately available at
the $A+A$ center-of-mass energy for $R_{AA}$ studies, Eq.~(\ref{rab}), 
it may be preferable to study ratios 
of two quantities measured at the same energy in $A+B$ collisions where $B = p$,
d, or $A$.  In this case, we utilize $R_{CP}$, the ratio of $A+B$ cross sections
in central relative to peripheral collisions, 
\begin{eqnarray} 
R_{CP}(y) = \frac{T_{AB}(b_P)}{T_{AB}(b_C)} 
\frac{d\sigma_{AB}(b_C)/dy}{d\sigma_{AB}(b_P)/dy} \, \, ,
\end{eqnarray} 
where $b_C$ and $b_P$ correspond to the central and peripheral values of the
impact parameter.  Indeed, shadowing may best be probed by
$R_{CP}$ measurements in asymmetric systems
since the most peripheral collisions are a good approximation to nucleon-nucleon
collisions.  The same rapidity shift is common to both central and
peripheral collisions.  We note, however, that an experimental measurement will
not be able to define a precise impact parameter but will instead define impact
parameter bins of finite width.  Thus any comparison of calculations to data
must be integrated over the width of the impact parameter bin which will average
the impact parameter dependence of the shadowing over the bin width.  Our
calculations include a width of $0.2R_A$ for the impact parameter bins.

\begin{figure}[htbp]
\centering
\includegraphics[width=0.75\textwidth]{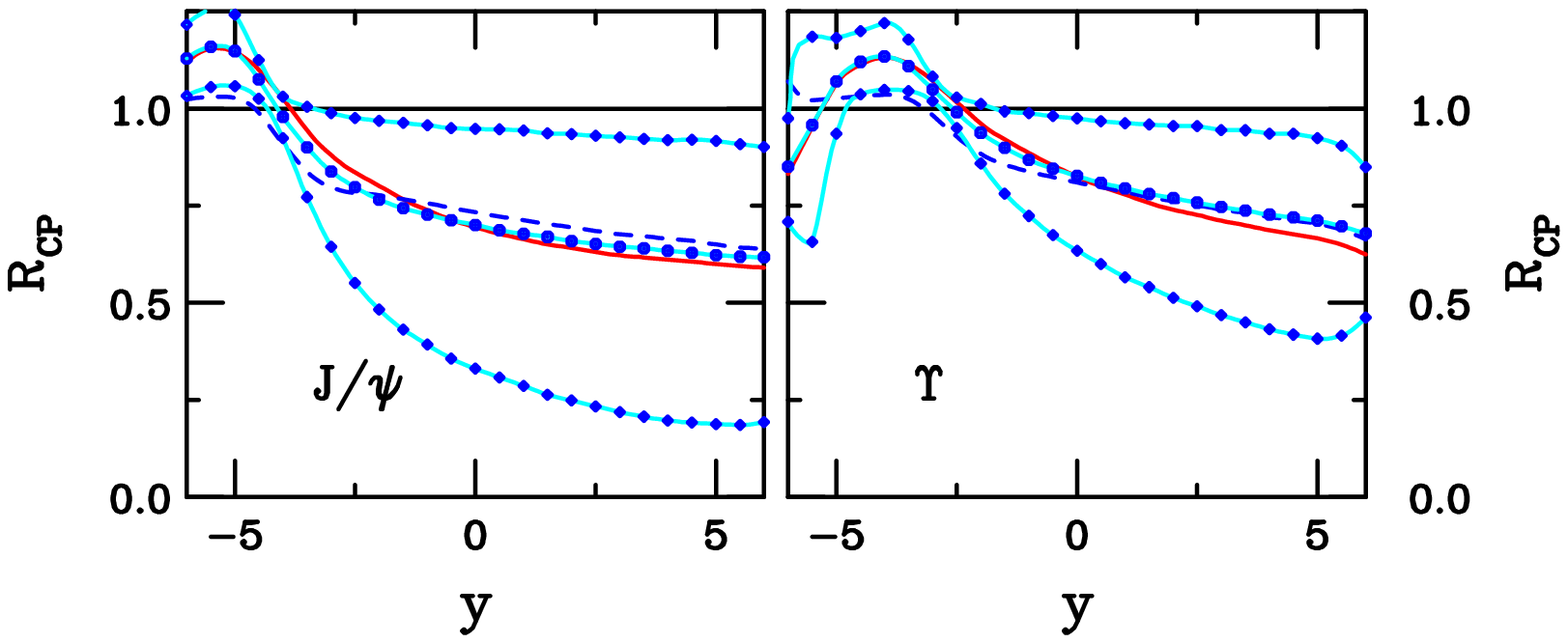}
\caption[]{(Color online)
The central-to-peripheral ratios, $R_{CP}$, as a function of rapidity
for $b_P \approx R_A$ relative to $b=0$ for d+Pb collisions at 
$\sqrt{s_{_{NN}}} = 6.2$
TeV.  The calculations are with CTEQ6 
and employ the EKS98 (solid),
nDSg (dashed) and EPS09 (solid curves with
symbols) shadowing parameterizations.
}
\label{fig19}
\end{figure}

In fact, studying $R_{CP}$ in $p+A$ and d$+A$ collisions 
could provide a direct measure
of shadowing if absorption is negligible since higher-order corrections
unrelated to shadowing cancel in the ratio \cite{KVpsi}.
As an example of an asymmetric system, Fig.~\ref{fig19} presents 
$R_{CP}(y)$ for d+Pb collisions with $b_C = 0$
and $b_P \approx R_A$\footnote{We do not show $R_{CP}$ for $p+A$ collisions.}.  
As expected, the resulting $R_{CP}(y)$ are very similar to
the impact-parameter averaged (d$+A)/(p+p)$ ratios shown in Fig.~\ref{fig11}.
Since $R_{CP}(y)$ with $b_P \approx 2R_A$ are similar to those in
Fig.~\ref{fig19}, they are not shown.

\begin{figure}[htbp]
\begin{center}
\includegraphics[width=0.5\textwidth]{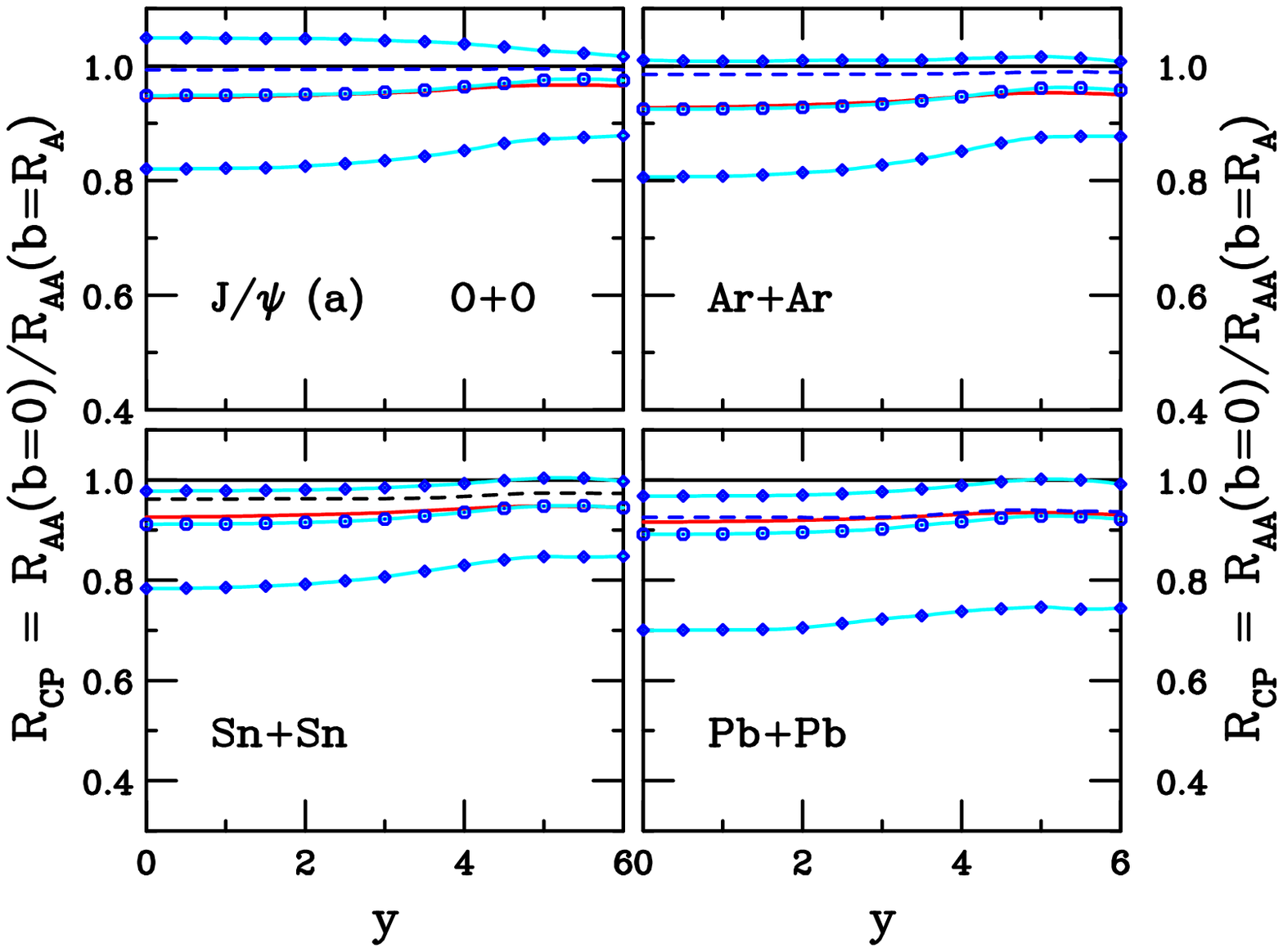}
\\
\includegraphics[width=0.5\textwidth]{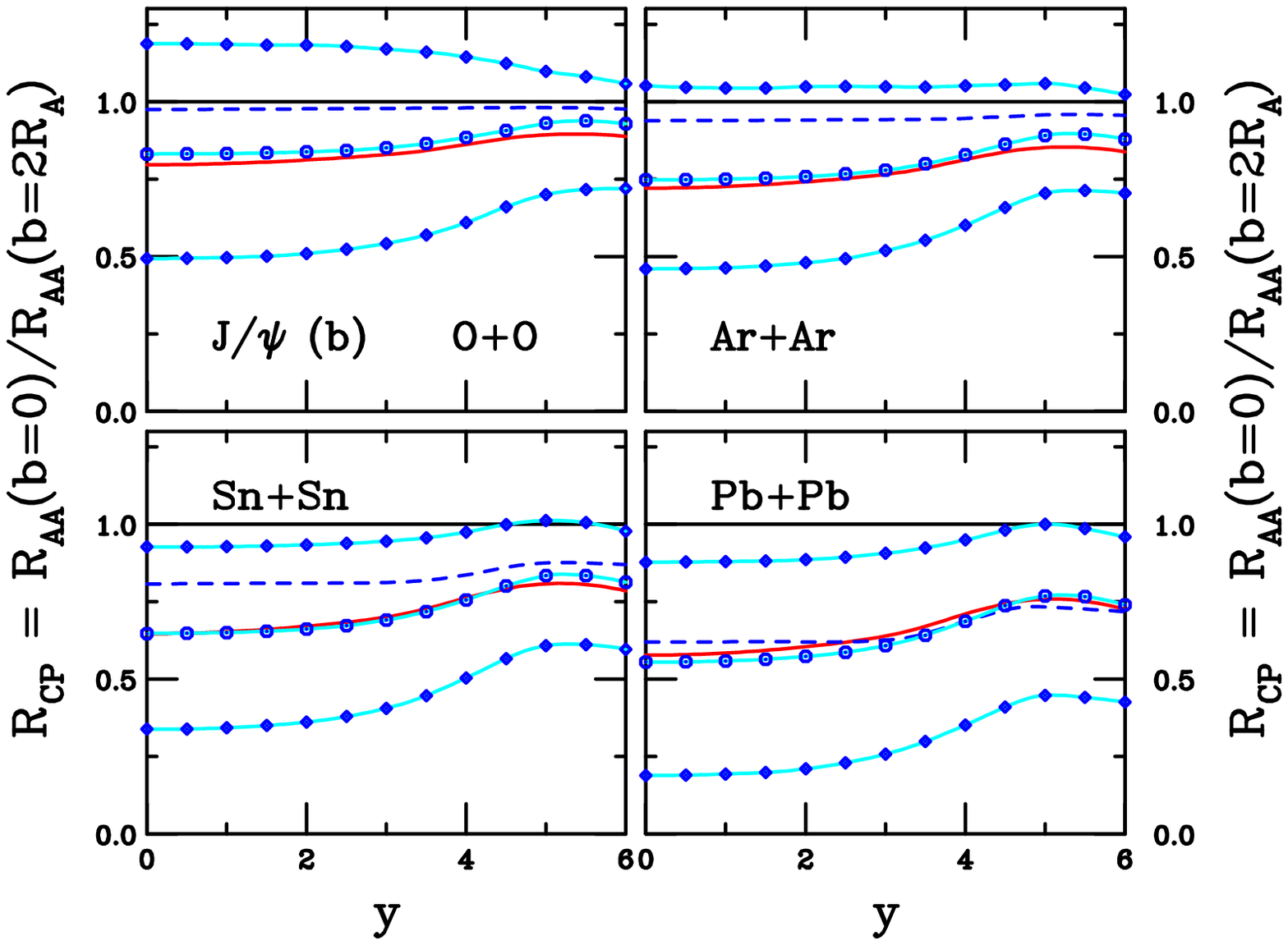}
\end{center}
\caption[]{(Color online) The central-to-peripheral ratios, $R_{CP}(y)$,
for $b = R_A$ (a) and $b = 2R_A$ (b) 
relative to $b=0$.  The 
effect of shadowing on $J/\psi$ production is 
shown for O+O at $\sqrt{s_{_{NN}}} = 7$ TeV (upper left), Ar+Ar at 
$\sqrt{s_{_{NN}}} = 6.3$ TeV 
(upper right), Sn+Sn at $\sqrt{s_{_{NN}}} = 6.14$ TeV (lower left) and Pb+Pb at
$\sqrt{s_{_{NN}}} = 5.5$ TeV (lower right).  The calculations are with CTEQ6 
and employ the EKS98 (solid),
nDSg (dashed) and EPS09 (solid curves with symbols) 
shadowing parameterizations.
}
\label{fig16}
\end{figure}

\begin{figure}[htbp]
\begin{center}
\includegraphics[width=0.5\textwidth]{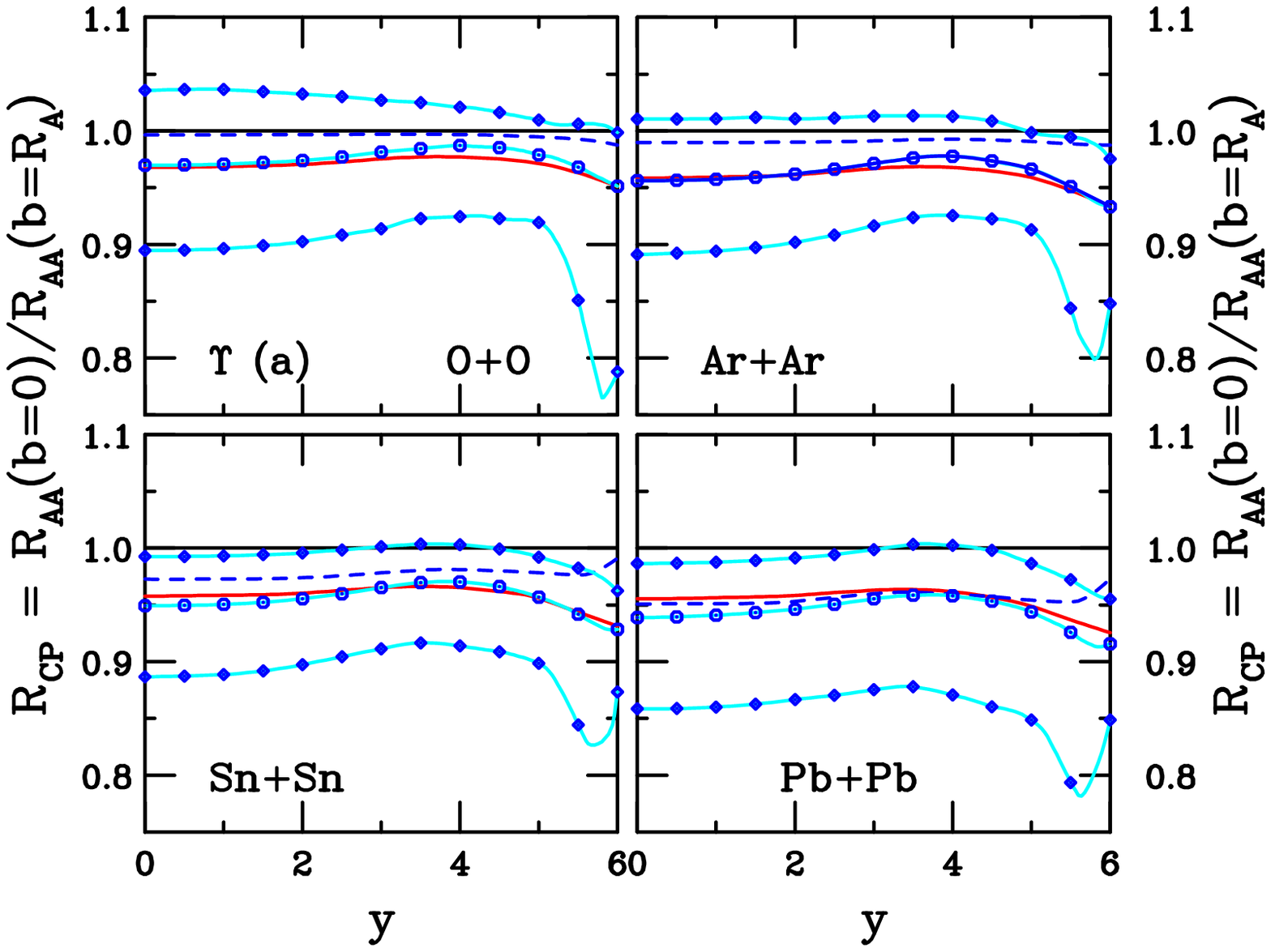}
\\
\includegraphics[width=0.5\textwidth]{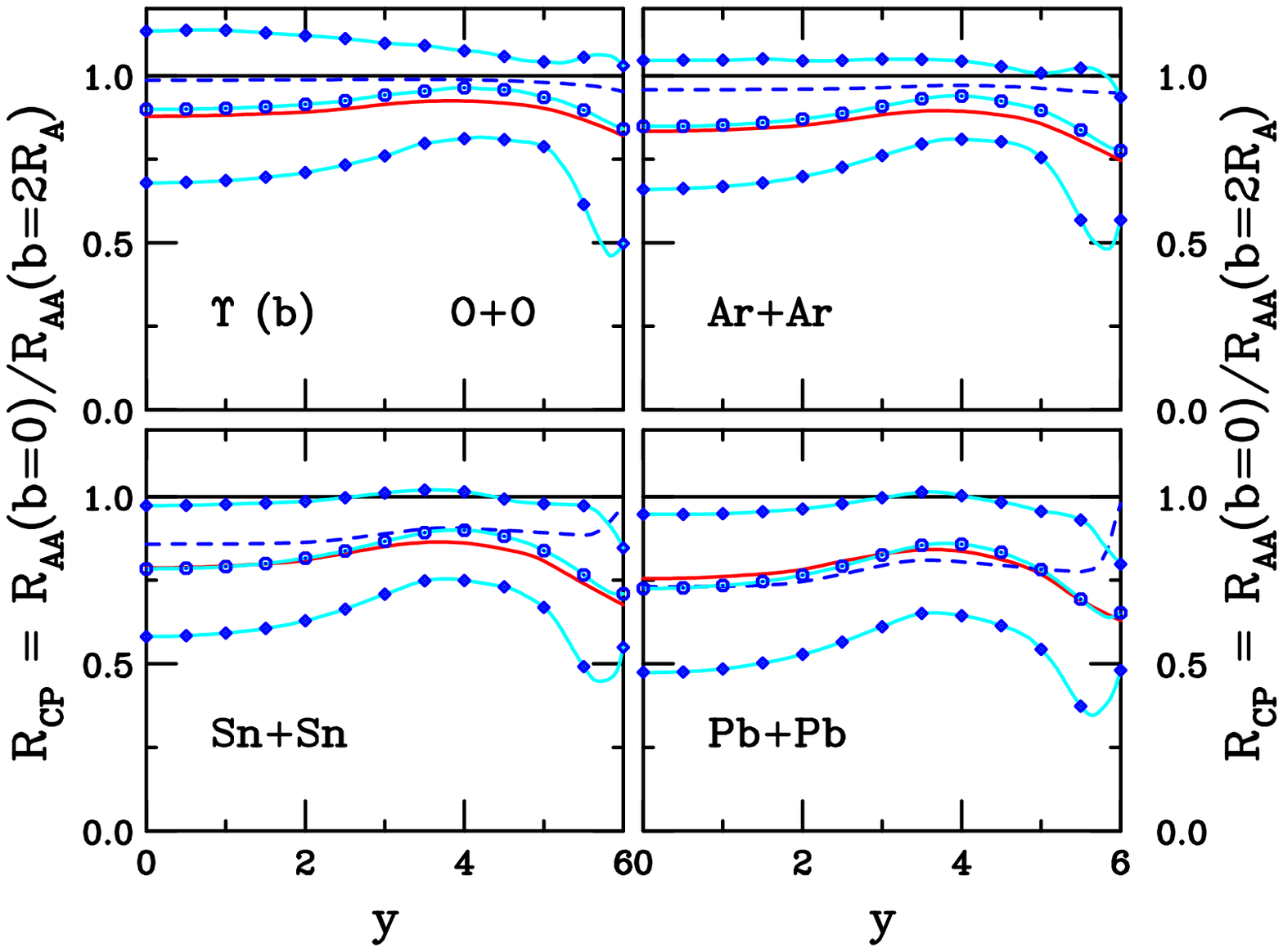}
\end{center}
\caption[]{(Color online) The central-to-peripheral ratios, $R_{CP}(y)$,
for $b = R_A$ (a) and $b = 2R_A$ (b) relative to $b=0$.  The 
effect of shadowing on $\Upsilon$ production is 
shown for O+O at $\sqrt{s_{_{NN}}} = 7$ TeV (upper left), Ar+Ar at 
$\sqrt{s_{_{NN}}} = 6.3$ TeV 
(upper right), Sn+Sn at $\sqrt{s_{_{NN}}} = 6.14$ TeV (lower left) and Pb+Pb at
$\sqrt{s_{_{NN}}} = 5.5$ TeV (lower right).  The calculations are with CTEQ6 
and employ the EKS98 (solid),
nDSg (dashed) and EPS09 (solid curves with symbols) 
shadowing parameterizations.
}
\label{fig20}
\end{figure}

Figures~\ref{fig16} and \ref{fig20} show the values of $R_{CP}$
for $b_P \approx R_A$ and $2R_A$ relative to $b_C=0$ in the four $A+A$ systems 
studied for the $J/\psi$ (Fig.~\ref{fig16}) and the $\Upsilon$ 
(Fig.~\ref{fig20}).  Since the change
in $R_{AA}(N_{\rm part})$ between $b_C=0$ and $b_P \approx R_A$ is small 
(see Fig.~\ref{fig15}), these ratios are almost independent of rapidity 
and give $R_{CP}$ close to unity.  
On the other hand, the weaker shadowing effect at $b_P
\approx 2R_A$ produces a stronger rapidity dependence and
a lower $R_{CP}$.  Note that, as in Fig.~\ref{fig19},
$R_{CP}(y)$ for $b_P \approx 2R_A$ is similar to $[(A+A)/(p+p)_{y>0}$ in 
Fig.~\ref{fig13}.  Thus, if no other medium effects are present, 
it is possible to trace the shadowing effect
rather accurately by determining $R_{CP}$ for sufficiently narrow centrality
bins.

\section{Summary}

We have provided a survey of the quarkonium total cross sections to 
next-to-leading order in the color evaporation model for all $A+B$ combinations
and energies at the LHC.  We have included initial-state shadowing, employing 
several parameterizations  of the nuclear modifications of the parton densities,
but assumed final-state absorption is negligible.  If the nuclear
absorption of quarkonium production can indeed be ignored at LHC energies,
it may be possible to use the different mass scales for $J/\psi$ and $\Upsilon$
production to study the scale dependence of the gluon density 
in the nucleus as well as in the proton.  
There are considerable uncertainties in the predictions due to the incomplete
knowledge of the nuclear gluon distribution.  Indeed, at midrapidity, the
range of the EPS09 $(p+A)/(p+p)$ ratios differs by a factor of two.

To illustrate the range of predictions for the different systems, we 
have calculated $(p+A)/(p+p)$ and (d$+A)/(p+p)$ ratios from the most naive (both
systems at the same energy) to the most realistic (the $p+p$ reference at 14 TeV
and the rapidity shift of $p+A$ interactions in the equal-speed frame).
The most naive ratios are most straightforward for extracting the nuclear gluon 
distributions.  It is still possible to use the most realistic ratios 
using a combination of experimental cuts on the rapidity distributions when
$\Delta y_{\rm cm}^{iA} \neq 0$ and modeling the appropriate $x$ values 
for comparing $p+p$
collisions at $\sqrt{s} = 14$ TeV with lower energy $p+A$ collisions.  
As is clear from the RHIC analyses \cite{rhicii}, the $A+A$ studies require a
good understanding
of the nuclear gluon distribution to extract hot and dense matter effects.
Thus the more realistic d$+A$ 
scenario shown in Fig.~\ref{fig12} would be preferable for determining 
nuclear effects on the parton densities both because of the relatively
similar d$+A$ and $A+A$ center-of-mass energies as well as the smaller 
rapidity shift relative to $p+A$ 
collisions in the equal-speed frame.  

To more cleanly extract the parton densities at LHC energies, it would be
preferable to have $e+p$ and $e+A$ data at the appropriate $x$ and $\mu^2$ range
of the LHC data.  (The HERA $x$ range reaches to approximately the value
appropriate for $J/\psi$ production in 5.5 TeV/nucleon collisions at 
midrapidity.  Unfortunately, the $\mu^2$ probed at these $x$ values is smaller
than the $J/\psi$ mass scale.)  
So far, the nDIS data is not available at small enough $x$
values and, simultaneously, large enough $\mu^2$ to be relevant for quarkonium
production at high energies.  While electron-proton collisions, as studied at
HERA, would be useful for obtaining the baseline in $p+p$, they are not 
sufficient for defining the modification of the nuclear gluon distributions
for $p+A$ collisions, $e+A$ studies are needed.  

The shadowing parameterizations used in our study exhibit a wide
range of behavior for the nuclear gluon density at low $x$, outside the
current range of the fits from fixed-target nDIS
data at higher $x$ and low $\mu^2$.  If nuclear data
were available from high energy
$e+A$ collisions, the nuclear gluon densities could be
more precisely pinned down by global analyses of
the scale dependence of the nuclear structure functions.  In hadroproduction,
direct photon or open charm production, dominated by gluon-induced processes
but without the  additional complexities of nuclear absorption, could be
utilized to study the nuclear gluon density.  
Any new $e+A$ data before an electron ring is available at the LHC will be
at lower energies than previously available at HERA, reducing the potential
overlap of the low $x$ range between an electron-ion collider and the LHC. 

We note that, since we have assumed
absorption is negligible at the LHC and include no other cold nuclear matter
effect, the uncertainties on the ratios can be obtained from the EPS09
bands shown in the figures.  However, if other effects are incorporated, a more
extensive error analysis, including the uncertainties on these other effects,
is necessary.

Finally, we note that the central-to-peripheral ratio, $R_{CP}$, may be useful
for extracting the shadowing effect at a given collision energy
if the experimental resolution of the impact parameter bins is 
narrow enough.  Indeed, $R_{CP}$ measurements 
may be a superior method of studying asymmetric systems
since very peripheral collisions are a good approximation to nucleon-nucleon
collisions.  This ratio is advantageous
because it can be made at the same collision energy with a common rapidity
shift.

\section*{Acknowledgements}

The numerical values of the ratios shown in this paper are available from
the author.

We thank K. J. Eskola, H. Paukkunen and C. Salgado for
providing the EPS09 files and for discussions.
This work was performed under the auspices of the U.S.\
Department of Energy by Lawrence Livermore National Laboratory under
Contract DE-AC52-07NA27344 and was also supported in part by the
National Science Foundation Grant NSF PHY-0555660.

\end{document}